\documentclass{aa}  

\usepackage{graphicx}
\usepackage{txfonts}
\usepackage{subcaption}
\usepackage[colorlinks=true,citecolor=blue, linkcolor=cyan]{hyperref}%
\usepackage{color}

\begin{document}

 \title{Optical constraints on the coldest metal-poor population % WISEA\,J153429.75-104303.3 (a.k.a. "the Accident") 
   }

   \author{
  J.-Y.\ Zhang
  \inst{1,2,3}   \fnmsep\thanks{\email{jzhang@iac.es}}
%              \and
%      P.\ Esparza \inst{4}
%        \and     F.\ Gracia \inst{1}
%       \and     J.\ L.\ Rasilla \inst{1}       
%            \and
%                   T. Masseron \inst{1,2}  
%     \and  A.\ J.\ Burgasser \inst{5} 
%                   J. Olivares \inst{1,2}                 
     \and  N. Lodieu \inst{1,2}
     \and  E.\ L.\ Mart\'in \inst{1,2}
     \and M. R. Zapatero Osorio \inst{4}
     \and V. J. S. B\'ejar \inst{1,2}
     \and V.\ D. Ivanov \inst{3}
     \and H.\ M.\ J. Boffin \inst{3}
     \and T. Shahbaz \inst{1,2}
     \and Ya.\ V.\ Pavlenko \inst{5,1}
     \and R. Rebolo \inst{1,2}
     \and B. Gauza \inst{6}
     \and N. Sedighi \inst{1,2}
     \and C. Quezada \inst{7,8,3}
 %     \newline
  %     whoelse
        }

   \institute{Instituto de Astrof\'isica de Canarias (IAC), 
Calle V\'ia L\'actea s/n, E-38200 La Laguna, Tenerife, Spain %\\
       %\email{jzhang@iac.es}
       \and
       Departamento de Astrof\'isica, Universidad de La Laguna (ULL), E-38206 La Laguna, Tenerife, Spain
     %\and
      % Centro de Astrobiolog\'ia (CSIC-INTA), Ctra. Ajalvir km 4, 28850, Torrej\'on de Ardoz, Madrid, Spain\\
     \and European Southern Observatory, Karl-Schwarzschild-Str. 2, 85748 Garching bei München, Germany
      \and
       Centro de Astrobiolog\'ia (CAB), CSIC-INTA, Camino Bajo del Castillo s/n, 28692 Villanueva de la Ca\~nada, Madrid, Spain
    %   \and
    %  Consejo Superior de Investigaciones Cient\'ificas (CSIC), E-28006 Madrid, Spain
    \and
        Main Astronomical Observatory, Academy of Sciences of the Ukraine, 27 Zabolotnoho, Kyiv 03143, Ukraine
       \and
    Janusz Gil Institute of Astronomy, University of Zielona Góra, Lubuska 2, 65-265 Zielona Góra, Poland
%      \and
 %     Departamento de Qu\'imica, Universidad de La Laguna (ULL), E-38206 La Laguna, Tenerife, Spain
 %     \and
 %     Center for Astrophysics and Space Science, University of California San Diego, 9500 Gilman Drive, La Jolla, CA 92092, USA
    \and 
    Instituto de Astrofísica, Pontificia Universidad Católica de Chile, Av. Vicuña Mackenna 4860, 782-0436 Macul, Santiago, Chile
    \and 
    Millennium Institute of Astrophysics, Av. Vicuña Mackenna 4860, 82-0436 Macul, Santiago, Chile
       }

   \date{\today{},\today{}}

% \abstract{}{}{}{}{} 
% 5 {} token are mandatory
 
  \abstract
   {The coldest metal-poor population made of T and Y dwarfs are archaeological tracers of our Galaxy because they are very old and have kept the pristine material. The optical properties of these objects are important to characterise their atmospheric properties.}
   {We aim at characterising further the optical properties of ultracool metal-poor population with deep far-red optical images and parallax determinations.
   }
   {We collected deep optical imaging of 12 metal-poor T dwarf candidates and the only potential metal-poor Y dwarf, a.k.a., the Accident using the 10.4-m Gran Telescopio Canarias, the 8.2-m European Southern Observatory Very Large Telescope, and the Dark Energy Survey. We have been monitoring the positions of five metal-poor T dwarf candidates using Calar-Alto 3.5-m telescope for two years to infer their distances. We compared these objects with a known subdwarf benchmark and solar-metallicity dwarfs in colour-magnitude and colour-colour diagrams, as well as with state-of-the-art theoretical ultracool models.}
   {We solve trigonometric parallaxes of the five metal-poor T dwarf candidates. We obtain $z'$-band photometry for the other 12 metal-poor T dwarf candidates, increasing the sample of T subdwarfs with optical photometry from 12 to 24. We report a 3-$\sigma$ limit for the Accident in five optical bands. %The absolute $z'$ magnitude of four metal-poor T dwarfs with trigonometric parallaxes are fainter than their solar-metallicity counterparts. 
   We confirm three more T subdwarfs and show that the Accident is subluminous compared to the current Y dwarf limit. Additionally, we propose two more Y subdwarf candidates. We emphasise that the $z_{PS1}-W1$ colour combining with the $W1-W2$ colour could break the metallicity-temperature degeneracy for T and possibly for Y dwarfs. The $z_{PS1}-W1$ colour shifts redward when metallicity decreases for a certain temperature, which is not predicted by models. The Accident has the reddest $z_{PS1}-W1$ colour among our sample. The $z_{PS1}-W1$ colour will be useful to search for other examples of this cold and old population in upcoming and existing deep optical and infrared large-area surveys.
   }
   {}

  \maketitle

\begin{table*}[htbp]
      \caption[]{List of metal-poor T and Y dwarf candidates in our work separated by astrometric and photometric samples. They are ordered by right ascension. The coordinates are in degrees under equinox J2000\@. MJDs are the Modified Julian dates of these coordinates. MJD = JD $-$ 2400000.5. If not specified, all the information is from the same references as those of the targets. In the rest of the paper, we subsequently use W\textit{hhmm} as abbreviations for these objects.}
         \label{tg}
         
For astrometry:
\\         
 \resizebox{\textwidth}{!}{
 \begin{tabular}{crrrcc}
            \hline
            \noalign{\smallskip}
           Full name  &  $\alpha$, $\delta$ (deg)
            %,ICRS)
            &  MJD  & $\mu_{\alpha\cos\delta,\ \delta}$ (mas/yr) & Sp.T. & Parallax (mas) \\
            
            \hline
 $^{(1)}$WISEA\,J000430.66$-$260402.3
 & $^{(12)}$1.127859, $-$26.067652 & $^{(12)}$57234.57 & $^{(12)}$$+1\pm4$, $-244\pm4$ & sdT$2.0$,$^{(12)}$T$4.0$  & $^{(12)}$$33.1\pm4.7$
           \\       
    $^{(1)}$WISEA\,J030119.39$-$231921.1 & $^{(12)}$45.331221, $-$23.322665 & $^{(12)}$57262.58  & $^{(12)}$$+234\pm3$, $-119\pm4$ & sdT$1.0$& $^{(12)}$$37.3\pm3.8$ 
           \\
           %WISE0414$-$5854 & & &  &  &  & &\\
    $^{(2)}$WISEA\,J042236.95$-$044203.5 &65.656712, $-$4.702669\ &58467.54 & $+1150\pm53$, $-695\pm55$  & T$8.0\pm1.0$  & 
           \\
        $^{(2)}$WISEA\,J155349.96$+$693355.2 & 238.445876, +69.567907& 58358.98 & $-1684\pm56$, $+1348\pm53$ & $^{(2)}$T$5.0\pm1.0$, $^{(3)}$sdT4.0  & 
           \\
        %WISE2207$-$5036 &  &  &  &  & & &\\
    $^{(3)}$CWISE\,J221706.28$-$145437.6 &  $^{(13)}$334.278402, $-$14.912140 & $^{(13)}$59826.03 & $^{(13)}$$+1257\pm44$, $-986\pm44$ & esdT$5.5\pm1.2$ &  \\
           
            \hline
            \end{tabular}}

\vspace{5mm}
For photometry:
\\
 \resizebox{\textwidth}{!}{
 \setlength{\tabcolsep}{2pt}
 \begin{tabular}{crrrccc}
            \hline
            \noalign{\smallskip}
           Full name  &  $\alpha$, $\delta$ (deg)
            %,ICRS)
            &  MJD  & $\mu_{\alpha}\cos\delta,\ \mu_{\delta}$ (mas/yr) & Sp.T. & Parallax (mas) & Notes \\
            \hline
        $^{(4)}$CWISEP\,J015613.24+325526.6   & 29.0568626, +32.9235107 & 58419.89 & $+1138\pm84$, $-373\pm84$ & sdT  & &
           \\  
        $^{(5)}$CWISE\,J021948.68+351845.3   & $^{(14)}$34.9528878, +35.3125796 & $^{(14)}$57095.63 & $+673.2\pm46.4$, $-504.4\pm57.0$ & T$9.5\pm1.5$ & $^{(15)}$$57.33\pm0.04$ & Ross\,19B
           \\  
$^{(4)}$CWISEP\,J050521.29$-$591311.7
 & 76.3403687, $-$59.2215888 &  58448.04 & $+461\pm36$, $-1003\pm35$ & T6.5  & &
           \\             
$^{(6)}$CWISE\,J052306.42$-$015355.4
 & $^{(14)}$80.7767575, $-$1.8987457 & $^{(14)}$57392.63 & $^{(14)}$$+511.6\pm53.6$, $-39.9\pm55.9$ & esdT$0.0\pm1.0$ & &
           \\  
 $^{(7)}$WISEA J071121.36$-$573634.2 & $^{(14)}$107.8390823, $-$57.6092955 & $^{(14)}$56557.08 & $+18\pm10$, $+990\pm90$  & sdT$0\pm1$ &  &
           \\  
            $^{(3)}$CWISE\,J073844.52$-$664334.6
 & $^{(14)}$114.6855096, $-$66.7262825 & $^{(14)}$57258.48 & $^{(14)}$$+768.4\pm32.0$, $-445.1\pm29.8$ & esdT$5.5\pm1.2$ & &
           \\  
    $^{(4)}$CWISEP J090536.35+740009.1 & 136.4033148, +74.0008111 & 58494.02 & $+465\pm67$, $-1490\pm71$ & T6.0 & 
    \\
    $^{(1)}$WISEA\,J101944.62$-$391151.6
 & $^{(14)}$154.9352802, $-$39.1974770 & $^{(14)}$57156.08 & $-472.2\pm28.0$, $+222.7\pm26.4$ & sdT$3.0$   & &
           \\  
    $^{(8)}$WISEA\,J153429.75$-$104303.3
 & 233.621873, $-$10.721775 & 57645.36  & $-1253.1\pm8.9$, $-2377.0\pm7.0$ & esdY?   & $38.5\pm8.4$ & The Accident
        \\
    $^{(9)}$WISEA\,J200520.35+542433.6
 & $^{(16)}$301.333016, +54.408501 & $^{(16)}$56658.00 & $^{(16)}$$-1154.4\pm1.2$, $-900.4\pm1.2$ & T8  & $^{(17)}$$59.91\pm0.55$& Wolf 1130C 
           \\      
$^{(10,19)}$WISE\,J201404.13+042408.5  & $^{(18)}$303.516175, +4.402821 & $^{(18)}$57326.30 & $^{(18)}$$-611.3\pm1.5$, $+313.4\pm1.5$ & $^{(19)}$T6.5 & $^{(12)}$$45.5\pm2.4$&
           \\  
 $^{(11)}$WISE\,J210529.08$-$623558.7 & $^{(14)}$316.3716241, $-$62.5999392 & $^{(14)}$55961.60 & $^{(14)}$$765.0\pm15.7$, $-1507.2\pm15.0$ & T1.5  & & 
\\ 
           $^{(2)}$WISEAR\,J220746.67$-$503631.7
 & 331.9449276, $-$50.6122455  & 58503.19 & $+108\pm99$, $-1380\pm98$ & T7.5  & &
           \\      
            \hline
         \end{tabular}}
        
    \tablebib{
    (1): \citet{greco2019neowise};
    (2): \citet{meisner2020extremecoldBD};
    (3): \citet{meisner2021esdT}; 
    (4): \citet{meisner2020Yspitzer};
    (5): \citet{schneider2021Ross19B}
    (6): \citet{brooks2022esdT};
    (7): \citet{kellog2018_W0711}
         (8): \citet{kirkpatrick2021accident};
         (9): \citet{mace2013Wolf1130C};
         (10): \citet{kirkpatrick2012furtherY}
         (11): \citet{luhman2014WISEhighpm}
         (12): \citet{williambest2020_L0-8UKIRTparallax};
         (13): \citet{zhangjerry2023optical_sdT};
         (14): CatWISE, \citet{marocco2021catwise};
         (15): Parallax of the primary Ross\,19A \citep{schneider2021Ross19B};
         (16): \citet{kirkpatrick2019parallax184TY};
         (17): Parallax of the primary Wolf\,1130A \citep{mace2018Wolf1130};
         (18): \citet{kirkpatrick2021census20pc};
         (19): \citet{mace2013WISE_Tpopulation}.
}
%{\raggedright \textbf{Note}: }

   \end{table*}

\section{Introduction}

Substellar objects formed at early times of our Milky Way are valuable tracers of the original chemical composition of our Galaxy. With masses below the hydrogen burning limit, which ranges from $0.075M_\odot$ for solar metallicity \citep{chabrier2023BDnewHlimit} to $0.092M_\odot$ for zero metallicity \citep{saumon1994cia}, they could not establish a stable hydrogen fusion and, thus, they retain most of their pristine metal-poor material. Due to their old age, they have emitted away most of their gravitational contraction energy. They have been cooling down to extremely low temperatures, and hence have very late spectral types: T type \citep{burgasser2002dT_spec_classification, burgasser2003dT_optical} and Y type \citep{delorme2008TY_J0059,cushing2011dY_WISE}. They normally possess high proper motions and high radial velocities,  which are typical kinematics associated with the halo or the thick-disk population of our Milky Way \citep{gizis1997pm_subdwarf,zhang2017six_sdL_classification}.

Since the first two unambiguous brown dwarfs, Teide 1 and Gliese 229B, were discovered three decades ago \citep{rebolo1995discovery, nakajima1995discovery}, more than a thousand of substellar objects with a wide range of properties have been found. On the other hand, the first metal-poor brown dwarfs, or subdwarfs, were announced eight years later with the discovery of the first L subdwarf \citep{burgasser2003esdL7_2M0532}. Later on, dozens of L subdwarfs have been found and classified \citep{sivarani2009sdL,cushing2009haloL,burgasser2009optNIRsdL,lodieu2010GTCsdL,kirkpatrick2014allwise_sd,zhang2017six_sdL_classification,zhangzenghua2018subL}. Few metal-poor T dwarf candidates have been revealed in the last ten years \citep{burningham2010cool_binary_sdL,scholz2010ULAS1416,murray_burningham2011blueT, mace2013Wolf1130C, pinfield2014subdwarf,  burningham2014T6.5, kellog2018_W0711, Schneider2020W0414_W1810, greco2019neowise,zhangzenghua2019metal-poor_sdT, meisner2020extremecoldBD, meisner2021esdT,schneider2021Ross19B, brooks2022esdT, burgasser2024esdT}. The classification of the T subdwarf population is still under development. The only metal-poor Y dwarf candidate to date, known as 'the Accident', was announced just recently by \citet{kirkpatrick2021accident}.

Most cold substellar objects, whether metal-poor or not, are primarily characterised in the infrared (IR) wavelengths where their spectral energy distributions peak, in accordance with their temperatures and Planck's radiation law. Our previous work \citep{zhangjerry2023optical_sdT} demonstrates that the far-red optical window can provide a unique window for constraining the metallicity of this cold population, while \citet{martin2024Yoptical} shows that the coldest population of Y dwarfs are actually brighter in the optical than what models predict.

In this paper, we present the results from two-year ground-based parallax measurements of five metal-poor T dwarf candidates. We expand our previous work \citep{zhangjerry2023optical_sdT} by obtaining $z$-band photometry of an extra 12 metal-poor T dwarf candidates, increasing the size of the metal-poor T dwarf sample from 12 to 24\@.  We also present photometric constraints in five optical bands on the Accident.  Section~\ref{obssec} describes the sample selection, the observation logs, and data reduction procedures. Section~\ref{results} analyses and discusses the data. Section~\ref{conclusion} provides a summary and gives a prospective of the impact of this research.

\section{Observations and data reduction}
\label{obssec}

\subsection{Astrometry}

\subsubsection{Sample}
We selected six metal-poor T dwarf candidates to measure their trigonometric parallax: WISE0004 and WISE0301 \citep{greco2019neowise}, WISE0422 and WISE1553 \citep{meisner2020extremecoldBD}, WISE2217 \citep{meisner2021esdT}, and WISEA\,J181006.18$-$101000.5 \citep{Schneider2020W0414_W1810}. They all have $z$-band photometry from our previous work \citep{zhangjerry2023optical_sdT}. WISE1810 was measured to be the closest metal-poor brown dwarf with a distance to our Solar system of $8.9^{+0.7}_{-0.6}$ pc \citep{lodieu2022W1810}. In this chapter we focus on the remaining five objects, with their details listed in the upper part of Table~\ref{tg}. 

\subsubsection{Astrometric observation details}
\label{obs_astdetail}

We used Omega2000 \citep{omega2000,omega2000_cryo,charact_omega2000} on the 3.5-m telescope at the Calar Alto Observatory in Andalucía, Spain. Omega2000 is an imager in the near-infrared (NIR) range equipped with a single $2048\times2048$ HAWAII-2 HgCdTe detector, covering a field of view (FoV) of $15\farcm5\times15\farcm4$. Its optics has very low dispersion, which allows accurate astrometry.

We carried out a observing campaign with a baseline of two years from 2021 to 2022 (Table~\ref{astrometry_obstable}). We observed in the $J$ band, under a monthly cadence, with a maximum seeing of 1\farcs2, a maximum airmass of 1.8, and no constraint on the moon phase (programme numbers H20-3.5-020, F21-3.5-010, 22A-3.5-010, 22B-3.5-010; PI N. Lodieu). We set 9 to 10-point dithering pattern (and repeated
it as necessary) for sky subtraction in the NIR, with different single exposure times for each observing block (OB) depending on the brightness of the object. The total exposure times were adjusted several times according to the results obtained in the previous semester.

%For the faintest source W2217, we added another epoch from GTC/OSIRIS $z'$-band photometry from \citet{zhangjerry2023optical_sdT}.
%and another one in the $Ks$ band obtained from GTC/EMIR (programme GTC37-23B, PI J.-Y. Zhang, excecuted on 25th July 2024 under visitor mode).  

\begin{table}[htbp]
      \centering
      \caption[]{Logs of the Omega2000 observations of five metal-poor T dwarf candidates. The data with notes 'bad' or 'no detection' are not used to solve the astrometry.}
         \label{astrometry_obstable}
         \footnotesize
         \begin{tabular}{ccccc}
            \hline
            \noalign{\smallskip}
            Name  &   MJD  & Fil.  & Exp. & Notes  \\
            \hline
W0004 & 59396.16  & $J$ & 30s$\times$18 &
           \\
    &   59423.15  & $J$ & 30s$\times$18 &
           \\
     & 59478.01  & $J$ &  30s$\times$9 &
           \\
     & 59506.89  & $J$ & 30s$\times$9 &
           \\
     & 59774.15  & $J$ & 30s$\times$9 &
           \\
     & 59802.10  & $J$ & 30s$\times$9 &
           \\
     & 59829.01  & $J$ & 30s$\times$9 &
           \\
     & 59856.89  & $J$ & 30s$\times$9 &
           \\
           \hline
W0301 & 59242.78  & $J$ & 30s$\times$9  &
           \\
    & 59269.78  & $J$ & 30s$\times$9  &
           \\
     & 59423.16  & $J$ & 30s$\times$18 &
           \\
     & 59478.15  & $J$ & 30s$\times$9 &
           \\
     & 59507.07  & $J$ & 30s$\times$9 &
           \\
     & 59598.77  & $J$ & 30s$\times$9 &
           \\
     & 59626.78  & $J$ & 30s$\times$9 &
           \\
     & 59802.17  & $J$ & 30s$\times$9 &
           \\
    & 59830.09  & $J$ & 30s$\times$9 &
           \\
           \hline
W0422 & 59242.78   & $J$ & 150s$\times$10 & bad
           \\
     & 59269.79  & $J$ & 150s$\times$10 &
           \\
     & 59297.82  & $J$ &  150s$\times$10 &
           \\
     & 59480.16  & $J$ & 150s$\times$20 &
           \\ 
     & 59507.09  & $J$ & 150s$\times$20 &
           \\
     & 59598.79  & $J$ & 150s$\times$20 &
           \\
     & 59626.81  & $J$ & 150s$\times$20 & bad
           \\ 
     & 59830.16  & $J$ & 150s$\times$20 &
           \\ 
     %& 59844.19 &GTC/OSIRIS & $z'$ &  &
     %      \\ 
           \hline
W1553 & 59242.78  & $J$ & 150s$\times$10 &
           \\
    & 59297.82  & $J$ & 150s$\times$10 &
           \\
     & 59325.14  & $J$ & 150s$\times$10 & bad
           \\
     & 59395.96  & $J$ & 50s$\times$100 &
           \\
     & 59506.83  & $J$ & 150s$\times$20 &
           \\
     & 59571.22  & $J$ & 150s$\times$20 &
           \\
     & 59598.22  & $J$ & 150s$\times$20 &
           \\
     & 59627.19  & $J$ & 150s$\times$20 &
           \\
    & 59716.08  & $J$ & 150s$\times$20 &
           \\
    %& 59725.02 &GTC/OSIRIS & $z'$ &  &
    %       \\
     & 59746.06  & $J$ & 150s$\times$20 &
           \\
     & 59772.85  & $J$ & 150s$\times$20 &
           \\
     & 59801.87  & $J$ & 150s$\times$20 &
           \\
     & 59828.86  & $J$ & 150s$\times$20 &
           \\
     & 59856.82  & $J$ & 150s$\times$20 &
           \\
           \hline
W2217 & 59396.12  & $K_s$ & 30s$\times$60 & no det.
           \\
      & 59423.01  & $J$ & 150s$\times$20 & bad
           \\
    & 59477.97  & $J$ & 150s$\times$20 &
           \\
     & 59480.04  & $J$ & 150s$\times$20 &
           \\
     & 59506.85  & $J$ & 150s$\times$20 &
           \\
     & 59716.14  & $J$ & 150s$\times$20 &
           \\
     & 59746.12  & $J$ & 150s$\times$20 &
           \\
     & 59774.11  & $J$ & 150s$\times$20 &
           \\
     & 59802.06  & $J$ & 150s$\times$20 & bad
           \\
    % & 59826.03 &GTC/OSIRIS & $z'$ & 50s$\times$40 & 
    %       \\
     & 59828.90  & $J$ & 150s$\times$20 & bad
           \\
     & 59856.85  & $J$ & 150s$\times$20 &
           \\
%     & 60517.10 &GTC/EMIR & $K_s$ & 5s$\times$273 &
%           \\
            \hline
         \end{tabular}
   \end{table}

%\begin{table}[htbp]
%      \centering
%      \caption[]{Our astrometry solutions of five metal-poor T dwarf candidates.}
%         \label{astrometry_result}
         
%         \begin{tabular}{cccc}
%            \hline
%            \noalign{\smallskip}
%            Name  &  $\mu_{\alpha\cos{\delta}}$ (mas/yr)  & $\mu_{\delta}$ (mas/yr) & $\varpi$ (mas) \\
%            \hline
%W0004 & $-5.5\pm12.2$ & $-248.0\pm9.7$ & $32.3\pm9.8$ 
%           \\
%W0301 &  $252.9\pm16.7$& $125.5\pm16.6$& $26.4\pm9.8$
%           \\
%W0422 & $+1286.4\pm35.2$ & $-708.2\pm50.4$ & $46.6\pm24.2$
%           \\
%W1553 & $-1526.7\pm12.4$ & $+1258.4\pm12.7$ & $38.5\pm8.4$ 
%           \\
%W2217 & $+1352.4\pm22.8$ & $-843.8\pm23.5$ & $48.0\pm13.2$ 
%           \\
%            \hline
%         \end{tabular}
%   \end{table}

\subsubsection{Astrometric data reduction}

We reduced the $J$-band images using the Image Reduction and Analysis Facility \citep[IRAF;][]{tody1986iraf, tody1993iraf}. We bias and flat-corrected all the individual exposures using the master bias and master sky flat frames of the same night. We created the sky frame for each dithering point by median combining all the images of the other dithering positions in the same dithering cycle using the task \textit{imcombine}, with scale parameter set to 'mode'. We subtracted the sky from each image using the task \textit{imarith}. All the sky-subtracted images of the same epoch were then averaged via \textit{imcombine}, taking into account the offsets. The pixels with the highest and the lowest deviating counts were rejected. 

We used the task \textit{daofind} with a threshold of 10 $\sigma$, an average full width half maximum (FWHM) of point source, and an average background fluctuation of each image to extract the point sources with high signal-to-noise-ratio (S/N). These high-S/N point sources are key to determine the shift, rotation, magnification of the field, as well as the distortion. Our targets, on the other hand, are normally faint, so we used \textit{imcentroid} to locate the target if it was not extracted with the 10-$\sigma$ threshold. We then used the task \textit{xyxymatch} to match all the extracted high-S/N sources (about a hundred) between the reference epoch and each other epoch. In the next step, we used \textit{geomap} to fit the pixel-level transformation between epochs, which applied a third-order polynomial in x and y, and computed linear terms and distortions terms separately. It is an iterative process so sources deviate from the fit (e.g., defects, high-parallax stars, sources at the edges) will be discarded. We then used \textit{geoxytran} to transform the target pixel coordinates of the different epochs to the reference frame of the reference epoch (0.0 year in Fig.~\ref{fig:parallax}). At the end, we obtained the absolute differences of the target coordinates in pixels between epochs, and we multiplied them with the pixel scale.  The pixel scale of Omega2000 obtained for the images of W1810 \citep[$449.45\pm0.45$ mas/pix;][]{lodieu2022W1810} was used for all the five sources studied in this paper. The errors in both axes are the quadratic sums of the centroid error and the residuals from the fitting of the task \textit{geomap}. We note that our measurements are free from uncertainties associated with individual-epoch astrometry. No correction from relative to absolute parallax was applied since we statistically created good references using sufficient amount of bright distant sources. We verified for all the five targets, most of their reference sources have Gaia parallaxes averaging about 1 mas with approximately 1 mas of standard deviation. The corrections would be much smaller than the error budgets.

The difference of the coordinates in RA $\alpha$ and Dec $\delta$ should follow
\[
    \Delta\alpha\cdot\cos\delta = \mu_{\alpha\cos\delta} \left(t_i-t_0\right) + \varpi \left(f^\alpha_i-f^\alpha_0\right) + k_\alpha;
\]
\[
    \Delta\delta = \mu_{\delta} \left(t_i-t_0\right) + \varpi \left(f^\delta_i-f^\delta_0\right) + k_\delta,
\]
where $t$ is the time in year, $0$ and $i$ stand for the reference epoch and the $i$-th epoch, $\mu$ is the proper motion in mas/yr, $\varpi$ is the parallax in mas, $f^\alpha$ and $f^\delta$ are the parallax factors for the right ascension and declination. The parallax factors were computed using the Earth geocenter as obtained from the JPL DE441 solar system ephemeris \citep{park2021JPL_DE441}, and object coordinates from Table~\ref{tg}. $k$ is a small offset in both axes that allows better statistical solutions. We solve the astrometry by applying a maximum likelihood estimation method. We visualise the fitting in Fig.~\ref{fig:parallax}, and the results including the reduced chi-square values $\chi^2_{\nu}$ are given in Table~\ref{astrometry_result}. We report three completely new parallaxes for metal-poor T dwarf candidates: W0422, W1553, and W2217. We note that W0422's astrometry solution is rather poor and has a high  $\chi^2_{\nu}$ value. Our results for the two brightest objects, W0004 and W0301, are consistent at the 1-$\sigma$ level with the values obtained by \citet{williambest2020_L0-8UKIRTparallax} shown in Table~\ref{tg}. For these two objects, we will use our parallax determinations for the remainder of the analysis, noting that using \citet{williambest2020_L0-8UKIRTparallax} values will not alter the following conclusions.

\begin{table}[h]
      \centering
      \caption[]{The astrometry solutions of five metal-poor T dwarf candidates.}
         \label{astrometry_result}
         
         \begin{tabular}{ccccc}
            \hline
            \noalign{\smallskip}
            Name  &  $\mu_{\alpha\cos{\delta}}$ (mas/yr)  & $\mu_{\delta}$ (mas/yr) & $\varpi$ (mas) & $\chi^2_\nu$\\
            \hline
W0004 & $-6\pm12$ & $-248\pm10$ & $32\pm10$ & 0.47
           \\
W0301 &  $253\pm17$& $126\pm17$& $26\pm10$ & 0.85
           \\
W0422 & $+1286\pm35$ & $-708\pm50$ & $47\pm24$ & 4.21
           \\
W1553 & $-1527\pm12$ & $+1258\pm13$ & $39\pm8$ & 0.94
           \\
W2217 & $+1352\pm23$ & $-844\pm24$ & $48\pm13$ & 0.64
           \\
            \hline
         \end{tabular}
   \end{table}

   \begin{table*}[htbp]
      \centering
      \caption[]{Logs of
new optical observations of the Accident and the other 11 metal-poor T dwarf candidates.}
         \label{obs_pho_table}
         
         \begin{tabular}{ccccccc}
            \hline
            \noalign{\smallskip}
            Name  &   MJD & Seeing & Moon &  Instrument & Filters  & On-source integration time   \\
            \hline
%  The accident & 60379.20 & 0\farcs8 & Dark &GTC/HiPERCAM & $u'g'r'i'z'$ & 60.9s$\times$72$=$4384.8s & possible light contamination
           %\\
  The Accident & 60466.07 & 0\farcs8 & Dark &GTC/HiPERCAM & $u'g'r'i'z'$ & 60.5s$\times$124$=$7502s 
           \\
  The Accident & 60473.98 & 0\farcs8 & Grey-Dark &GTC/HiPERCAM & $u'g'r'i'z'$ & 60.9s$\times$180$=$10962s  
 \\
W0156 & 60565.35 & 0\farcs6 & Dark &VLT/FORS2 & $z_{\mathrm{Gunn}}$ & 120s$\times$24=2880s 
           \\
Ross\,19B & 60565.30 & 0\farcs5 & Dark &VLT/FORS2 & $z_{\mathrm{Gunn}}$ & 120s$\times$24=2880s 
           \\
W0505 & 60578.21 & 0\farcs7 & Grey &VLT/FORS2 & $z_{\mathrm{Gunn}}$ & 120s$\times$20=2400s 
           \\
W0523 & 60227.20 & 0\farcs8 & Grey &GTC/OSIRIS+ & $z'$ & 60s$\times$48=2880s 
           \\
W0711 & 60565.38 &  0\farcs6 & Dark &VLT/FORS2 & $z_{\mathrm{Gunn}}$ & 60s$\times$8=480s 
           \\
W0738 & 60407.13 & 0\farcs6 & Dark &VLT/FORS2 & $z_{\mathrm{Gunn}}$ & 120s$\times$16=1920s 
           \\
W0905 & 60606.19 & 1\farcs0 & Bright & GTC/OSIRIS+ & $z'$ & 60s$\times$48=2880s 
\\
W1019 & 60468.06 & 0\farcs7 & Dark &VLT/FORS2 & $z_{\mathrm{Gunn}}$ & 60s$\times$8=480s 
           \\
Wolf\,1130C & 60210.97 & 0\farcs8 & Grey &GTC/OSIRIS+ & $z'$ & 60s$\times$48=2880s  
\\
W2014 & 60562.22 &  0\farcs9 & Dark &VLT/FORS2 & $z_{\mathrm{Gunn}}$ & 120s$\times$16=1920s 
           \\
W2207 & 60622.16 & 0\farcs7 & Dark &VLT/FORS2 & $z_{\mathrm{Gunn}}$ & 120s$\times$24=2880s 
\\
            \hline
         \end{tabular}
   \end{table*}

\subsection{Optical photometry}
\subsubsection{Sample}

Our sample for far-red optical photometry consists of 12 metal-poor T dwarf candidates from the literature, complementing the 12 metal-poor T dwarf candidates in our previous work \citep{zhangjerry2023optical_sdT}:

\begin{itemize}
 \item W0156, W0505, and W0905 were identified because of their red $J-ch2$ and blue $ch1-ch2$ Spitzer colours \citep{meisner2020Yspitzer}. A red $J-ch2$ colour suggests low temperature \citep{Schneider2015HubbleWISE_BD,meisner2023coldoldBD}. W0505 has a large reduced proper motion and a high tangential velocity \citep{meisner2023coldoldBD}. W0905 also has a high proper motion \citep{meisner2020Yspitzer}.
 
  \item W0219 (Ross\,19B) was discovered by the Backyard Worlds: Planet 9 citizen science project \citep{kuchner2017backyard} to be a comoving companion to Ross\,19A, which is an M3.5 subdwarf with a metallicity of [Fe/H]$=-0.40\pm0.12$~dex \citep{schneider2021Ross19B}. Ross\,19B has an estimated spectral type of T$9.5\pm1.5$ \citep{schneider2021Ross19B}, though spectroscopic confirmation is still lacking.
  
\item W0523 was discovered by \citet{goodman2021W0523} and \citet{brooks2022esdT} and was classified as an extreme T subdwarf candidate based on its distinctive infrared colours and high proper motion. Its locus in the $W1-W2$ versus $J-W2$ colour-colour diagram is similar to those of extreme T subdwarf W1810 and WISEA\,J041451.67$-$585456.7 \citep{Schneider2020W0414_W1810}.

 \item W0711 was discovered by the AllWISE2 Motion Survey as a possible thick-disk early-T subdwarf, with a tentative low metallicity \citep{kellog2018_W0711}.
 
\item W0738 was discovered by \citet{meisner2021esdT}, based on its large motion and also colours similar to those of the two extreme T subdwarf W1810 and W0414 \citep{Schneider2020W0414_W1810}.

\item W1019 was identified as an early T subdwarf together with W0004 and W0301, by \citet{greco2019neowise} based on their blueish NIR spectrum.

\item The benchmark late-type T8 subdwarf W2005 (Wolf 1130C) belongs to the triple system Wolf 1130. Wolf 1130A is a M subdwarf and Wolf 1130B is an ultramassive white dwarf, tidally locked with Wolf 1130A. This system is old ($>$\,3.7 Gyr) and metal-poor, with a subsolar iron abundance derived from the M subdwarf Wolf\,1130A, although with a discrepancy from $-0.62\pm0.10$ to $-1.22\pm0.24$ dex \citep{woolf_wallerstein2006feh_M_molecule,mace2013Wolf1130C,newton2014feh447dM,mace2018Wolf1130}. 

 \item W2014 was discovered by \citet{mace2013WISE_Tpopulation} in WISE and it has a suppressed K-band spectrum. Its IR colour just falls outside the extreme T subdwarf IR colour-colour criteria \citep{meisner2023coldoldBD}. 

\item W2105 was discovered as a thick-disk or halo early T dwarf, although its $YJH$-band spectrum does not show an obvious signature of low metallicity \citep{luhman2014WISEhighpm}. \citet{meisner2023coldoldBD} shows that it lies outside but next to the extreme T subdwarf IR colour-colour criteria. It is worthwhile to have the optical photometry done to see its optical-IR colour behaviour. 

\item W2207 was identified by \citet{meisner2020extremecoldBD} as potential subdwarf owing to its high kinematics, and red $J-ch2$ colour.
\end{itemize} 

We also include the enigmatic brown dwarf WISEA\,J153429.75$-$104303.3 \citep[a.k.a. "The Accident", ][]{meisner2020extremecoldBD,kirkpatrick2021accident}. Thanks to its extremely high proper motion, it was recently found by citizen scientist Dan Caselden in the  Near-Earth Object Wide-field Infrared Survey Explorer Reactivation Mission \citep[NEOWISE;][]{mainzer2014neowise} data. It possesses peculiar NIR colours compared to normal Y dwarfs discovered so far. With robust parallax measurements (distance $16.3^{+1.4}_{-1.2}$ pc), its absolute $J$, $W2$, and $ch2$ magnitudes are in line with the coldest known Y dwarfs but the $W1$ and $ch1$ bands are abnormally bright \citep{kirkpatrick2021accident}. It is suspected to be the first and the only metal-poor Y dwarf to date: it is cold, with an effective temperature of $T_{\mathrm{eff}}=$ 400 -- 550 K,  while the T/Y transition temperature is $T_{\mathrm{eff}}\approx 485$ K for solar-metallicity dwarfs \citep{leggett2021coldestSED}; and is low in metallicity, likely to be lower than $- 1.0$ dex \citep{kirkpatrick2021accident,meisner2023coldoldBD}. All metal-poor T and Y dwarf candidates with optical photometry are also listed in Table~\ref{tg}.

\subsubsection{Photometric observation details}
\label{obs_pho_detail}
We downloaded co-added $z$-band images of W0505, W2105, and W2207 from the Dark Energy Survey \citep[DES;][]{abbott2021DESdr2} archive. However, there are no detection for W0505 or W2207.%, so we adopted the single-epoch DES 3-$\sigma$ $z$-band limit for these two targets, which is about 23.4 mag \citep[under the AB system, ][]{oke1974ABsystem}, as inferred from the 10-$\sigma$ limit 22.1 mag \citep{morganson2018DESpipeline,abbott2021DESdr2}. 
We requested deeper imaging of these two objects, described in the following paragraphs. For W2105, the motion is clearly seen in the different epochs. We adopted the DR1 $z_{DES}$ magnitude \citep{abbott2019DESdr1}, since in the DR2 catalogue W2105 was recognised as three different objects in the multi-epoch co-added images. 

We collected $z$-band images of three metal-poor T dwarf candidates in the northern hemisphere:  W0523, W0905 and Wolf\,1130C with the upgraded Optical System for Imaging and low-Intermediate-Resolution Integrated Spectroscopy (OSIRIS$+$) on the GTC (programmes GTC13-23B \& GTC31-24B; PI J.-Y. Zhang). OSIRIS$+$ is an imager and spectrograph for the optical wavelength range, an upgraded version of OSIRIS \citep{cepa2000osiris}, located in the Cassegrain focus of GTC. It is equipped with a $4096\times4096$ deep-depleted e2v CCD231-842 (astro-2 coating) detector, with an unvignetted field of view of $7\farcm8\times7\farcm8$ and a $2\times2$-binned pixel scale of 0\farcs254/pix. 

We also collected $z$-band images of eight objects: W0156, Ross\,19B, W0505, W0711, W0738, W1019, W2014 and W2207 using the FOcal Reducer/low dispersion Spectrograph 2 \citep[FORS2; ][]{appenzeller1998fors} mounted on the Cassegrain focus of the 8.2-m Antu Unit Telescope (UT1) of the European Southern Observatory (ESO) Very Large Telescope (VLT) at Cerro Paranal, Chile, through ESO programme 113.2688 (PI J.-Y. Zhang). FORS2 is an optical imager, polarimeter and spectrograph. It is equipped with a mosaic of two $2048\times4096$ MIT/LL CCID-20, backside illuminated, AR coated CCDs: chip 1 and chip 2. We used the high resolution collimator which produces an FoV of $4\farcm25\times4\farcm25$ and a pixel scale of 0\farcs125/pix with the standard $2\times2$ binned readout modes. We always put the target on the chip 1 of FORS2. We used the $z_{\mathrm{Gunn}}$ filter, which is similar to the Sloan $z'$ filter. 

We requested a four-point dithering pattern (only one cycle) with several contiguous exposures at each point. For the GTC we set the single exposure time to 60s and for the VLT we set it to 120s to avoid saturation because of the sky background. Except for W0711 and W1019, we set an single exposure time of 60s on the VLT and without dithering to save the overheads, since they are clearly detected in a single exposure.

We collected photometry of the Accident simultaneously in the $u'g'r'i'z'$ bands using the High PERformance CAMera \citep[HiPERCAM;][]{dhillon2021hipercam} on the 10.4-m Gran Telescopio Canarias (GTC) at Roque de los Muchachos Observatory on the island of La Palma, Spain, programme GTC51-24A (PI J.-Y. Zhang). 
HiPERCAM is a quintuple-beam, high-speed optical imager equipped with four dichroic beamsplitters and five custom-made Teledyne e2v CCD231-42 detectors. Each detector has an FoV of $2\farcm8\times1\farcm4$, or $3\farcm1$ in diagonal and a pixel scale of 0\farcs081/pix without binning.
We requested in total five OBs under service mode of GTC. The total on-source exposure time is 5.1h. We set a four-point dithering pattern with 15 exposures of 60s at each point for each OB, with a $2\times2$ binning. The dithering was set specially for the sky subtraction in the $z$ band while saving as much overheads as possible. Our weather constraints were set to be seeing better than 0\farcs9, dark nights, and clear sky.  %The fisrt OB was executed on the second-half of the night on March , 2024 but with bad quality and re-observation was requested by the PI. 
The first two OBs were executed on the first-half of the night on June 4, 2024 and three more OBs were executed on the first-half of the night on June 11, 2024. All the photometric observations details are listed in Table~\ref{obs_pho_table}.

\subsubsection{Photometric data reduction}
\label{pho_datareduction}

The individual HiPERCAM exposures were bias and flat-field corrected by the HiPERCAM pipeline. For each band, we created a supersky frame by median combining a third of all the frames (around 100 frames) with lowest counts and without the position offsets using the \textit{imcombine} task of IRAF. The zero offset was set to be the mode value of all the counts. The supersky was subtracted from each frame using the task \textit{imarith}. We stacked all the sky subtracted frames and rejected the two highest- and two lowest pixels using \textit{imcombine}, with the zero offset set to be the mode value. The dithering position offsets were manually measured through the \textit{imexam} task.

The individual OSIRIS$+$ images were bias and flat-field corrected manually using IRAF. The individual FORS2 exposures were bias and flat-field corrected by the  FORS pipeline run in the EsoReflex environment \citep{freudling2013esoreflex}. First, the sky frame of each position in the four-point dithering pattern was created by mode-scaling and median-combining all the images at the other three positions, but two-thirds of the highest counts were rejected. The corresponding sky frame was then subtracted from all the images. Finally, all the sky-subtracted images were aligned 
and average-combined, but the two highest and two lowest pixels were rejected.

\subsubsection{Object recognition}
\label{recognition}
We created the world coordinate system (WCS) for all reduced images using \textit{Astrometry.net} \citep{dustin2010astrometry.net}, which extracts stars and solves the WCS by matching subsets of four stars to the pre-computed 4200 series index with the 2MASS catalogue as reference. For W2207, because there are not many bright 2MASS stars in the field, we built our own index file based on the Dark Energy Survey \citep[DES;][]{abbott2021DESdr2} catalogue. We projected the expected positions of the objects in the images based on their proper motions. Table~\ref{tg} lists the proper motions from the literature.  We drew a circle with a radius equal to the error centred on the expected position for each target, as shown in Fig.~\ref{fig:projection}.

\begin{table}[htbp]
\centering
      \caption[]{Revised astrometry of W0156 and W0711 with the astrometry rms, the centroid errors, and the pixel sizes.}
         \label{pm_corr}
         W0156 at MJD 60565.35
         \\
         \vspace{1mm}
         \begin{tabular}{cc}
            \hline
            $\alpha$, $\delta$ (deg)& 29.058675, 32.923258
            \\
            rms$_{X, Y}$ (pix) & 1.274, 0.920
            \\
            err$_{X, Y}$ (pix) & 0.239, 0.224
            \\
            scale (mas/pix) & 127.67 
            \\
            $\mu_{\alpha}\cos\delta,\ \mu_{\delta}$ (mas/yr) & $+1020\pm 29$, $-155\pm 22$
            \\
            \hline
         \end{tabular}
\vspace{2mm}

         W0711 at MJD 60565.38
         \\
         \vspace{1mm}
         \begin{tabular}{cc}
            \hline
            $\alpha$, $\delta$ (deg)& 107.840390, $-$57.606015
            \\
            rms$_{X, Y}$ (pix) & 0.589, 0.725
            \\
            err$_{X, Y}$ (pix) & 0.030, 0.029
            \\
            scale (mas/pix) & 125.53 
            \\
             $\mu_{\alpha}\cos\delta,\ \mu_{\delta}$ (mas/yr)&  $+230\pm 9$, $+1076\pm 11$
            \\
            \hline
         \end{tabular}
   \end{table}

\begin{table*}[htbp]
        \caption[]{New optical photometry and published IR photometry of the Accident, and other 12 metal-poor T dwarf candidates. If not specified otherwise, the reference for the IR photometry will be the same as the one for the object in Table~\ref{tg}. The optical photometry is in the AB magnitude system \citep{oke1974ABsystem}. The $JHK$ photometry uses the MKO filters and is in the Vega magnitude system.}
        %Pan-STARRS AB magnitudes $z$ were calculated according to Table~\ref{PS1_SDSS}.  Published NIR photometry (from the same references shown in Table~\ref{tg} are also listed.
         \label{phot_result}
    \resizebox{\textwidth}{!}{
     \setlength{\tabcolsep}{2pt}
    \centering
    \begin{tabular}{ccccccccccccccc}  
    \hline        
        Name & $u'$ & $g'$ & $r'$ & $i'$ & $z'$ & $z_{DES}$ & $z_{PS1}$ & $Y$ & $J$ & $H$ & $K$ & $W1$ & $W2$\\ \hline
        The Accident & >28.7 & >27.9  & >27.6 & >27.3 & >26.2 & & >26.7 & $^{(1)}$>21.79 & $^{(2)}$$24.5\pm0.3$  & >18.58  &  $K_s$ >17.85   & $18.18\pm0.19$  & $16.15\pm0.08$\\ 
        W0156 &  &  &  &   & $24.70\pm0.24$ &  & $25.42\pm0.31$ & $^{(2)}$$21.94\pm0.06$ & $21.46\pm0.30$  &   &   & $18.81\pm0.42$  & $16.06\pm0.08$  \\ 
        Ross\,19B&  &  &  &   &$25.10\pm0.33$ &  &  $25.80\pm0.40$ & $^{(2)}$$21.86\pm0.06$ & $21.14\pm0.02$  &   &   & $18.62\pm0.20$  & $15.81\pm0.06$ \\ 
        W0505&  &  &  &   & $25.24\pm0.36$ &  $^*$>23.4& $26.09\pm0.36$ &  & $^{(2)}$$20.74\pm0.07$   & $^{(2)}$$20.62\pm0.08$  & $K_s$ >18.38  & $^{**}$$17.64\pm0.06$  & $^{**}$$16.13\pm0.05$ \\ 
        W0523&  &   &  &   & $22.10\pm0.03$ &  & $22.74\pm0.15$ &  &  $^{(2)}$$19.14\pm0.05$  &  &   &$17.27\pm0.06$  & $15.91\pm0.06$\\ 
        W0711 &  &  &  &   & $19.48\pm0.03$ &  & $20.09\pm0.12$  & &   &   &   &  $15.09\pm0.03$ & $14.63\pm0.04$ \\ 
        W0738&  &   &  &   & $24.29\pm0.19$ &   & $25.23\pm0.20$ &  &  $^{(2)}$$21.37\pm0.14$  & $^{(2)}$$20.73\pm0.14$  & $^{(2)}$$21.44\pm0.34$    &  $17.22\pm0.04$ & $15.82\pm0.04$  \\ 
        W0905 & & & & & $24.08\pm0.09$ & & $24.97\pm0.13$ &  & $20.45\pm0.14$ & & & $18.25\pm0.16$ & $16.46\pm0.09$
        \\
        W1019&  &   &  &   & $19.27\pm0.01$ &  & $20.16\pm0.08$&  & $^{(3)}$$16.03\pm0.10$   & $^{(3)}$$15.77\pm0.13$  & $^{(3)}$$K_s\ 15.73\pm0.27$  &  $15.65\pm0.04$ & $14.22\pm0.04$  \\ 
        Wolf 1130C&  &   &  &   & $22.64\pm0.05$ &   &  $23.50\pm0.19$ &  &  $19.64\pm0.09$  & $19.57\pm0.07$ &   & $^{(4),\dagger}$$17.16\pm0.04$ & $^{(4)}$$15.14\pm0.02$\\ 
        W2014 &  &  &  &   & $21.52\pm0.03$&  & $22.36\pm0.5$ &  & $^{(5)}$$18.01\pm0.02$  & $^{(5)}$$18.71\pm0.30$  &  $^{(5)}$$17.97\pm0.29$ &  $^{(4)}$$16.66\pm0.04$ & $^{(4)}$$14.83\pm0.03$ \\ 
        W2105&  &   &  &   & &  $19.49\pm0.01$ & $20.19\pm0.02$ & $^{(2)}$$17.96\pm0.03$ &  $^{(3)}$$16.85\pm0.14$ & $^{(2)}$$15.91\pm0.03$ & $^{(3)}$$15.31\pm0.15$  & $^{(2)}$$14.82\pm0.01$  & $^{(2)}$$13.93\pm0.02$  \\ 
        W2207&  &   &  &   & $24.23\pm0.10$ & $^*$>23.4 & $25.20\pm0.12$ &  & >20.36  & >18.98 & $K_s$ >18.38  & $^{(4)}$$18.88\pm0.27$  & $^{(4)}$$16.01\pm0.07$ \\ 
        \hline
    \end{tabular}}
          \tablebib {
          (1): \citet{meisner2023accident_Y}
          (2): \citet{meisner2023coldoldBD};
     (3): 2MASS point source catalogue \citep{cutri2003_2MASSpoint};
     (4): Profile-fit photometry including motion from CatWISE catalogue (w1mpro\_pm and w2mpro\_pm column) \citep{marocco2021catwise};
     (5): \citet{best2021ucd25pc}
    }

    \textbf{Notes:} *: Single-epoch DES 3-$\sigma$ $z$-band limit under the AB system, as inferred from the 10-$\sigma$ limit 22.1~mag \citep{morganson2018DESpipeline,abbott2021DESdr2}; **: From the WISE and DES images we notice that the WISE photometry may be slightly contaminated by a background source.
    $\dagger$: There are two CatWISE detections with similar proper motion values. We adopted the $W1, W2$ values of the one with more profile-fit flux measurements with S/N no less than 3 in both bands (column names w1NM and w2NM).
\end{table*}

We detected all the objects at their expected positions, except for W0156, W0711, and the Accident. We recognised W0711 using the DESI Legacy Survey DR10 image, and we attribute the faint source next to the expected position of W0156 as its true position.  We indicated W0156 and W0711 using red arrows in Fig.~\ref{fig:projection}. We revised their proper motions following the same routine as we did in \citet{zhangjerry2023optical_sdT}. The IRAF \textit{imcentroid} task was used to provide the centroid error in both axes. The astrometry residuals of both axes are the standard error of the mean (SEM) of the pixel deviation between the index position and field position of the matched reference stars (match\_weight > 0.99) in the corr.fits file generated by \textit{Astrometry.net}. The revised proper motion results are shown in Table~\ref{pm_corr}.

\subsubsection{Photometric measurement}
\label{phot_measure}

We performed aperture photometry on these 11 detected metal-poor T dwarf candidates using the \textit{Astropy} package \textit{photutils}. The aperture radius was fixed to be 1\arcsec, and the background residual after the sky subtraction is from the median absolute deviation within an annulus with an inner and outer radius of 3\farcs5 and 5\farcs5, respectively. W0523, W1019 and W2014 were close to other background sources, hence, we set a smaller aperture with a radius of 0\farcs5 for these three objects only. For these three, we did a consistency check using a programme \textit{daofun}\footnote{\url{https://pypi.org/project/daofun/0.2.0/}} (Quezada et al. in prep.) which is an interactive adaption of \textit{daophot} \citep{stetson1987daophot} to easily go through the \textit{daophot} subroutines to perform a point-spread-function (PSF) photometry. Using the GUI we automatically picked brighter and isolated stars over the background noise and check by eye with the GUI each point spread function to then create the PSF model. The PSF photometry is based on the ALLSTAR subroutine of \textit{daophot} also included in the daofun GUI, we activate the variable PSF parameter, set the FWHM of stars by 4 pixels, the fitting radius by 6 pixels and the PSF radius by 12 pixels as the GUI recommended parameters.

For these 11 detected metal-poor T dwarf candidates, we did differential photometry using an extra set of several nearby sources from The Panoramic Survey Telescope and Rapid Response System \citep[Pan-STARRS;][]{chanbers2016panstarrs}, SkyMapper Southern Sky Survey \citep{onken2024skymapper_arxiv}, and the DES with $z$ magnitudes fainter than 16~mag in the AB system). 

Since there are no detection of the Accident in any of the HiPERCAM bands, we estimated a 3-$\sigma$ detection limit by conservatively assuming a 3-$\sigma$ signal would have a total flux inside the 1.5-FWHM-radius (1\farcs2) aperture of $3\times\sigma_{sky}\times\sqrt{N_{pix}}$ where $\sigma_{sky}$ is the fluctuation or standard deviation inside the aperture, and $N_{pix}$ is the pixel number inside the aperture. For the $g'r'i'z'$ bands we did differential photometry using an extra set of several nearby Pan-STARRS sources with magnitudes fainter than 16~mag. The 3-$\sigma$ limits set in this way in the $g'r'i'z'$ bands are very well consistent with the aperture photometry on the faintest sources (S/N less than 10 $\sigma$) in the field.  For the $u'$ band of HiPERCAM we calculated the limit using the zero point value of 28.17~mag provided by the GTC\footnote{\url{https://www.gtc.iac.es/instruments/hipercam/hipercam.php}}, since there is no $u'$-band survey in this region. 

For the T and Y dwarfs, a correction for each spectral type must be applied to convert Sloan $z'$ magnitudes, or DES $z_{DES}$ magnitudes to Pan-STARRS $z_{PS1}$ magnitudes, due to significant difference between filter profiles longwards of  9300$\AA{}$. Following the procedure of \citet{zhangjerry2023optical_sdT}, we applied the offsets calculated from T dwarf standards and Y dwarf models. Table~\ref{phot_result} lists the $u'g'r'i'z'$ and $z_{DES}$ photometry directly determined from the observations and the $z_{PS1}$ after the correction. Other IR photometric measurements from the literature are also listed in that table.

\begin{figure*}[htbp]
    \centering
    \includegraphics[width=\linewidth]{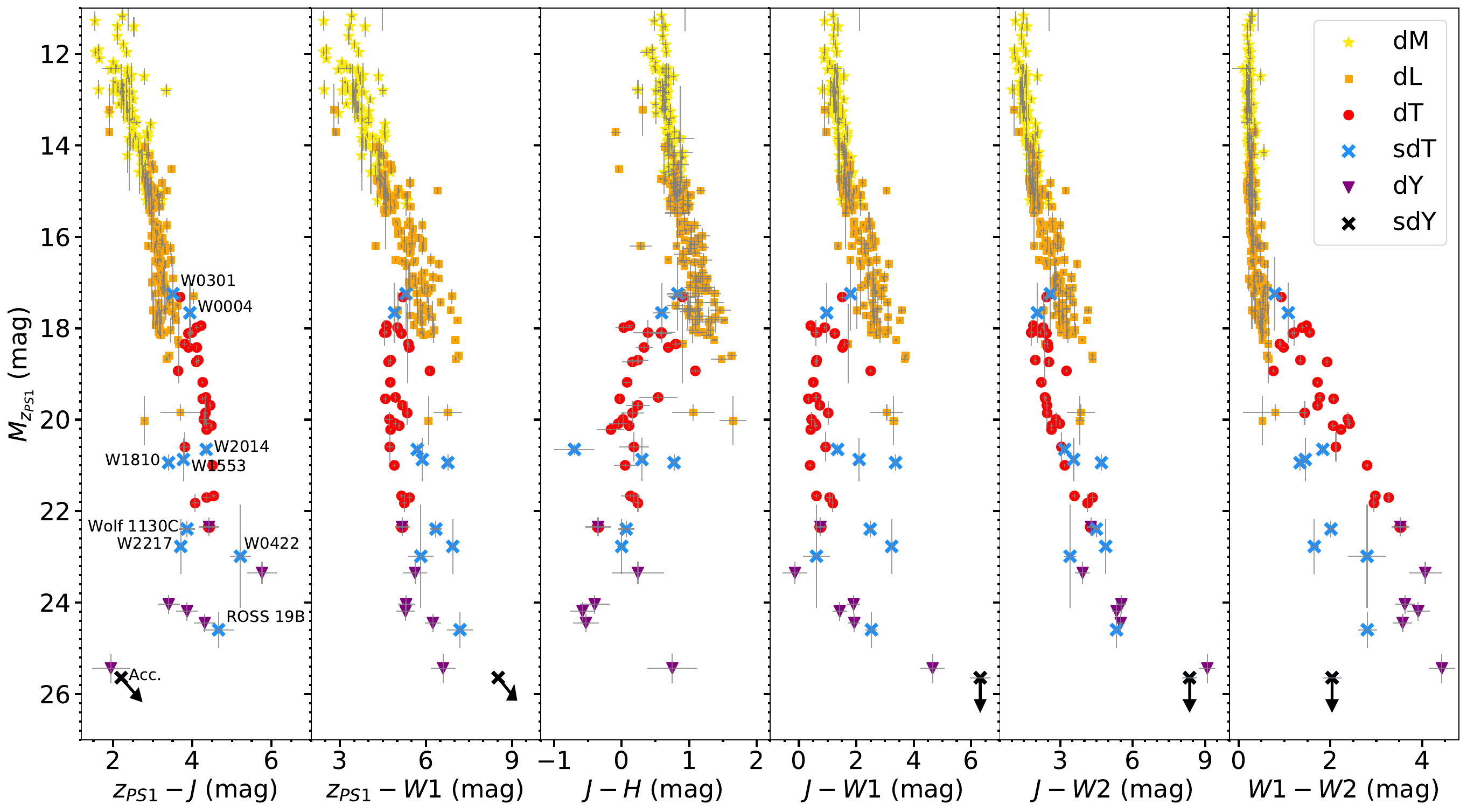}
    \caption{Colours vs. absolute $z_{PS1}$ magnitude diagrams of metal-poor T dwarf candidates (blue crosses), using trigonometric parallaxes obtained from this work and from the literature, $z_{PS1}$ photometry from this work and from \citet{zhangjerry2023optical_sdT} and infrared photometry from the literature. The parallax of W1810 comes from \citet{lodieu2022W1810}. For the Accident (black cross and arrow), the parallax is from \citep{kirkpatrick2021accident}. For comparison, field M, L, T dwarfs with trigonometric parallaxes from Pan-STARRS1 3$\pi$ survey \citep{best2018PS1_3pi} are plotted as yellow stars, orange squares, and red dots, respectively. Y dwarfs with $z_{PS1}$ photometry (corrected from $z'$) from \citet{lodieu2013Yoptical} and \citet{martin2024Yoptical}, and parallaxes from \citet{kirkpatrick2019parallax184TY} are plotted as purple triangles.  W1217 is a T9+Y2 system so it is plotted as a mixture of red dot and purple triangle.}
    \label{CMDs}
\end{figure*}

\begin{figure*}[htbp]
    \centering
    \includegraphics[width=\textwidth]{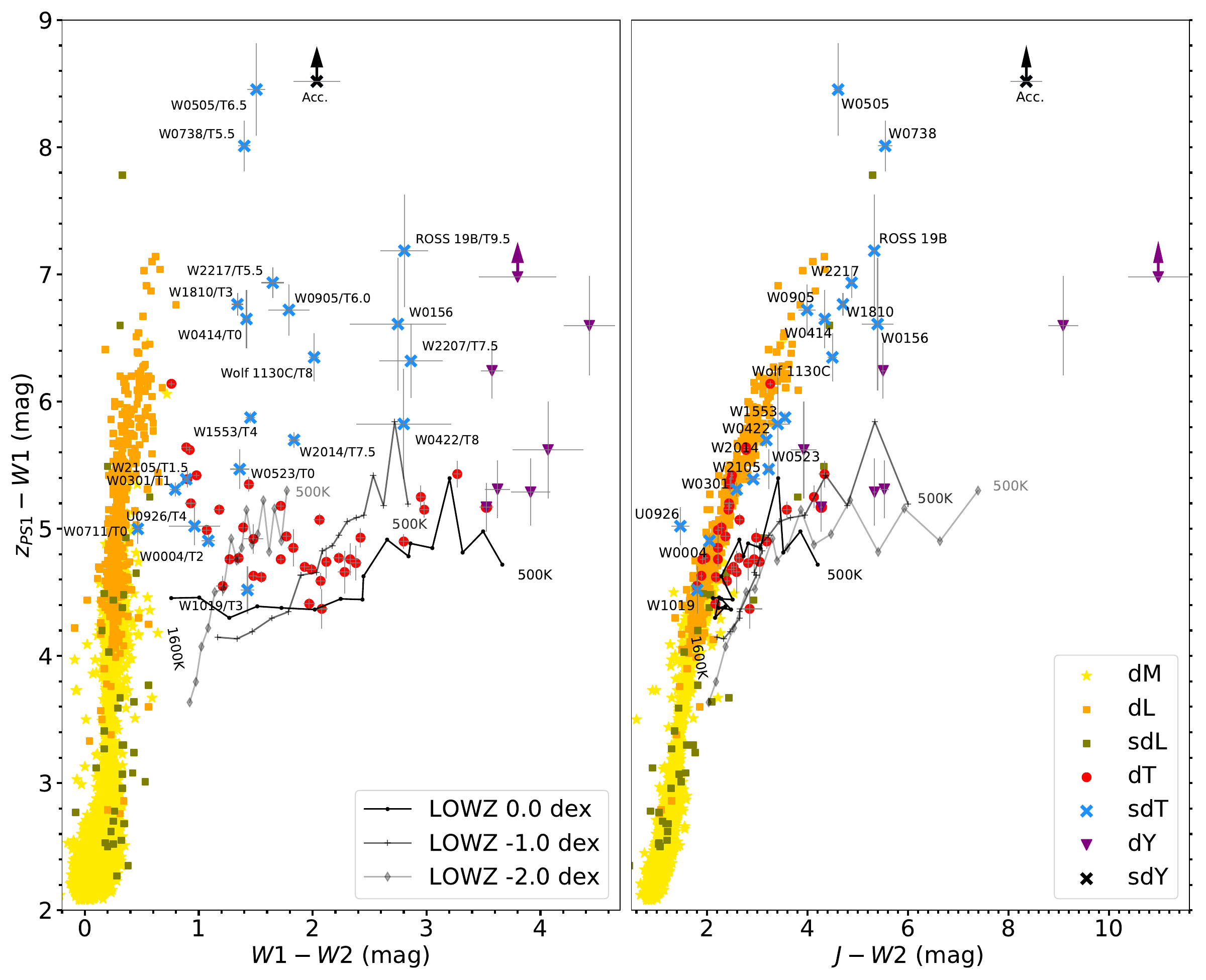}
    \caption{Updated $W1-W2$ vs $z_{AB}-W1$ and $J-W2$ vs $z_{AB}-W1$  colour--colour diagrams with more metal-poor T, Y dwarf candidates compared to the one of \citet{zhangjerry2023optical_sdT}. All the metal-poor T dwarf candidates are labeled with blue crosses and arrows and the Accident is labeled with the black cross and arrow. We overplotted solar-metallicity M, L, and T sequences from Pan-STARRS (yellow stars, orange squares and red dots, respectively), L subdwarfs (olive squares), four Y0 dwarf, a Y2 dwarf, the Y4 dwarf (purple triangles), and a T9+Y0 system (red dot mixed with purple triangle). All Y dwarfs are with CatWISE photometry.  Error bars are included for all sources, except for M and L dwarfs. We also show three iso-metallicity curves from the low-metallicity theoretical model LOWZ \citep{meisner2021esdT} with parameters $\log g=5.0$, $\log_{10}K_{zz}=2$, solar C/O ratio 0.55 in both diagrams. From left to right, the effective temperature of the model decreases from 1600 K to 500 K, with a step of 100 K and 50K above and under 1000K, respectively.}
    \label{colourcolour}
\end{figure*}

\begin{figure*}[htbp]
    \centering
    \includegraphics[width=\textwidth]{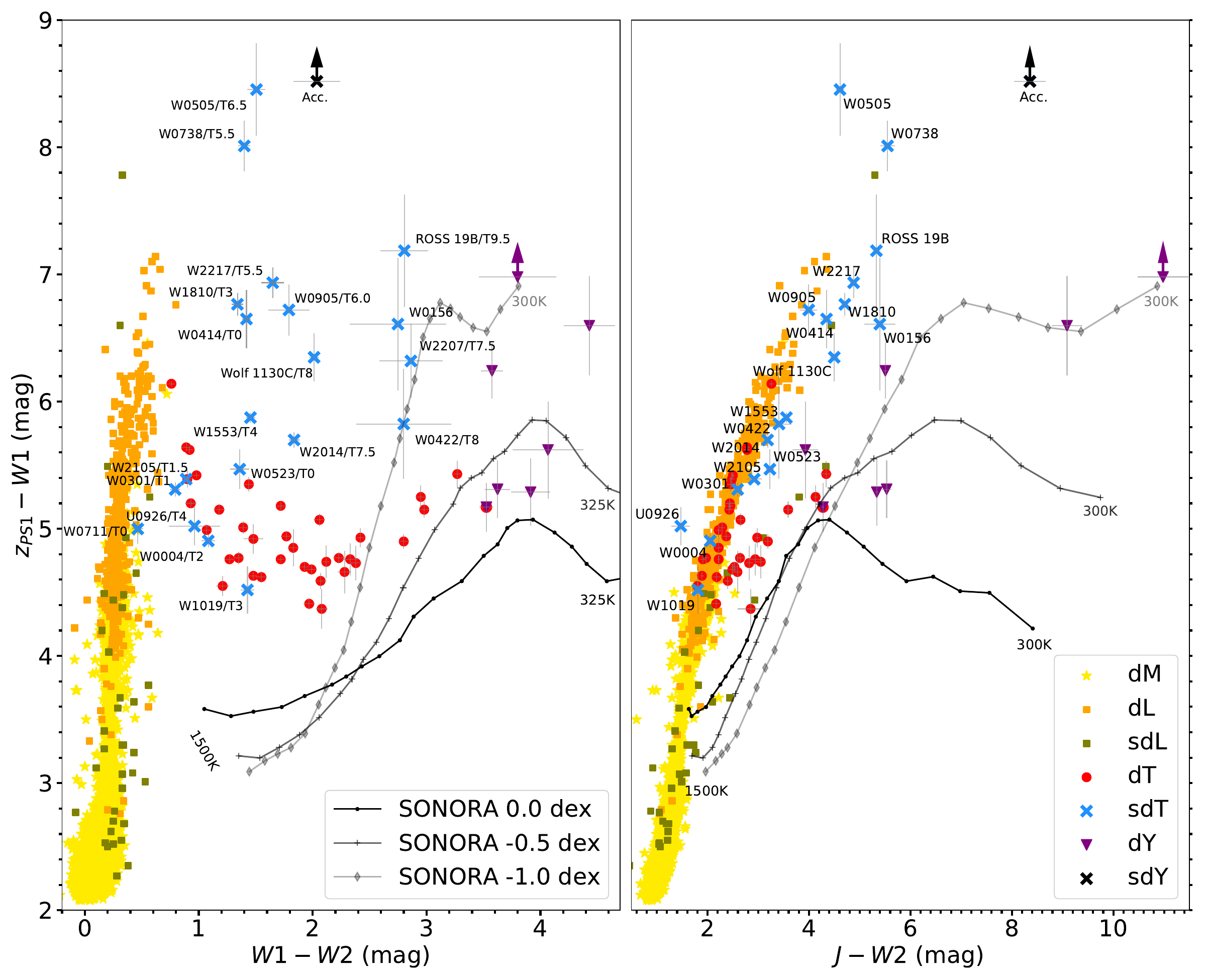}
    \caption{The same as Fig.~\ref{colourcolour} but with three iso-metallicity curves from the theoretical model SONORA Elf Owl with parameters $\log g=5.0$, $\log_{10}K_{zz}=2$, solar C/O ratio in both diagrams. From left to right, the effective temperature of the model decreases from 1500 K to 300 K, with a step of 100 K, 50 K and 25 K at the temperature above 1000 K, between 600 K - 1000 K and below 600 K, respectively.}
    \label{colourcolour_sonora}
\end{figure*}

\section{Discussion}
\label{results}

\subsection{Colour--magnitude diagrams}
\label{HR}

With trigonometric parallaxes, we could derive the absolute magnitudes of subdwarfs and position them in colour-magnitude diagrams. For the distance of two companions, i.e., Wolf\,1130C and Ross\,19B, we used well-determined parallaxes from their primaries, Wolf\,1130A and Ross\,19A, respectively. %For the distance of Wolf\,1130C, we used the trigonometric parallax of the primary Wolf\,1130A of 16.69$\pm$0.15 pc located at a projected separation of 3150 au \citep{mace2013Wolf1130C,mace2018Wolf1130}. For the distance of Ross\,19B, we also used the one of its comoving M subdwarf primary Ross\,19A. \textcolor{red}{you should give its metallicity here because you discuss the metallicity of Wolf1130.} 
Both objects have well-constrained spectroscopic metallicities derived from their primary. Hence, they are benchmarks in our sample.
%from the primary Wolf\,1130A.  Its low metallicity was spectroscopically confirmed, hence it is a benchmark for our study.

We plotted several colours of eight metal-poor T dwarf candidates and the Accident against their absolute $z_{PS1}$ magnitudes in Fig.~\ref{CMDs}. For comparison, we also put another benchmark object, the extreme T subdwarf W1810, in the diagrams. W1810 was claimed to be a subdwarf because it has a bolometric luminosity significant lower than that of an early-T dwarf, aligning instead with the luminosity expected for solar-metallicity T8-9 dwarfs \citep{lodieu2022W1810}. It is spectroscopically classified as an esdT3 dwarf lately \citep{burgasser2024esdT}. Subdwarfs appear intrinsically fainter than their solar-metallicity counterparts with the same spectral type, and the absolute $z_{PS1}$ magnitude is a proxy for the luminosity \citep{sanghi2023hawaii_parallax_ucd}.  We over-plotted solar-metallicity field M, L, T dwarfs with trigonometric parallaxes from Pan-STARRS1 \citep{best2018PS1_3pi}. We also overlaid several colder objects with $z'$-band magnitudes from GTC/OSIRIS  
\citep{lodieu2013Yoptical,martin2024Yoptical} and trigonometry parallaxes \citep{kirkpatrick2019parallax184TY}. These include four Y0 dwarfs: WISE\,J041022.71$+$150248.4, WISEP\,J173835.52$+$273258.9, WISE\,J205628.91$+$145953.2 \citep{cushing2011dY_WISE}, and WISE\,J014656.66$+$423410.0 \citep{kirkpatrick2012furtherY}; a Y2 dwarf WISE\,J182831.08$+$265037.7 \citep{cushing2011dY_WISE}; as well as
%, which is the coldest among our sample of solar-metallicity dwarfs. 
a cold binary system, WISE\,J121756.90$+$162640.8, which consists of a T9 dwarf and a Y0 dwarf \citep{liu2012W1217_TYbinary}, supposed to be overluminous compared to single T9 and Y0 dwarfs. We use motion-corrected photometry from CatWISE catalogue \citep{marocco2021catwise} for all Y dwarfs.

From Fig.~\ref{CMDs}, we could infer the following points:
\begin{itemize}
    \item The benchmark extreme T subdwarf W1810 has an absolute $z_{PS1}$ magnitude $\sim$2~mag fainter than that of solar-metallicity early-T dwarfs. It also has bluer $z_{PS1}-J$ and $W1-W2$ colours and redder $z_{PS1}-W1$, $J-H$, $J-W1$ and $J-W2$ colours.
    \item The benchmark Wolf\,1130C is most probably a T subdwarf. Its absolute $z_{PS1}$ magnitude is located at the solar-metallicity T/Y boundary which is $\sim$1~mag fainter than that of the T8 dwarfs, confirming its low metallicity (from $-0.6$ to $-1.2$ dex). Its colours compared with the solar-metallicity sequence follow the same trend as W1810, except in $J-H$ and $J-W2$, where it lies on the sequence.
    \item W1553 has an absolute $z_{PS1}$ magnitude slightly fainter than that of solar-metallicity mid-T dwarfs. Its colours fall in between W1810 and the solar-metallicity sequence in all the diagrams. It is most probably a T subdwarf but may not have a metallicity as low as W1810, which has been shown via NIR spectroscopy \citep{meisner2021esdT}.
    \item W2217 is most probably a T subdwarf. It again lies below the faintest T dwarf in the solar-metallicity sequence, suggesting it could be the coldest or most metal-poor object among the three. Its colours follow W1810, except in the $J-H$ and $J-W2$.
    \item W0422's absolute magnitude is 1.5-$\sigma$ fainter than that of the faintest solar-metallicity T dwarf, if we adopt the astrometric solution despite its high  $\chi^2_{\nu}$ value. It suggests a possible colder nature of W0422, or a probable slightly low metallicity. However, its colours are not aligned with W1810, W1553, W2217 and Wolf\,1130C except in the $W1-W2$.
    \item Ross\,19B lies significantly below the T dwarf sequence. It inherits the metallicity of $-0.4$ dex from its primary but without a spectroscopic confirmation of the secondary itself. Comparing with Wolf\,1130C, which is also a late-type subdwarf and with a much lower metallicity, the difference between the absolute $z_{PS1}$ magnitude of Ross\,19B and those of the T dwarf sequence seems to be too large. This suggests, either Ross\,19B is a colder object, or less likely that it has a lower metallicity than the value from its comoving primary. 
    \item W0004 and W0301 lie on top of the sequence of early-T dwarfs in all the six colour-magnitude diagrams, which suggests that they are not so metal-poor, as corroborated by their optical spectra \citep{zhangjerry2023optical_sdT} and slightly blue NIR spectra \citep{greco2019neowise}, or they could be equal-mass binaries.
    \item W2014 has a slightly red $z_{PS1}-W1$ and $J-W1$ colours and a slightly blue $W1-W2$ colour. It lies on top of the sequence of solar-metallicity mid-T dwarfs in the diagram of $z_{PS1}-J$ and $J-W2$, which suggests that its metallicity is not extremely low. It has a bluer $J-H$ colour than the sequence with a 2-$\sigma$ significance.
    \item The Accident is below the faintest solar-metallicity Y dwarf in all diagrams, arguing in favour of its low-metallicity and cold nature.
\end{itemize}

In summary, we confirm the subluminosity of three more T subdwarfs: Wolf\,1130C, W1553, and W2217. These objects seem to follow the extreme T subdwarf W1810, lying to the blue of the sequence in the following colours: $z_{PS1}-J$ and $W1-W2$; and to the red in the colours $z_{PS1}-W1$ and $J-W1$. We also have another three possible subdwarfs: Ross\,19B, W0422, and W2014 in these colour-magnitude diagrams. We observe that the Y subdwarf candidate, the Accident, has an upper limit of the absolute $z_{PS1}$ magnitude fainter than the current limit for Y dwarfs.

$J-H$ and $J-W2$ colours do not seem to be good metallicity indicators. Solar-metallicity objects have larger relative $J-H$ colour dispersions probably due to smaller wavelength interval between two bands.  With respect to the $J-W2$ colour, most of metal-poor objects lie on the solar-metallicity sequence. We are going to discuss more about the colours in the next subsection.

\subsection{Colour--colour diagrams}
\label{secccd}

As an update to Fig.\,1 in our previous work \citep{zhangjerry2023optical_sdT}, Fig.~\ref{colourcolour} and \ref{colourcolour_sonora} present the colour--colour diagrams of $W1-W2$ versus $z_{PS1}-W1$ and $J-W2$ versus $z_{PS1}-W1$ for all the objects, with 8508 solar-metallicity M dwarfs, 800 L dwarfs, and 42 T dwarfs from Pan-STARRS1 with and without trigonometric parallaxes \citep{best2018PS1_3pi}, four Y0 dwarfs W0146, W0410, W1738, W2056, a Y2 dwarf W1828, and the T9$+$Y0 binary W1217. We added the $z_{PS1}-W1$ limit of the Y4 dwarf WISE\,J085510.83$-$071442.5 \citep{luhman2014W0855} which is the coldest brown dwarf discovered so far \citep{beamin2014W0855_260K,luhman2024W0855}.  There are also 39 L subdwarfs with photometric errors smaller than 0.2~mag in $z$, $J$, $W1$, and $W2$ bands  \citep{zhangzenghua2018subL} %\textcolor{red}{how many of these have parallaxes ?}. 
Three objects, WISEA\,J001354.40+063448.1, WISEA\,J083337.81+005213.8, and ULAS J092605.47+083516.9, %ULAS J131610.28+075553.0, and  ULAS J131943.77+120900.2 
included in the previous work are kept \citep{zhangjerry2023optical_sdT}.

We compared the observations with synthetic photometry derived using filter profiles downloaded from the Spanish Virtual Observatory (SVO) service \citep{rodrigo2012SVOfilter,rodrigo2020SVOfilter}, and theoretical spectra from the low-metallicity ultracool LOWZ models \citep{meisner2021esdT} and the SONORA Elf Owl substellar atmosphere models for T and Y dwarfs \citep{mukherjee2024sonora,mukherjee2023sonora_T,mukherjee2023sonora_Y}. We used the following parameters: a high surface gravity $\log$(g)\,=\,5.0, a medium vertical mixing $\log_{10}K_{zz}$\,=\,2 and a solar C/O ratio of 0.55 for both the LOWZ and the SONORA model. The LOWZ models provide spectra down to metallicity $-$2.5 dex while the SONORA models only reach to $-$1.0 dex. The solar-metallicity field dwarf sequence more or less follows the trend of the two sets of models. 

Both the LOWZ and the SONORA models predict bluer $W1-W2$ colours and redder $J-W2$ colours for lower metallicity objects in the 500--1000~K temperature range. This is clear for the Accident, since it has a significantly bluer $W1-W2$ colour and a redder $J-W2$ colour than field Y dwarfs. For the rest of metal-poor T dwarf candidates, they do not tend to concentrate in a smaller $W1-W2$ interval.  %Following the model, we speculate that for 500--1300~K objects, if the metallicity is lower than a certain value, there would be an upper limit for the $W1-W2$ colour. %\textcolor{red}{NL: this colour interval is an observational bias because this is the region where the WISE team has identified the first 2 esdT and then they kept looking in a limited region in w1-w2.}

The LOWZ model predicts a $z_{PS1}-W1$ colour between 4 to 5.5~mag for objects between 500~K and 1600~K (black line in Fig.~\ref{colourcolour}). The LOWZ models predict that for low metallicity ($-$2.0 dex) the temperature gradient is mainly responsible for the redward shift of the $z_{PS1}-W1$ colour from 4 to 5~mag. The SONORA model predicts a much stronger temperature dependence than the LOWZ model. The SONORA model also predicts a stable $z_{PS1}-W1$ colour for hotter objects no matter the metallicity, which contradicts the observations.

\begin{figure}[htbp]
    \centering
    \includegraphics[width=0.49\textwidth]{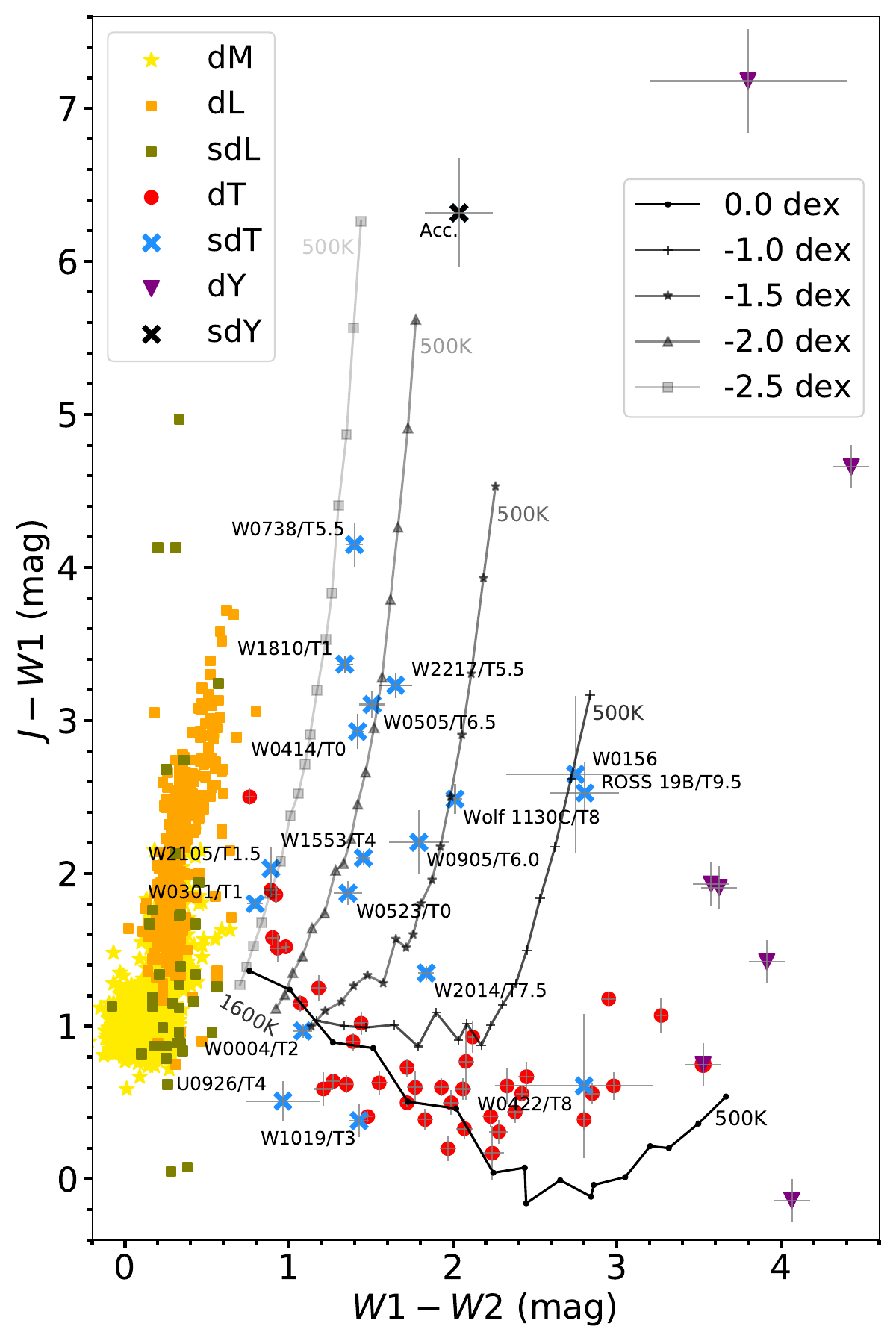}
    \caption{$W1-W2$ vs $J-W1$ colour--colour diagrams of all metal-poor T dwarf candidates (blue crosses and arrows) in Table~\ref{tg} and in \citet{zhangjerry2023optical_sdT}; and the Accident (the black cross and arrow). We also included solar-metallicity M, L, and T sequences from Pan-STARRS as yellow stars, orange squares, and red dots, respectively, along with L subdwarfs (olive squares), four Y0 dwarf, a Y2 dwarf, the Y4 dwarf (purple triangles), and a T9$+$Y0 system (red dot mixed with purple triangle). Error bars are included for all sources, except for M and L dwarfs. We also show five iso-metallicity curves from the low-metallicity theoretical model LOWZ with parameters $\log(g)$\,=\,5.0, $\log_{10}K_{zz}$\,=\,2, solar C/O ratio 0.55\@. From left to right of the five curves, the effective temperatures of the model decrease from 1600~K to 500~K, with a step of 100~K and 50~K above and under 1000K, respectively.}
    \label{colourcolour2}
\end{figure}

%In the left panel of Fig.~\ref{colourcolour} and \ref{colourcolour_sonora}, the metal-poor objects segregate into three groups of different $z_{PS1}-W1$ colours with decreasing metallicity, tentatively \textbf{named} Group I, II, and III by \citet{zhangjerry2023optical_sdT}. %Each group contains objects from early- to late-type, justifying the statement that the colour is not significantly affected by the effective temperature. %This claims that  the $z_{PS1}-W1$ colour could be a good low-metallicity indicator, breaking the degeneracy between metallicity and temperature. 
We infer that the LOWZ models under-estimate the suppression of the $z_{PS1}$-band flux, caused by the low metallicity; and also does the SONORA models, but only for hotter objects or early-T dwarfs.

%The column of the metal-poor candidates with similar $W1-W2$ colour around 1.4~mag, from bottom to the top are W0523, W1553, W0505, W0414, W1810, W2217, W0738, having spectral types of T0, T4, T6.5, T0, T1, T5.5, T5.5 does not have any signs of temperature order, if we assume that the spectral type is a monotonic function of the effective temperature for extreme subdwarfs, unless the current spectral classification has to be revised. 

%Having the sample size increased, we quantitatively test and verify the robustness of this indicator in the next Section~\ref{z-w1}. 

On the other hand, the behaviour of the $z_{PS1}-W1$ colour of subdwarfs is similar to that of the $J-W1$ colour. Unlike the $z_{PS1}-W1$ vs. $W1-W2$ diagram, the observations agree with model predictions in the $J-W1$ vs. $W1-W2$ diagram (we only plot the LOWZ models in Fig.~\ref{colourcolour2}). The LOWZ and SONORA models predict that the $J-W1$ colours get significant redder for objects cooler than about 1000~K when metallicity decreases, and the temperature is the main driver for the metal-poor objects to have the $J-W1$ colours shifted redward. However, the hotter end of both models (1300--1600~K) exhibits the same problem in the $J-W1$ colours as in $z_{PS1}-W1$. They also do not accurately reproduce the observed L subdwarfs, as well as the extreme cold solar-metallicity Y2 and Y4 dwarfs. Solar-metallicity T dwarfs exhibit greater dispersion across the spectral type in the $J-W1$ colour (around 2 mag) compare to the $z_{PS1}-W1$ colour (around 1.5 mag)%, likely due to the narrower wavelength range between the $J$ and $W1$ passbands. 
%This increased dispersion makes Group II harder to distinguish (the intermediate-metal-poor T dwarf group). %We also question whether the increase of the $J-W1$ colour is driven by the temperature gradient in lower-metallicity objects. 
In Fig.~\ref{colourcolour2}, the spectral types of each metal-poor object are shown, but no clear temperature gradient is evident.

%The latter is a slightly worse metallicity indicator because solar-metallicity T dwarfs exhibit greater dispersion in the $J-W1$ colour due to the narrower wavelength interval, which may complicate the distinction of Group II (the intermediate T subdwarf group). The LOWZ model actually predicts the trend of drastic reddening of $J-W1$ colour for objects cooler than $\lesssim$\,1000~K. The hot end of the model (1300--1600~K) does not reproduce those L subdwarfs. We question the reddening of the $J-W1$ colour caused by the temperature gradient in lower metallicity objects. Here in Fig.~\ref{colourcolour2}, we have also marked the spectral type for each metal-poor objects: there is no clear temperature gradient. \textcolor{red}{NL: i know what you want to say but you need to rephrase. it is poor English as it is written}

\subsection{$z_{PS1}-W1$ colour vs metallicity}
\label{z-w1}

Our previous work categorised the few candidates into three groups with decreasing metallicity from $\simeq0$ to $\simeq-1.5$~dex while $z-W1$ colours increasing from 5 to 7~mag \citep{zhangjerry2023optical_sdT}. With a larger sample we actually see a continuity between the groups and even extensions bluer than 5~mag and much redder than 7~mag.

To see the $z_{PS1}-W1$ colour trend against the metallicity more clearly, we assigned a coarse metallicity value to each source in our sample. With the exception of Wolf\,1130C and Ross\,19B, which have spectroscopically determined metallicities derived from their M dwarf primaries, the metallicities of the remaining objects are estimated based on metallicity subclass classifications obtained through spectroscopy or photometry. This classification of subdwarfs has been a long debated topic and it has not been fully established for T subdwarfs yet because of lack of objects across the whole spectral type range. We adopted the maximum interval of the metallicity value for subclasses published by different research groups for M and L subdwarfs \citep{gizis1997pm_subdwarf,lepine2007sdM,zhang2017six_sdL_classification,lodieu2019sdM}: i.e.,\ subdwarfs have metallicities mainly between $-$0.3 and $-$1.0 dex, extreme subdwarfs between $-$1.0 and $-$1.7 dex, and ultra subdwarfs have below $-$1.7 dex. We also adopted a metallicity range of $+$0.3 to $-$0.3 dex for solar-metallicity field T/Y dwarfs.

 \begin{table}[htbp]
 \centering
      \caption[]{Adopted metallicity [Fe/H] ranges for metal-poor T/Y dwarf candidates. Objects are ordered by RA\@. The `Spec.' column indicates whether the object has been spectroscopically observed. The `Sublum.' column indicates if the object appears subluminous in the colour-magnitude diagrams in Fig.\ \ref{CMDs}.}
         \label{feh_table}
         
 \begin{tabular}{ccccc}
            \hline
            \noalign{\smallskip}
           Name  & [Fe/H]$_{\mathrm{max}}$ & [Fe/H]$_{\mathrm{min}}$ & Spec. & Sublum.
           \\
            \hline
    T, Y dwarfs &  $+0.3$ &  $-0.3$ & - & -
           \\   
    W0004 &  $-0.3$ &  $-1.0$ & Y & N
           \\   
    $^{(1)}$W0013 &  $-0.5$ &  $-1.5$ & N & -
           \\
    W0156 &  $-0.28$ &  $-0.52$ & N & -
           \\
    W0301 &  $-0.3$ &  $-1.0$ & Y & N
           \\     
    $^{(2)}$Ross\,19B & $-0.28$ &  $-0.52$ & N & Y?
           \\ 
    $^{(3)}$W0414 & $-0.5$ &  $-1.5$ & Y & -
    \\
    W0422 & $-0.3$ &  $-1.7$ & N & Y?
           \\
 W0505 & $-1.0$ &  $-1.7$ & N & -
           \\             
 $^{(4)}$W0523 & $-0.5$&  $-1.5$ & N & -
           \\ 
    W0711&  $-0.3$ &  $-1.0$ & Y & -
           \\ 
 W0738 & $-0.5$ &  $-2.0$ & N & -
           \\ 
 W0905 & $-0.3$ & $-1.7$  & N & -
           \\  
 $^{(1)}$W0833 & $-0.5$ &  $-1.5$ & N & -
           \\   
 ULAS0926 & $-0.3$ &  $-1.0$ & N & -
           \\   
 W1019  & $-0.3$ &  $-1.0$ & Y & -
           \\  
 $^{(5)}$The Accident & $-1.0$ & - & N & Y?
        \\
 $^{(6)}$W1553  & $-0.5$ &  $-1.5$& Y & Y
           \\
 $^{(7)}$W1810  & $-1.0$ & $-2.0$ & Y & Y
           \\
Wolf\,1130C  &  $^{(8)}$$-0.52$&  $^{(9)}$$-1.46$ & Y & Y
           \\
 W2014 &  $-0.3$ &  $-1.0$ & Y & Y?
           \\ 
 W2105 & $0.0$ & $-1.0$ & Y & -
\\ 
 W2207 & $-0.3$ & $-1.7$ & N & -
           \\      
 W2217 & $-0.5$ & $-2.0$ & N & Y
    \\
            \hline
         \end{tabular}
        
    \tablebib{
    (1): \citet{pinfield2014subdwarf};
    %(2): \citet{meisner2023coldoldBD};
    (2): From the metallicity of the primary \citep{schneider2021Ross19B};
    (3): \citet{Schneider2020W0414_W1810}; 
    (4): \citet{brooks2022esdT}; 
    (5): \citet{kirkpatrick2021accident};
    (6): \citet{meisner2021esdT}; 
    (7): \citet{lodieu2022W1810}; 
    (8): From the metallicity of the primary \citep{woolf_wallerstein2006feh_M_molecule};
    (9) From the metallicity of the primary \citep{newton2014feh447dM}.
    Rest: from our assignation, see section~\ref{z-w1}.
    }
    
   \end{table}

For each individual object, we assign a corresponding metallicity based on the arguments below. A summary of the adopted metallicity for each object is given in Table~\ref{feh_table}. 

 \begin{figure*}[htbp]
    \centering
    \includegraphics[width=\linewidth]{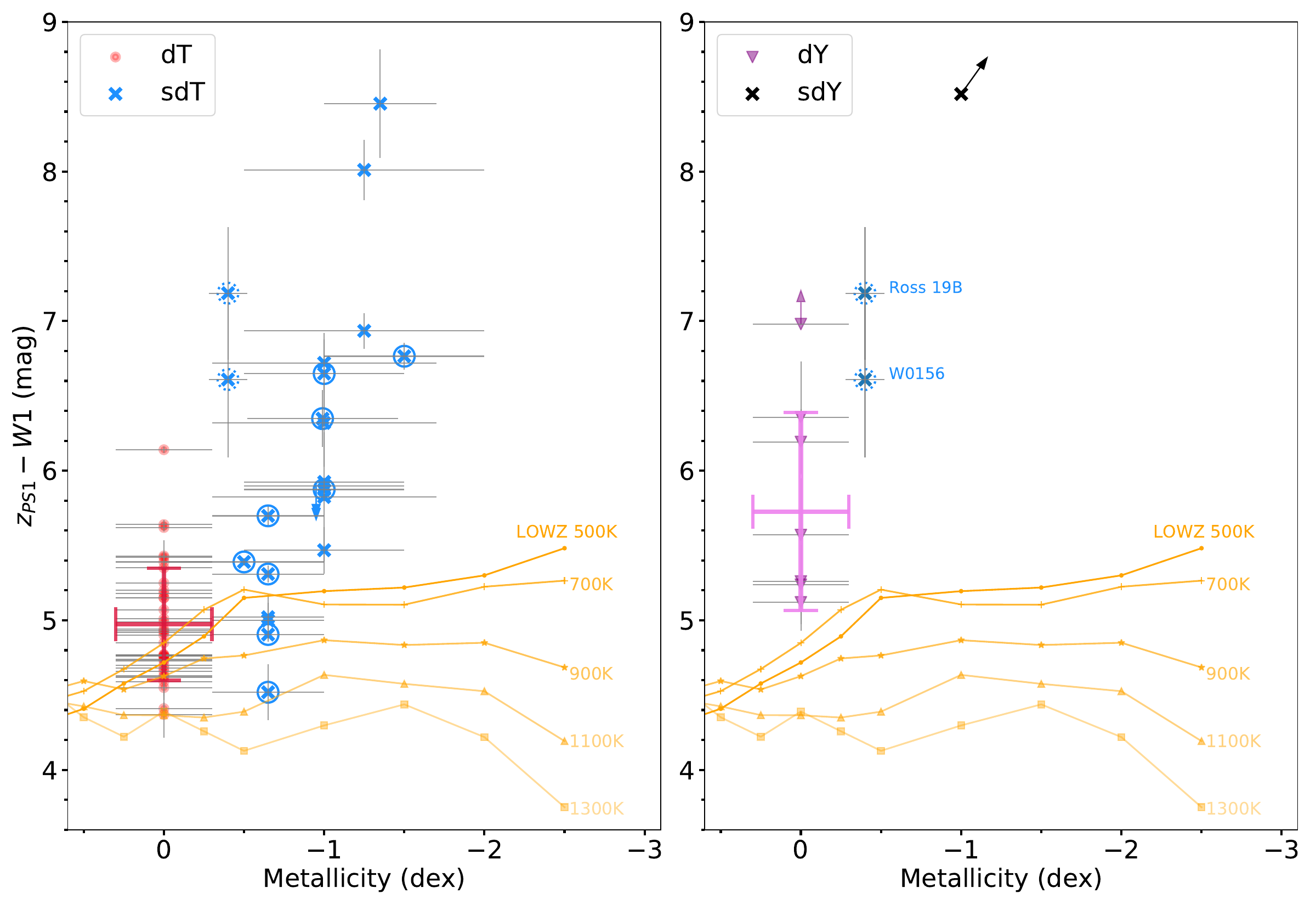}
    \caption{$z_{PS1}-W1$ colour against metallicity for T and Y dwarfs. The vertical error bars are the photometric errors and the horizontal bars are the adopted metallicity ranges (Table~\ref{feh_table}). The blue and black crosses are metal-poor T and Y dwarf candidates, respectively. Some upper limit arrows are plotted next to the points to avoid overlapping. The solid blue circle indicates that the object has an NIR spectrum. The two very-late-type metal-poor T dwarf candidates or possible metal-poor Y dwarf candidates Ross\,19B and W0156 are indicated by dotted blue circles and are plotted in both panels. The thick red and purple crosses demonstrate the average $z_{PS1}-W1$ value and dispersion of solar-metallicity T and Y dwarfs, respectively. %We drew two blue dashed lines in the left panel at $z_{PS1}-W1$\,=\,5.5 and 6.3 mag to indicate the separation of Group I, II, and III in Fig.~\ref{colourcolour} for metal-poor T dwarfs. 
    %We used an empirical exponential relation to fit the points between the colour and the metallicity. 
    Five isothermal curves from 500~K to 1300~K from LOWZ model with parameters $\log(g)$\,=\,5.0, $\log_{10}K_{zz}$\,=\,2, C/O ratio of 0.55 are shown in orange colour.} %We drew a dashed grey line indicating the synthetic $z_{PS1}-W1$ colour of the Accident using its JWST NIRSpec spectrum \textcolor{red}{NL: IMPORTANT this mention should be removed}.}
    \label{indicator}
\end{figure*}

\begin{itemize}
    \item For the secondaries of comoving pairs, we assumed that all components of a multiple system have a similar chemical composition. There are different measurement of metallicity of the primary Wolf\,1130A: $-0.62\pm0.10$ dex \citep{woolf_wallerstein2006feh_M_molecule}, $-0.70\pm0.12$ dex \citep{mace2018Wolf1130}, $-0.64\pm0.17$ dex \citep{rojas-ayala2012metallicity_Kband_M} but updated to $-1.22\pm0.24$ dex \citep{newton2014feh447dM}. We adopted a maximum 1-$\sigma$ range from $-0.52$ to $-1.46$ dex for Wolf\,1130C. 
    \item We adopted the iron abundance [Fe/H] of the primary Ross\,19A \citep[$-0.40\pm0.12$ dex;][]{schneider2021Ross19B} for Ross\,19B. 
    \item \citet{greco2019neowise} used NIR spectroscopy and classified W0004, W0301, W1019 as subdwarfs. They do not appear as very metal-poor objects and they are not subluminous in the colour-magnitude diagrams in Fig.~\ref{CMDs}.  We assigned them a metallicity [Fe/H] from $-0.3$ to $-1.0$ dex. \citet{meisner2020extremecoldBD} classified W0422 and W2207 as subdwarfs but did not specify the metal class. Hence, we adopted a possible range from normal subdwarfs to extreme subdwarfs from $-0.3$ to $-1.7$ dex for these two objects. 
    \item \citet{brooks2022esdT} listed W0505 as an extreme T subdwarf based on its infrared colours, hence, we adopted a metallicity from $-1.0$ to $-1.7$ dex. 
    \item \citet{meisner2021esdT} reported a spectral analysis for W1553, but different models provided different metallicity values, from $-0.5$ dex to $\lesssim-1.5$ dex. We adopted that range. 
    \item \citet{meisner2023coldoldBD} assigned W0156 a metallicity between $-0.4$ to $-0.5$ dex because of the adjacent locus to the Ross\,19B and the LOWZ model track of $-0.5$ dex in the colour space. We adopted the same metallicity as that of Ross\,19B. 
    \item For WISEA\,J041451.67$-$585456.7, we adopted a range of metallicities from $-$0.5 to $-$1.5 dex \citep{Schneider2020W0414_W1810}. For WISEA\,J181006.18$-$101000.5, we used the range derived by \citet{lodieu2022W1810}, from $-$1.0 to $-$2.0 dex. 
    \item \citet{kellog2018_W0711} classified W0711 as a subdwarf using its NIR spectrum, its low metallicity is consistent with the thick-disk or halo kinematics. We adopted the metallicity range of subdwarfs, from $-$0.3 to $-$1.0 dex.
    \item We classify W0738 and W2217 \citep{meisner2021esdT} as extreme T subdwarfs based on their large motions and similar colours to the two known extreme T subdwarfs, W1810 and W0414. So we adopted the metallicity range of W1810 and W0414, i.e.\ from $-$0.5 to $-$2.0 dex for W0738 and W2217. 
    \item For W0013 and W0833, \citet{pinfield2014subdwarf} estimated a metallicity range from $-$0.5 to $-$1.5 dex, based on their peculiar $H-W2$ colour. %For ULAS1316, \citet{burningham2014T6.5} assigned it as subdwarf and speculated it should be as metal poor as Wolf 1130C. We adopted $-0.65\pm0.35$ dex. 
    \item \citet{murray_burningham2011blueT} discussed the blue NIR colours and possible halo kinematics of ULAS0926, so we classify it as a subdwarf with a metallicity from $-0.3$ to $-1.0$ dex. 
    \item W0905 and W2014 just falls outside the fiducial colour-colour criteria of extreme subdwarf region proposed by \citet{meisner2023coldoldBD}. However, W2014 shows a normal spectrum in the $Y$-, $J$-, and $H$-bands. Only its $K$-band was suppressed \citep{mace2013WISE_Tpopulation}. Besides, W2014 appears inline with those T subdwarfs only in some of the colour-magnitude diagrams in Fig.~\ref{CMDs}, and has a $J-H$ colour abnormality. We hence assigned a wide metallicity range from subdwarf to extreme subdwarf to W0905, from $-$0.3 to $-$1.7 dex; but only a subdwarf metallicity to W2014, from $-$0.3 to $-$1.0 dex.
    \item The kinematics of W2105 shows that it could be a thick disk/halo dwarf, but the $J$- and $H$-band spectra do not show features of low metallicity \citep{luhman2014WISEhighpm}, and \citet{meisner2023coldoldBD} excluded it from  the list of extreme T subdwarf candidates because its colours put it outside of the locus where extreme T subdwarfs lie. Therefore, we adopted a metallicity from solar 0.0 dex to the most-metal-poor subdwarfs, $-1.0$ dex.
\end{itemize}

 We plot the $z_{PS1}-W1$ colour and metallicity of solar-metallicity field T dwarfs and metal-poor T dwarf candidates in the left panel of Fig.~\ref{indicator}. We highlighted objects with spectra as circles. We also plot the mean colour and the standard deviation of the colour of solar-metallicity field T dwarfs. The temperature effect for the solar-metallicity T dwarfs on this colour is less than 1~mag (from 4.5 to 5.5~mag). 
 %The metallicity groups I, II, III for the metal-poor T dwarfs, first introduced by \citet{zhangjerry2023optical_sdT}, are now separated by the blue dashed lines in \textbf{Fig.~\ref{indicator}}. 
 %We adopted $z_{PS1}-W1$\,=\,5.5 and 6.3 mag as the separation of groups I and II, and groups II and III, respectively. Group I is a mixture of solar-metallicity dwarfs and slightly-metal-poor subdwarfs, with similar $z_{PS1}-W1$ colours. 
 The $z_{PS1}-W1$ colour does not change significantly at least until the metallicities of W0004, W0301, and W1019. Once the metallicity gets lower than $\approx -$1.0~dex, the $z_{PS1}-W1$ colour shoots up and gets drastically redder.

We identify Ross\,19B and W0156 as two outliers (dotted blue circles) in the left panel of Fig.~\ref{indicator} with mildly low metallicities but unusually red $z_{PS1}-W1$ colours. %Their metallicities appear to be well constrained.  
Since the metallicity of W0156 was constrained by Ross\,19B \citep{meisner2023coldoldBD}, it is not surprising that W0156 aligns with Ross 19B as an outlier. From this point forward, we will focus only on Ross\,19B, although our conclusions are applicable to both objects.

Ignoring the possibility of wrong photometry, in the colour-magnitude diagrams, Ross\,19B appears extremely subluminous compared to solar-metallicity T dwarf counterparts. It would likely be a colder object, a Y subdwarf, sharing the same chemical component with the primary. It is not contradictory to the photometric spectral type determination: T9.5$\pm$1.5 \citep{schneider2021Ross19B}. We put both Ross\,19B and W0156 into the right panel of Fig.~\ref{indicator}, with the mean colour and the standard deviation of the colour of solar-metallicity field Y dwarfs, and the colour limit of the Accident. The colours of both Ross\,19B and W0156 stay between the average colour of the solar-metallicity Y dwarfs and that of the Accident, which is consistent with their mildly low metallicities derived from Ross\,19A. However, Ross\,19 system has a projected separation of 9900 au \citep{schneider2021Ross19B}. For such a loosely bounded system, it is hard for it to survive dynamically from the Galactic tides given its old age \citep[7.2$^{+3.6}_{-3.2}$ Gyr;][]{schneider2021Ross19B}. For a binary with a semi-major axis 10\,000 au, the simulation for a system with total mass of 1$M_\odot$ yields a very low probability of surviving for 7 Gyr \citep{weinberg1987binary_fate}. We note that the Ross\,19 system has only a total mass of 0.4 $M_\odot$.

We provide another possible scenario in which Ross\,19A captured the B component which is not necessarily colder but has a lower metallicity. The absolute $z_{PS1}$ magnitude difference between the solar-metallicity sequence and Ross\,19B is larger than that between the solar-metallicity sequence and Wolf\,1130C in Fig.~\ref{CMDs}. Wolf\,1130C has a similar spectral type as Ross\,19B. We therefore hypothesise that Ross\,19B may have a much lower metallicity than Wolf\,1130C, and certainly lower than its primary, Ross\,19A. However, the chance for a metal-poor M dwarf to catch specifically a brown dwarf with even lower metallicity is not high either. 

It is challenging to assess which case is more probable. In summary, Ross\,19B could be colder and be among the metal-poor Y dwarf candidates, or it is more metal poor. The same applies to W0156.

Putting these two outliers aside, in the left panel of Fig.~\ref{indicator}, we see that objects with $z_{PS1}-W1$\,$\gtrsim$\,6~mag have metallicity lower or equal than $-1.0$~dex, i.e., extreme subdwarfs. Their spectral types spread from T0 to T8. In other words, the $z_{PS1}-W1$ colour is observationally a good extreme subdwarf indicator for T dwarfs, which was first proposed by \citet{zhangjerry2023optical_sdT}. This colour could be a metallicity indicator for Y dwarfs according to the right panel of Fig.~\ref{indicator} after adding Ross\,19B and W0156, but more observations are needed to confirm this hypothesis.
 
 %Hence, we developed an empirical relation, based on an exponential function, applicable for T, Y subdwarfs. On the other hand, via the inverse function the metallicity can be inferred by the colour 
%\begin{equation}
%    \mathrm{[Fe/H]}=-\frac{\log{\left[3.89\left(z_{PS1}-W1\right)-18.30\right]}}{1.43}.
%\end{equation}
%The fitting has a reduced $\chi^2$ value of 0.15.

We again plot the LOWZ models in both panels of Fig.~\ref{indicator}, using five isothermal lines with metallicity from $+$0.5 dex to $-$2.5 dex. The LOWZ models do predict that the $z_{PS1}-W1$ colour gets redder with decreasing metallicity until $-$1.0 dex for objects cooler than 900~K (equivalent to mid- to late-T dwarfs). As the temperature decreases, the colour becomes redder in a more pronounced manner.
%\textcolor{red}{NL: you always use reddening, which I understand, but have to be careful because we are talk talking about the reddening due to dust} 
However, the models fail to account for the very red colours observed at lower metallicities for both T and early-Y dwarfs (objects with temperature higher than 500~K). %The model also predicts slightly bluer $z_{PS1}-W1$ colours with decreasing metallicity until $-$1.0 dex for objects warmer than 900~K (i.e.\ early-T dwarfs). This effect was probably shown \textcolor{red}{NL: do you mean this model prediction is corroborated by observations ? i do not see such evidence, they lie in the same position as solar-metallicity T dwarfs} by W0004, W0301, W2105, W1019, ULAS0926 (from sdT1.0 to sdT4.0) in the $W1-W2$ vs. $z_{PS1}-W1$ diagram in Fig.~\ref{colourcolour}, as they all lie at the bottom of the normal T dwarf sequence. 

Our previous work attributed the redder $z_{PS1}-W1$ colours to the flux increase in the $W1$ band, due to the weakening of the methane absorption, and expected a saturation at some point \citet{zhangjerry2023optical_sdT}. However, at this moment we still do not appreciate any sign of saturation in the $z_{PS1}-W1$ colour, although in the case of the current most metal-poor T dwarf with spectroscopic metallicity, W1810, methane is absent or very weak \citep{lodieu2022W1810}. On the other hand, the absorption wings of alkali metals, especially the sodium Na{\small{I}} and potassium K{\small{I}} resonance doublets at optical wavelengths are broadened in high-gravity atmospheres due to the collision with helium He and molecular hydrogen H$_2$, contributing partly to the suppressed flux in the $z$ passband \citep{burrows2003NaK,pavlenko2007alkali,allard2016K_H2,phillips2020TYatm}. This absorption could be even strengthened due to higher gravity and denser atmospheres in metal-poor environment  \citep{allard2003Na_K,allard2023Na_He,allard2024K_He}. These two effects could be underestimated by the LOWZ model, explaining the current discrepancy with the observations.

  \begin{figure}[htbp]
    \centering
    \includegraphics[width=\linewidth]{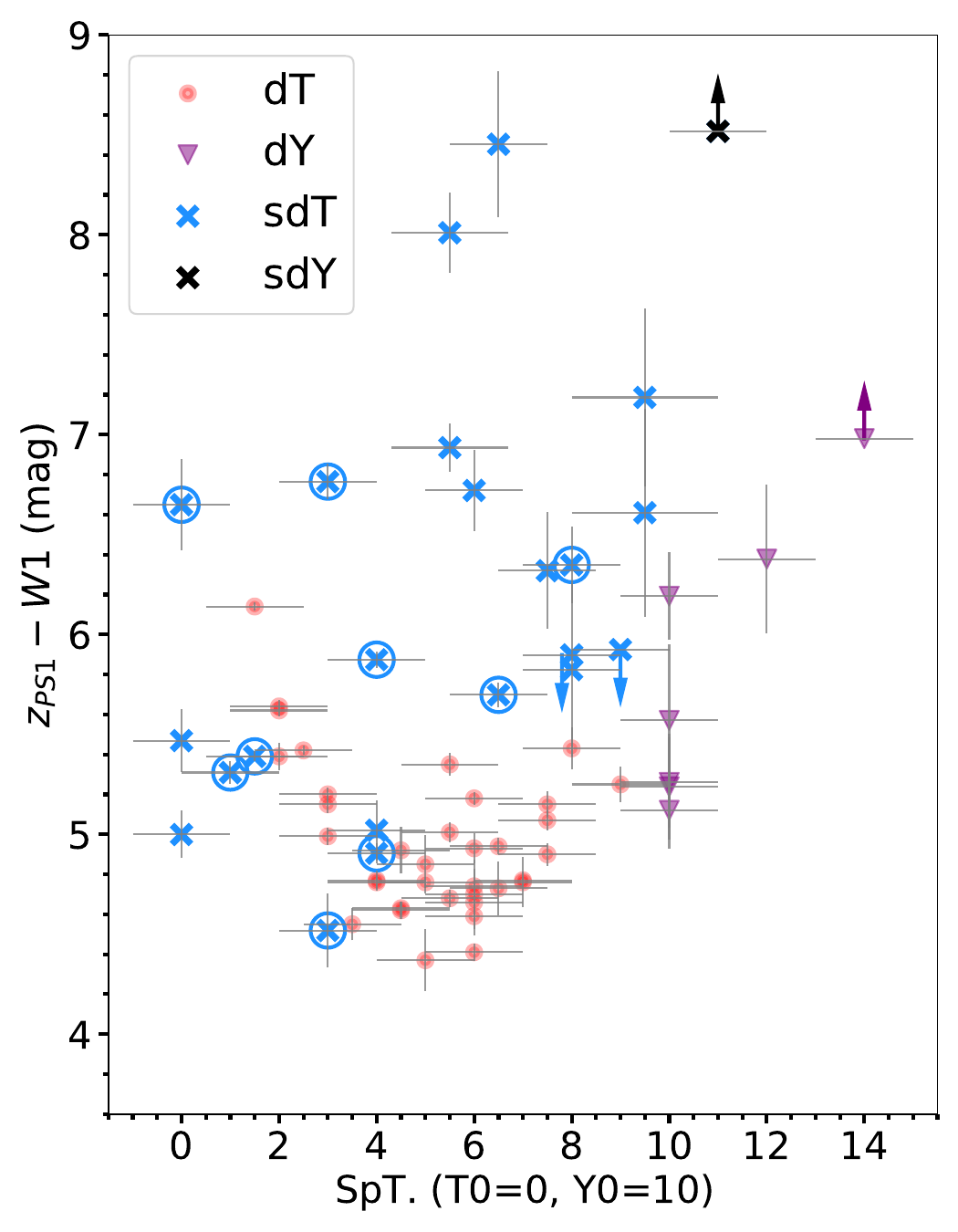}
    \caption{$z_{PS1}-W1$ colour against spectral type. The solid blue circle indicates that the object has an NIR spectrum. All the objects have a spectral type uncertainty of one subtype, unless specified differently in Table~\ref{tg}. We adopted the classification of W1810 as esdT3 in \citet{burgasser2024esdT}. W0156 was classified as a T subdwarf without a subtype, we assigned it with the same late spectral type as Ross\,19B. We assigned a Y0--Y2 range for the Accident, according to its effective temperature. %T dwarfs are separated into three groups, and Group III are supposed to be the most metal-poor T dwarfs. The $z_{PS1}-W1$ colours of Group III dwarfs show no dependence on the spectral type.
    }
    \label{indicator_temp}
\end{figure}

We further checked the temperature dependence of the $z_{PS1}-W1$ colour. We plotted the colour of all objects against their spectral type in Fig.~\ref{indicator_temp}. We see that the $z_{PS1}-W1$ colour gets bluer until mid-T dwarfs and then gets redder for solar-metallicity dwarfs. The range of colours remains in a small interval between 4.5 to 5.5~mag. %The trend is similar to the trend in the $z_{PS1}-W1$ vs $W1-W2$ colour-colour diagram in Fig.~\ref{colourcolour} \textcolor{red}{NL: what trend?}. 
We do not see a clear spectral type dependence for those T subdwarf candidates, and we can conclude that the differences in the effective temperature of solar-metallicity T dwarfs will not lead to a extreme red $z_{PS1}-W1$ colours. Thus, the $z_{PS1}-W1$ colour can reliably serve as an indicator for extreme T subdwarfs. 

%\textcolor{red}{z-J color relation, }

%\subsection{Y subdwarf}

\subsection{Most metal-poor objects}

We see that W0505 and W0738 have exceptionally red $z_{PS1}-W1$ colours of 8.01~mag and 8.45~mag, respectively. There may exist a population in the colour-colour diagram with $z_{PS1}-W1$ colours redder than 8.0~mag, represented by ultra subdwarfs (usdTs) with metallicity [Fe/H] below $-$1.7 dex. However, the WISE photometry of W0505 could be slightly affected by faint background sources, hence the true $z_{PS1}-W1$ colour of W0505 could be slightly bluer.

%By its lower limit on the ground-based $z_{PS1}-W1$ colour, our relation predicts an exceptionally low metallicity [Fe/H] $\lesssim -1.7$ dex for the Accident, which falls into the ultra subdwarf category \textcolor{red}{NL: refer to the table with the adopted metallicity. In the table, you give only -1.0 dex for the Accident, in disagreement with the text in this section. In the figure, the dashed line representing the Accident should contemplate the range of metallciity given in the text and table}. We synthesised the $z_{PS1}-W1$ colour of the Accident using the very recent low-resolution ($R\approx100$) prism spectrum (A. Meisner et al. in prep.) covering 0.6 to 5.3 microns from the Near-Infrared Spectrograph (NIRSpec) of the James Webb Space Telescope (JWST). We infer $z_{PS1}-W1$\,=\,9.6~mag, which yields $z_{PS1}$\,=\,27.8~mag. This colour and our relation suggest a metallicity of [Fe/H] $\approx -1.9$ dex, which agrees with the result from the LOWZ model fitting using only the IR photometry: $\approx -2.0$ dex \citep{kirkpatrick2021accident}.  The Accident would be the most metal-poor ultracool object, and the coldest metal-poor object up to date. \textcolor{red}{NL: is it a new claim or was it already claimed by Kirkpatrick and Meisner in their papers?}

The Accident's extreme low luminosity also points towards a very low effective temperature. The Accident has the reddest $z_{PS1}-W1$ colour among all the coldest subdwarf candidates, and a blue $W1-W2$ colour suggesting that it could have a very low metallicity.  \citet{martin2024Yoptical} observed a sharp transition towards bluer $z-J$ colours from 5 to 2~mag from late-T to Y2 dwarfs, respectively, as seen in the colour-magnitude diagram in Fig.~\ref{CMDs}. The transition agrees with the SONORA solar-metallicity model \citep{marley2021SONORA}. We attribute this effect to the weakening of the very broad K{\small{I}} resonance doublet in the Y dwarf atmosphere, with alkali elements being locked into molecules at low temperatures \citep{lodders1999alkali} and carried into deeper layer by rainout processes \citep{burrows&sharp1999rainout}.  The Accident has a lower limit of 2.2~mag on its $z_{PS1}-J$, which is consistent with those of solar-metallicity Y dwarfs at the moment. %From JWST, we inferred a synthetic $z_{PS1}-J$ colour of 2.5~mag.  
%This means in the cold metal-poor atmosphere, besides there is stronger and broadened collision-induced alkali absorption in the optical band, the $J$ band could also suffer a stronger collision-induced absorption from molecular hydrogen H$_2$ in the mean time.

%-----------------------------------------------------------------

\section{Conclusions}
\label{conclusion}

We obtained ground-based trigonometry parallaxes for five metal-poor T dwarf candidates and complemented them with those from the literature. We acquired $z$-band photometry for another 12 metal-poor T dwarf candidates and five-band optical photometry for the Accident, the only metal-poor Y dwarf candidate to date. 

We demonstrated the subdwarf nature of W1553, Wolf\,1130C, W2217, and possibly Ross\,19B, W0422, and W2014 using their loci in the optical-to-infrared colour-magnitude diagrams comparing with their solar-metallicity counterparts and the extreme T subdwarf W1810. We showed that the Y dwarf candidate, the Accident, is subluminous than current Y dwarf limit. We argue that W0004, and W0301 are not so metal-poor or they could be binary candidates based on their loci on the solar-metallicity sequence in the colour-magnitude diagrams. It is worthwhile to take an observation with adaptive optics or interferometry to spacially resolve these binary candidates. 

We provided updated optical-IR colour-colour diagrams, effectively doubling the size of the previous metal-poor T dwarf sample. We strengthened our previous discovery about the relation between the low metallicity and the red $z_{PS1}-W1$ colour for T dwarfs, and possibly extended it to Y dwarfs, which is not predicted by the state-of-art models.  Combining the three colours $z_{PS1}-W1$, $J-W1$, and $W1-W2$ may break the metallicity-temperature degeneracy for the coldest populations, and is useful to %determine a coarse metallicity for ultracool dwarfs and to 
identify future extreme- or even ultra- metal-poor T and Y dwarfs. We confirmed the candidacy of W0505 and W0738 to the population of extreme-metal-poor T dwarfs, and we propose that Ross\,19B and W0156 could be Y subdwarf candidates.
%We gave a metallicity estimation of the Accident of $-1.9$ dex using the relation and synthetic colour from its JWST spectrum. This empirical relation derived in this work is preliminary because it does not take into account any dependence with effective temperature, although it is seen to be little. 
%More optical photometric measurements are required over a wider range of metallicity to further constrain the metallicity indicator and disentangle any potential temperature-metallicity degeneracy of this population. 

We claim that NIR spectroscopy is highly desirable to determine the metallicity of the extreme T subdwarf candidates W2217, W0505, and W0738. However, due to their faintness, this remains challenging with current ground-based instrumentation but could be achievable from space using the JWST. A more practical approach would be obtaining $K$-band photometry for W0505 and W2217. We also discuss the importance of the spectroscopic determination of the metallicity and temperature of Ross\,19B as another benchmark object and an additional anchor point in the diagrams presented in this work. The analysis of the JWST spectrum of the Accident \citep[programme GO 3558, PI A. Meisner;][]{meisner2023jwst_proposal_acc} is of extreme importance since it could let us to synthesise the $z_{PS1}$ photometry of the only Y subdwarf candidate, and measure its metallicity and other physical parameters.

In the current/future deep optical surveys like the ESA \textit{Euclid} mission, Vera Rubin LSST and Roman space mission, we could launch a tailored survey to search for a specific ultracool population with a given metallicity and effective temperature \citep{solano2021virtual,martin2021ch4nh3}. The main objective is to increase the samples of metal-poor T and Y dwarfs taking advantage of the complementarity between optical and NIR photometry, as well as astrometric measurements thanks to the repeated observations of Vera Rubin LSST and the \textit{Euclid} deep fields.

\begin{acknowledgements}
We thank our referee for providing an insightful report with detailed comments.
Funding for this research was provided by the Agencia Estatal de Investigación del Ministerio de Ciencia e Innovación (AEI-MCINN) under grants PID2019-109522GB-C53 and PID2022-137241NB-C41 as well as the European Union (ERC, SUBSTELLAR, project number 101054354). 
BG acknowledges support from the Polish National Science Center (NCN) under SONATA grant No. 2021/43/D/ST9/0194.
Based on observations made with the Gran Telescopio Canarias (GTC), installed at the Spanish Observatorio del Roque de los Muchachos of the Instituto de Astrofísica de Canarias, on the island of La Palma. 
This work is partly based on data obtained with the instrument HiPERCAM, built by the Universities of Sheffield, Warwick and Durham, the UK Astronomy Technology Centre, and the Instituto de Astrofísica de Canarias. Development of HiPERCAM was funded by the European Research Council, and its operations and enhancements by the Science and Technology Facilities Council. 
This work is partly based on data obtained with the instrument OSIRIS, built by a Consortium led by the Instituto de Astrofísica de Canarias in collaboration with the Instituto de Astronomía of the Universidad Autónoma de México. OSIRIS was funded by GRANTECAN and the National Plan of Astronomy and Astrophysics of the Spanish Government. 
This work is partly based on data obtained with the instrument EMIR, built by a Consortium led by the Instituto de Astrofísica de Canarias. EMIR was funded by GRANTECAN and the National Plan of Astronomy and Astrophysics of the Spanish Government.
Based on observations collected at Centro Astronómico Hispano en Andalucía (CAHA) at Calar Alto, operated jointly by Instituto de Astrofísica de Andalucía (CSIC) and Junta de Andalucía.
Based on observations collected at the European Southern Observatory under ESO programme 113.2688.
This research has made use of the Spanish Virtual Observatory (https://svo.cab.inta-csic.es) project funded by MCIN/AEI/10.13039/501100011033/ through grant PID2020-112949GB-I00.
This research has made use of the SVO Filter Profile Service "Carlos Rodrigo", funded by MCIN/AEI/10.13039/501100011033/ through grant PID2020-112949GB-I00.
This research has made use of data provided by Astrometry.net. 
% CDS+ADS
This research has made use of the Simbad and Vizier databases, and the Aladin sky atlas operated
at the centre de Donn\'ees Astronomiques de Strasbourg (CDS), and
of NASA's Astrophysics Data System Bibliographic Services (ADS).
This research has made use of the NASA/IPAC Infrared Science Archive, which is funded by the National Aeronautics and Space Administration and operated by the California Institute of Technology.
This work has used the Pan-STARRS1 Surveys 
(PS1) and the PS1 public science archive that have been made 
possible through contributions by the Institute for Astronomy, the University of Hawaii, the Pan-STARRS Project Office, the Max-Planck Society and its participating institutes, the Max Planck Institute for Astronomy, Heidelberg and the Max Planck Institute for Extraterrestrial Physics, Garching, The Johns Hopkins University, Durham University, the University of Edinburgh, the Queen's University Belfast, the Harvard-Smithsonian Center for Astrophysics, the Las Cumbres Observatory Global Telescope Network Incorporated, the National Central University of Taiwan, the Space Telescope Science Institute, the National Aeronautics and Space Administration under Grant No. NNX08AR22G issued through the Planetary Science Division of the NASA Science Mission Directorate, the National Science Foundation Grant No. AST-1238877, the University of Maryland, Eotvos Lorand University (ELTE), the Los Alamos National Laboratory, and the Gordon and Betty Moore Foundation. This publication makes use of data products from the Wide-field Infrared Survey Explorer, which is a joint project of the University of California, Los Angeles, and the Jet Propulsion Laboratory/California Institute of Technology, funded by the National Aeronautics and Space Administration. This project used public archival data from the Dark Energy Survey (DES). Funding for the DES Projects has been provided by the U.S. Department of Energy, the U.S. National Science Foundation, the Ministry of Science and Education of Spain, the Science and Technology Facilities Council of the United Kingdom, the Higher Education Funding Council for England, the National Center for Supercomputing Applications at the University of Illinois at Urbana–Champaign, the Kavli Institute of Cosmological Physics at the University of Chicago, the Center for Cosmology and Astro-Particle Physics at the Ohio State University, the Mitchell Institute for Fundamental Physics and Astronomy at Texas A\&M University, Financiadora de Estudos e Projetos, Fundação Carlos Chagas Filho de Amparo à Pesquisa do Estado do Rio de Janeiro, Conselho Nacional de Desenvolvimento Científico e Tecnológico and the Ministério da Ciência, Tecnologia e Inovação, the Deutsche Forschungsgemeinschaft and the Collaborating Institutions in the Dark Energy Survey. The Collaborating Institutions are Argonne National Laboratory, the University of California at Santa Cruz, the University of Cambridge, Centro de Investigaciones Enérgeticas, Medioambientales y Tecnológicas–Madrid, the University of Chicago, University College London, the DES-Brazil Consortium, the University of Edinburgh, the Eidgenössische Technische Hochschule (ETH) Zürich, Fermi National Accelerator Laboratory, the University of Illinois at Urbana-Champaign, the Institut de Ciències de l’Espai (IEEC/CSIC), the Institut de Física d’Altes Energies, Lawrence Berkeley National Laboratory, the Ludwig-Maximilians Universität München and the associated Excellence Cluster Universe, the University of Michigan, the National Optical Astronomy Observatory, the University of Nottingham, The Ohio State University, the OzDES Membership Consortium, the University of Pennsylvania, the University of Portsmouth, SLAC National Accelerator Laboratory, Stanford University, the University of Sussex, and Texas A\&M University.
%Based in part on observations at Cerro Tololo Inter-American Observatory, National Optical Astronomy Observatory, which is operated by the Association of Universities for Research in Astronomy (AURA) under a cooperative agreement with the National Science Foundation. 
The national facility capability for SkyMapper has been funded through ARC LIEF grant LE130100104 from the Australian Research Council, awarded to the University of Sydney, the Australian National University, Swinburne University of Technology, the University of Queensland, the University of Western Australia, the University of Melbourne, Curtin University of Technology, Monash University and the Australian Astronomical Observatory. SkyMapper is owned and operated by The Australian National University's Research School of Astronomy and Astrophysics. The survey data were processed and provided by the SkyMapper Team at ANU. The SkyMapper node of the All-Sky Virtual Observatory (ASVO) is hosted at the National Computational Infrastructure (NCI). Development and support of the SkyMapper node of the ASVO has been funded in part by Astronomy Australia Limited (AAL) and the Australian Government through the Commonwealth's Education Investment Fund (EIF) and National Collaborative Research Infrastructure Strategy (NCRIS), particularly the National eResearch Collaboration Tools and Resources (NeCTAR) and the Australian National Data Service Projects (ANDS).
This work made use of Astropy:\footnote{http://www.astropy.org} a community-developed core Python package and an ecosystem of tools and resources for astronomy \citep{astropy2013, astropy2018, astropy2022}. 
\end{acknowledgements}


%%%%%%%%%%%%%%%%%%%%%%%%%%%%%%%%%%%%%%%%%%%%
%%%%% Bibliography %%%%%
%%%%%%%%%%%%%%%%%%%%%%%%%%%%%%%%%%%%%%%%%%%%
%

\bibliographystyle{aa} % style aa.bst
\bibliography{bibliography} 

\begin{thebibliography}{106}
\expandafter\ifx\csname natexlab\endcsname\relax\def\natexlab#1{#1}\fi

\bibitem[{{Abbott} {et~al.}(2018){Abbott}, {Abdalla}, {Allam}, {Amara},
  {Annis}, {Asorey}, {Avila}, {Ballester}, {Banerji}, {Barkhouse}, {Baruah},
  {Baumer}, {Bechtol}, {Becker}, {Benoit-L{\'e}vy}, {Bernstein}, {Bertin},
  {Blazek}, {Bocquet}, {Brooks}, {Brout}, {Buckley-Geer}, {Burke}, {Busti},
  {Campisano}, {Cardiel-Sas}, {Carnero Rosell}, {Carrasco Kind}, {Carretero},
  {Castander}, {Cawthon}, {Chang}, {Chen}, {Conselice}, {Costa}, {Crocce},
  {Cunha}, {D'Andrea}, {da Costa}, {Das}, {Daues}, {Davis}, {Davis}, {De
  Vicente}, {DePoy}, {DeRose}, {Desai}, {Diehl}, {Dietrich}, {Dodelson},
  {Doel}, {Drlica-Wagner}, {Eifler}, {Elliott}, {Evrard}, {Farahi}, {Fausti
  Neto}, {Fernandez}, {Finley}, {Flaugher}, {Foley}, {Fosalba}, {Friedel},
  {Frieman}, {Garc{\'\i}a-Bellido}, {Gaztanaga}, {Gerdes}, {Giannantonio},
  {Gill}, {Glazebrook}, {Goldstein}, {Gower}, {Gruen}, {Gruendl}, {Gschwend},
  {Gupta}, {Gutierrez}, {Hamilton}, {Hartley}, {Hinton}, {Hislop}, {Hollowood},
  {Honscheid}, {Hoyle}, {Huterer}, {Jain}, {James}, {Jeltema}, {Johnson},
  {Johnson}, {Kacprzak}, {Kent}, {Khullar}, {Klein}, {Kovacs}, {Koziol},
  {Krause}, {Kremin}, {Kron}, {Kuehn}, {Kuhlmann}, {Kuropatkin}, {Lahav},
  {Lasker}, {Li}, {Li}, {Liddle}, {Lima}, {Lin}, {L{\'o}pez-Reyes}, {MacCrann},
  {Maia}, {Maloney}, {Manera}, {March}, {Marriner}, {Marshall}, {Martini},
  {McClintock}, {McKay}, {McMahon}, {Melchior}, {Menanteau}, {Miller},
  {Miquel}, {Mohr}, {Morganson}, {Mould}, {Neilsen}, {Nichol}, {Nogueira},
  {Nord}, {Nugent}, {Nunes}, {Ogando}, {Old}, {Pace}, {Palmese},
  {Paz-Chinch{\'o}n}, {Peiris}, {Percival}, {Petravick}, {Plazas}, {Poh},
  {Pond}, {Porredon}, {Pujol}, {Refregier}, {Reil}, {Ricker}, {Rollins},
  {Romer}, {Roodman}, {Rooney}, {Ross}, {Rykoff}, {Sako}, {Sanchez}, {Sanchez},
  {Santiago}, {Saro}, {Scarpine}, {Scolnic}, {Serrano}, {Sevilla-Noarbe},
  {Sheldon}, {Shipp}, {Silveira}, {Smith}, {Smith}, {Smith}, {Soares-Santos},
  {Sobreira}, {Song}, {Stebbins}, {Suchyta}, {Sullivan}, {Swanson}, {Tarle},
  {Thaler}, {Thomas}, {Thomas}, {Troxel}, {Tucker}, {Vikram}, {Vivas},
  {Walker}, {Wechsler}, {Weller}, {Wester}, {Wolf}, {Wu}, {Yanny}, {Zenteno},
  {Zhang}, {Zuntz}, {DES Collaboration}, {Juneau}, {Fitzpatrick}, {Nikutta},
  {Nidever}, {Olsen}, {Scott}, \& {NOAO Data Lab}}]{abbott2019DESdr1}
{Abbott}, T.~M.~C., {Abdalla}, F.~B., {Allam}, S., {et~al.} 2018, \apjs, 239,
  18

\bibitem[{{Abbott} {et~al.}(2021){Abbott}, {Adam{\'o}w}, {Aguena}, {Allam},
  {Amon}, {Annis}, {Avila}, {Bacon}, {Banerji}, {Bechtol}, {Becker},
  {Bernstein}, {Bertin}, {Bhargava}, {Bridle}, {Brooks}, {Burke}, {Carnero
  Rosell}, {Carrasco Kind}, {Carretero}, {Castander}, {Cawthon}, {Chang},
  {Choi}, {Conselice}, {Costanzi}, {Crocce}, {da Costa}, {Davis}, {De Vicente},
  {DeRose}, {Desai}, {Diehl}, {Dietrich}, {Drlica-Wagner}, {Eckert},
  {Elvin-Poole}, {Everett}, {Evrard}, {Ferrero}, {Fert{\'e}}, {Flaugher},
  {Fosalba}, {Friedel}, {Frieman}, {Garc{\'\i}a-Bellido}, {Gaztanaga},
  {Gelman}, {Gerdes}, {Giannantonio}, {Gill}, {Gruen}, {Gruendl}, {Gschwend},
  {Gutierrez}, {Hartley}, {Hinton}, {Hollowood}, {Honscheid}, {Huterer},
  {James}, {Jeltema}, {Johnson}, {Kent}, {Kron}, {Kuehn}, {Kuropatkin},
  {Lahav}, {Li}, {Lidman}, {Lin}, {MacCrann}, {Maia}, {Manning}, {Maloney},
  {March}, {Marshall}, {Martini}, {Melchior}, {Menanteau}, {Miquel}, {Morgan},
  {Myles}, {Neilsen}, {Ogando}, {Palmese}, {Paz-Chinch{\'o}n}, {Petravick},
  {Pieres}, {Plazas}, {Pond}, {Rodriguez-Monroy}, {Romer}, {Roodman}, {Rykoff},
  {Sako}, {Sanchez}, {Santiago}, {Scarpine}, {Serrano}, {Sevilla-Noarbe},
  {Smith}, {Smith}, {Soares-Santos}, {Suchyta}, {Swanson}, {Tarle}, {Thomas},
  {To}, {Tremblay}, {Troxel}, {Tucker}, {Turner}, {Varga}, {Walker},
  {Wechsler}, {Weller}, {Wester}, {Wilkinson}, {Yanny}, {Zhang}, {Nikutta},
  {Fitzpatrick}, {Jacques}, {Scott}, {Olsen}, {Huang}, {Herrera}, {Juneau},
  {Nidever}, {Weaver}, {Adean}, {Correia}, {de Freitas}, {Freitas},
  {Singulani}, {Vila-Verde}, \& {Linea Science Server}}]{abbott2021DESdr2}
{Abbott}, T.~M.~C., {Adam{\'o}w}, M., {Aguena}, M., {et~al.} 2021, \apjs, 255,
  20

\bibitem[{{Allard} {et~al.}(2003){Allard}, {Allard}, {Hauschildt}, {Kielkopf},
  \& {Machin}}]{allard2003Na_K}
{Allard}, N.~F., {Allard}, F., {Hauschildt}, P.~H., {Kielkopf}, J.~F., \&
  {Machin}, L. 2003, \aap, 411, L473

\bibitem[{{Allard} {et~al.}(2024){Allard}, {Kielkopf}, {Myneni}, \&
  {Blakely}}]{allard2024K_He}
{Allard}, N.~F., {Kielkopf}, J.~F., {Myneni}, K., \& {Blakely}, J.~N. 2024,
  \aap, 683, A188

\bibitem[{{Allard} {et~al.}(2023){Allard}, {Myneni}, {Blakely}, \&
  {Guillon}}]{allard2023Na_He}
{Allard}, N.~F., {Myneni}, K., {Blakely}, J.~N., \& {Guillon}, G. 2023, \aap,
  674, A171

\bibitem[{{Allard} {et~al.}(2016){Allard}, {Spiegelman}, \&
  {Kielkopf}}]{allard2016K_H2}
{Allard}, N.~F., {Spiegelman}, F., \& {Kielkopf}, J.~F. 2016, \aap, 589, A21

\bibitem[{{Appenzeller} {et~al.}(1998){Appenzeller}, {Fricke}, {F{\"u}rtig},
  {G{\"a}ssler}, {H{\"a}fner}, {Harke}, {Hess}, {Hummel}, {J{\"u}rgens},
  {Kudritzki}, {Mantel}, {Meisl}, {Muschielok}, {Nicklas}, {Rupprecht},
  {Seifert}, {Stahl}, {Szeifert}, \& {Tarantik}}]{appenzeller1998fors}
{Appenzeller}, I., {Fricke}, K., {F{\"u}rtig}, W., {et~al.} 1998, The
  Messenger, 94, 1

\bibitem[{{Astropy Collaboration} {et~al.}(2022){Astropy Collaboration},
  {Price-Whelan}, {Lim}, {Earl}, {Starkman}, {Bradley}, {Shupe}, {Patil},
  {Corrales}, {Brasseur}, {N{\"o}the}, {Donath}, {Tollerud}, {Morris},
  {Ginsburg}, {Vaher}, {Weaver}, {Tocknell}, {Jamieson}, {van Kerkwijk},
  {Robitaille}, {Merry}, {Bachetti}, {G{\"u}nther}, {Aldcroft},
  {Alvarado-Montes}, {Archibald}, {B{\'o}di}, {Bapat}, {Barentsen},
  {Baz{\'a}n}, {Biswas}, {Boquien}, {Burke}, {Cara}, {Cara}, {Conroy},
  {Conseil}, {Craig}, {Cross}, {Cruz}, {D'Eugenio}, {Dencheva}, {Devillepoix},
  {Dietrich}, {Eigenbrot}, {Erben}, {Ferreira}, {Foreman-Mackey}, {Fox},
  {Freij}, {Garg}, {Geda}, {Glattly}, {Gondhalekar}, {Gordon}, {Grant},
  {Greenfield}, {Groener}, {Guest}, {Gurovich}, {Handberg}, {Hart},
  {Hatfield-Dodds}, {Homeier}, {Hosseinzadeh}, {Jenness}, {Jones}, {Joseph},
  {Kalmbach}, {Karamehmetoglu}, {Ka{\l}uszy{\'n}ski}, {Kelley}, {Kern},
  {Kerzendorf}, {Koch}, {Kulumani}, {Lee}, {Ly}, {Ma}, {MacBride}, {Maljaars},
  {Muna}, {Murphy}, {Norman}, {O'Steen}, {Oman}, {Pacifici}, {Pascual},
  {Pascual-Granado}, {Patil}, {Perren}, {Pickering}, {Rastogi}, {Roulston},
  {Ryan}, {Rykoff}, {Sabater}, {Sakurikar}, {Salgado}, {Sanghi}, {Saunders},
  {Savchenko}, {Schwardt}, {Seifert-Eckert}, {Shih}, {Jain}, {Shukla}, {Sick},
  {Simpson}, {Singanamalla}, {Singer}, {Singhal}, {Sinha}, {Sip{\H{o}}cz},
  {Spitler}, {Stansby}, {Streicher}, {{\v{S}}umak}, {Swinbank}, {Taranu},
  {Tewary}, {Tremblay}, {de Val-Borro}, {Van Kooten}, {Vasovi{\'c}}, {Verma},
  {de Miranda Cardoso}, {Williams}, {Wilson}, {Winkel}, {Wood-Vasey}, {Xue},
  {Yoachim}, {Zhang}, {Zonca}, \& {Astropy Project Contributors}}]{astropy2022}
{Astropy Collaboration}, {Price-Whelan}, A.~M., {Lim}, P.~L., {et~al.} 2022,
  \apj, 935, 167

\bibitem[{{Astropy Collaboration} {et~al.}(2018){Astropy Collaboration},
  {Price-Whelan}, {Sip{\H{o}}cz}, {G{\"u}nther}, {Lim}, {Crawford}, {Conseil},
  {Shupe}, {Craig}, {Dencheva}, {Ginsburg}, {VanderPlas}, {Bradley},
  {P{\'e}rez-Su{\'a}rez}, {de Val-Borro}, {Aldcroft}, {Cruz}, {Robitaille},
  {Tollerud}, {Ardelean}, {Babej}, {Bach}, {Bachetti}, {Bakanov}, {Bamford},
  {Barentsen}, {Barmby}, {Baumbach}, {Berry}, {Biscani}, {Boquien}, {Bostroem},
  {Bouma}, {Brammer}, {Bray}, {Breytenbach}, {Buddelmeijer}, {Burke},
  {Calderone}, {Cano Rodr{\'\i}guez}, {Cara}, {Cardoso}, {Cheedella}, {Copin},
  {Corrales}, {Crichton}, {D'Avella}, {Deil}, {Depagne}, {Dietrich}, {Donath},
  {Droettboom}, {Earl}, {Erben}, {Fabbro}, {Ferreira}, {Finethy}, {Fox},
  {Garrison}, {Gibbons}, {Goldstein}, {Gommers}, {Greco}, {Greenfield},
  {Groener}, {Grollier}, {Hagen}, {Hirst}, {Homeier}, {Horton}, {Hosseinzadeh},
  {Hu}, {Hunkeler}, {Ivezi{\'c}}, {Jain}, {Jenness}, {Kanarek}, {Kendrew},
  {Kern}, {Kerzendorf}, {Khvalko}, {King}, {Kirkby}, {Kulkarni}, {Kumar},
  {Lee}, {Lenz}, {Littlefair}, {Ma}, {Macleod}, {Mastropietro}, {McCully},
  {Montagnac}, {Morris}, {Mueller}, {Mumford}, {Muna}, {Murphy}, {Nelson},
  {Nguyen}, {Ninan}, {N{\"o}the}, {Ogaz}, {Oh}, {Parejko}, {Parley}, {Pascual},
  {Patil}, {Patil}, {Plunkett}, {Prochaska}, {Rastogi}, {Reddy Janga},
  {Sabater}, {Sakurikar}, {Seifert}, {Sherbert}, {Sherwood-Taylor}, {Shih},
  {Sick}, {Silbiger}, {Singanamalla}, {Singer}, {Sladen}, {Sooley},
  {Sornarajah}, {Streicher}, {Teuben}, {Thomas}, {Tremblay}, {Turner},
  {Terr{\'o}n}, {van Kerkwijk}, {de la Vega}, {Watkins}, {Weaver}, {Whitmore},
  {Woillez}, {Zabalza}, \& {Astropy Contributors}}]{astropy2018}
{Astropy Collaboration}, {Price-Whelan}, A.~M., {Sip{\H{o}}cz}, B.~M., {et~al.}
  2018, \aj, 156, 123

\bibitem[{{Astropy Collaboration} {et~al.}(2013){Astropy Collaboration},
  {Robitaille}, {Tollerud}, {Greenfield}, {Droettboom}, {Bray}, {Aldcroft},
  {Davis}, {Ginsburg}, {Price-Whelan}, {Kerzendorf}, {Conley}, {Crighton},
  {Barbary}, {Muna}, {Ferguson}, {Grollier}, {Parikh}, {Nair}, {Unther},
  {Deil}, {Woillez}, {Conseil}, {Kramer}, {Turner}, {Singer}, {Fox}, {Weaver},
  {Zabalza}, {Edwards}, {Azalee Bostroem}, {Burke}, {Casey}, {Crawford},
  {Dencheva}, {Ely}, {Jenness}, {Labrie}, {Lim}, {Pierfederici}, {Pontzen},
  {Ptak}, {Refsdal}, {Servillat}, \& {Streicher}}]{astropy2013}
{Astropy Collaboration}, {Robitaille}, T.~P., {Tollerud}, E.~J., {et~al.} 2013,
  \aap, 558, A33

\bibitem[{{Bailer-Jones} {et~al.}(2000){Bailer-Jones}, {Bizenberger}, \&
  {Storz}}]{omega2000}
{Bailer-Jones}, C.~A., {Bizenberger}, P., \& {Storz}, C. 2000, in Society of
  Photo-Optical Instrumentation Engineers (SPIE) Conference Series, Vol. 4008,
  Optical and IR Telescope Instrumentation and Detectors, ed. M.~{Iye} \& A.~F.
  {Moorwood}, 1305--1316

\bibitem[{{Baumeister} {et~al.}(2003){Baumeister}, {Bizenberger},
  {Bayler-Jones}, {Kov{\'a}cs}, {R{\"o}ser}, \& {Rohloff}}]{omega2000_cryo}
{Baumeister}, H., {Bizenberger}, P., {Bayler-Jones}, C. A.~L., {et~al.} 2003,
  in Society of Photo-Optical Instrumentation Engineers (SPIE) Conference
  Series, Vol. 4841, Instrument Design and Performance for Optical/Infrared
  Ground-based Telescopes, ed. M.~{Iye} \& A.~F.~M. {Moorwood}, 343--354

\bibitem[{{Beam{\'\i}n} {et~al.}(2014){Beam{\'\i}n}, {Ivanov}, {Bayo},
  {Mu{\v{z}}i{\'c}}, {Boffin}, {Allard}, {Homeier}, {Minniti}, {Gromadzki},
  {Kurtev}, {Lodieu}, {Martin}, \& {Mendez}}]{beamin2014W0855_260K}
{Beam{\'\i}n}, J.~C., {Ivanov}, V.~D., {Bayo}, A., {et~al.} 2014, \aap, 570, L8

\bibitem[{{Best} {et~al.}(2020){Best}, {Liu}, {Magnier}, \&
  {Dupuy}}]{williambest2020_L0-8UKIRTparallax}
{Best}, W. M.~J., {Liu}, M.~C., {Magnier}, E.~A., \& {Dupuy}, T.~J. 2020, \aj,
  159, 257

\bibitem[{{Best} {et~al.}(2021){Best}, {Liu}, {Magnier}, \&
  {Dupuy}}]{best2021ucd25pc}
{Best}, W. M.~J., {Liu}, M.~C., {Magnier}, E.~A., \& {Dupuy}, T.~J. 2021, \aj,
  161, 42

\bibitem[{{Best} {et~al.}(2018){Best}, {Magnier}, {Liu}, {Aller}, {Zhang},
  {Burgett}, {Chambers}, {Draper}, {Flewelling}, {Kaiser}, {Kudritzki},
  {Metcalfe}, {Tonry}, {Wainscoat}, \& {Waters}}]{best2018PS1_3pi}
{Best}, W. M.~J., {Magnier}, E.~A., {Liu}, M.~C., {et~al.} 2018, \apjs, 234, 1

\bibitem[{{Brooks} {et~al.}(2022){Brooks}, {Kirkpatrick}, {Caselden},
  {Schneider}, {Meisner}, {Faherty}, {Casewell}, {Kuchner}, {Kuchner}, \&
  {Backyard Worlds: Planet 9 Collaboration}}]{brooks2022esdT}
{Brooks}, H., {Kirkpatrick}, J.~D., {Caselden}, D., {et~al.} 2022, \aj, 163, 47

\bibitem[{{Burgasser} {et~al.}(2002){Burgasser}, {Kirkpatrick}, {Brown},
  {Reid}, {Burrows}, {Liebert}, {Matthews}, {Gizis}, {Dahn}, {Monet}, {Cutri},
  \& {Skrutskie}}]{burgasser2002dT_spec_classification}
{Burgasser}, A.~J., {Kirkpatrick}, J.~D., {Brown}, M.~E., {et~al.} 2002, \apj,
  564, 421

\bibitem[{{Burgasser} {et~al.}(2003{\natexlab{a}}){Burgasser}, {Kirkpatrick},
  {Burrows}, {Liebert}, {Reid}, {Gizis}, {McGovern}, {Prato}, \&
  {McLean}}]{burgasser2003esdL7_2M0532}
{Burgasser}, A.~J., {Kirkpatrick}, J.~D., {Burrows}, A., {et~al.}
  2003{\natexlab{a}}, \apj, 592, 1186

\bibitem[{{Burgasser} {et~al.}(2003{\natexlab{b}}){Burgasser}, {Kirkpatrick},
  {Liebert}, \& {Burrows}}]{burgasser2003dT_optical}
{Burgasser}, A.~J., {Kirkpatrick}, J.~D., {Liebert}, J., \& {Burrows}, A.
  2003{\natexlab{b}}, \apj, 594, 510

\bibitem[{{Burgasser} {et~al.}(2024){Burgasser}, {Schneider}, {Meisner},
  {Caselden}, {Hsu}, {Gerasimov}, {Aganze}, {Softich}, {Karpoor}, {Theissen},
  {Brooks}, {Bickle}, {Gagn{\'e}}, {Artigau}, {Marsset}, {Rothermich},
  {Faherty}, {Kirkpatrick}, {Kuchner}, {Stevnbak Andersen}, {Beaulieu},
  {Colin}, {Gantier}, {Gramaize}, {Hamlet}, {Hinckley}, {Kabatnik}, {Kiwy},
  {Martin}, {Massat}, {Pendrill}, {Sainio}, {Sch{\"u}mann}, {Th{\'e}venot},
  {Walla}, {W{\k{e}}dracki}, {Backyard Worlds}, {:}, \& {Planet 9
  Collaboration}}]{burgasser2024esdT}
{Burgasser}, A.~J., {Schneider}, A.~C., {Meisner}, A.~M., {et~al.} 2024, arXiv
  e-prints, arXiv:2411.01378

\bibitem[{{Burgasser} {et~al.}(2009){Burgasser}, {Witte}, {Helling},
  {Sanderson}, {Bochanski}, \& {Hauschildt}}]{burgasser2009optNIRsdL}
{Burgasser}, A.~J., {Witte}, S., {Helling}, C., {et~al.} 2009, \apj, 697, 148

\bibitem[{{Burningham} {et~al.}(2010){Burningham}, {Leggett}, {Lucas},
  {Pinfield}, {Smart}, {Day-Jones}, {Jones}, {Murray}, {Nickson}, {Tamura},
  {Zhang}, {Lodieu}, {Tinney}, \& {Zapatero
  Osorio}}]{burningham2010cool_binary_sdL}
{Burningham}, B., {Leggett}, S.~K., {Lucas}, P.~W., {et~al.} 2010, \mnras, 404,
  1952

\bibitem[{{Burningham} {et~al.}(2014){Burningham}, {Smith}, {Cardoso}, {Lucas},
  {Burgasser}, {Jones}, \& {Smart}}]{burningham2014T6.5}
{Burningham}, B., {Smith}, L., {Cardoso}, C.~V., {et~al.} 2014, \mnras, 440,
  359

\bibitem[{{Burrows} \& {Sharp}(1999)}]{burrows&sharp1999rainout}
{Burrows}, A. \& {Sharp}, C.~M. 1999, \apj, 512, 843

\bibitem[{{Burrows} \& {Volobuyev}(2003)}]{burrows2003NaK}
{Burrows}, A. \& {Volobuyev}, M. 2003, \apj, 583, 985

\bibitem[{Cepa {et~al.}(2000)Cepa, Aguiar-Gonzalez, Gonzalez-Escalera,
  Gonzalez-Serrano, Joven-Alvarez, Cano, Rasilla, Rodriguez-Ramos,
  Gonz{\'a}lez, Duenas, {et~al.}}]{cepa2000osiris}
Cepa, J., Aguiar-Gonzalez, M., Gonzalez-Escalera, V., {et~al.} 2000, in Optical
  and IR Telescope Instrumentation and Detectors, Vol. 4008, SPIE, 623--631

\bibitem[{{Chabrier} {et~al.}(2023){Chabrier}, {Baraffe}, {Phillips}, \&
  {Debras}}]{chabrier2023BDnewHlimit}
{Chabrier}, G., {Baraffe}, I., {Phillips}, M., \& {Debras}, F. 2023, \aap, 671,
  A119

\bibitem[{{Chambers} {et~al.}(2016){Chambers}, {Magnier}, {Metcalfe},
  {Flewelling}, {Huber}, {Waters}, {Denneau}, {Draper}, {Farrow}, {Finkbeiner},
  {Holmberg}, {Koppenhoefer}, {Price}, {Rest}, {Saglia}, {Schlafly}, {Smartt},
  {Sweeney}, {Wainscoat}, {Burgett}, {Chastel}, {Grav}, {Heasley}, {Hodapp},
  {Jedicke}, {Kaiser}, {Kudritzki}, {Luppino}, {Lupton}, {Monet}, {Morgan},
  {Onaka}, {Shiao}, {Stubbs}, {Tonry}, {White}, {Ba{\~n}ados}, {Bell},
  {Bender}, {Bernard}, {Boegner}, {Boffi}, {Botticella}, {Calamida},
  {Casertano}, {Chen}, {Chen}, {Cole}, {Deacon}, {Frenk}, {Fitzsimmons},
  {Gezari}, {Gibbs}, {Goessl}, {Goggia}, {Gourgue}, {Goldman}, {Grant},
  {Grebel}, {Hambly}, {Hasinger}, {Heavens}, {Heckman}, {Henderson}, {Henning},
  {Holman}, {Hopp}, {Ip}, {Isani}, {Jackson}, {Keyes}, {Koekemoer}, {Kotak},
  {Le}, {Liska}, {Long}, {Lucey}, {Liu}, {Martin}, {Masci}, {McLean}, {Mindel},
  {Misra}, {Morganson}, {Murphy}, {Obaika}, {Narayan}, {Nieto-Santisteban},
  {Norberg}, {Peacock}, {Pier}, {Postman}, {Primak}, {Rae}, {Rai}, {Riess},
  {Riffeser}, {Rix}, {R{\"o}ser}, {Russel}, {Rutz}, {Schilbach}, {Schultz},
  {Scolnic}, {Strolger}, {Szalay}, {Seitz}, {Small}, {Smith}, {Soderblom},
  {Taylor}, {Thomson}, {Taylor}, {Thakar}, {Thiel}, {Thilker}, {Unger},
  {Urata}, {Valenti}, {Wagner}, {Walder}, {Walter}, {Watters}, {Werner},
  {Wood-Vasey}, \& {Wyse}}]{chanbers2016panstarrs}
{Chambers}, K.~C., {Magnier}, E.~A., {Metcalfe}, N., {et~al.} 2016, arXiv
  e-prints, arXiv:1612.05560

\bibitem[{{Cushing} {et~al.}(2011){Cushing}, {Kirkpatrick}, {Gelino},
  {Griffith}, {Skrutskie}, {Mainzer}, {Marsh}, {Beichman}, {Burgasser},
  {Prato}, {Simcoe}, {Marley}, {Saumon}, {Freedman}, {Eisenhardt}, \&
  {Wright}}]{cushing2011dY_WISE}
{Cushing}, M.~C., {Kirkpatrick}, J.~D., {Gelino}, C.~R., {et~al.} 2011, \apj,
  743, 50

\bibitem[{{Cushing} {et~al.}(2009){Cushing}, {Looper}, {Burgasser},
  {Kirkpatrick}, {Faherty}, {Cruz}, {Sweet}, \& {Sanderson}}]{cushing2009haloL}
{Cushing}, M.~C., {Looper}, D., {Burgasser}, A.~J., {et~al.} 2009, \apj, 696,
  986

\bibitem[{{Cutri} {et~al.}(2003){Cutri}, {Skrutskie}, {van Dyk}, {Beichman},
  {Carpenter}, {Chester}, {Cambresy}, {Evans}, {Fowler}, {Gizis}, {Howard},
  {Huchra}, {Jarrett}, {Kopan}, {Kirkpatrick}, {Light}, {Marsh}, {McCallon},
  {Schneider}, {Stiening}, {Sykes}, {Weinberg}, {Wheaton}, {Wheelock}, \&
  {Zacarias}}]{cutri2003_2MASSpoint}
{Cutri}, R.~M., {Skrutskie}, M.~F., {van Dyk}, S., {et~al.} 2003, {2MASS All
  Sky Catalog of point sources.}

\bibitem[{{Delorme} {et~al.}(2008){Delorme}, {Delfosse}, {Albert}, {Artigau},
  {Forveille}, {Reyl{\'e}}, {Allard}, {Homeier}, {Robin}, {Willott}, {Liu}, \&
  {Dupuy}}]{delorme2008TY_J0059}
{Delorme}, P., {Delfosse}, X., {Albert}, L., {et~al.} 2008, \aap, 482, 961

\bibitem[{{Dhillon} {et~al.}(2021){Dhillon}, {Bezawada}, {Black}, {Dixon},
  {Gamble}, {Gao}, {Henry}, {Kerry}, {Littlefair}, {Lunney}, {Marsh}, {Miller},
  {Parsons}, {Ashley}, {Breedt}, {Brown}, {Dyer}, {Green}, {Pelisoli},
  {Sahman}, {Wild}, {Ives}, {Mehrgan}, {Stegmeier}, {Dubbeldam}, {Morris},
  {Osborn}, {Wilson}, {Casares}, {Mu{\~n}oz-Darias}, {Pall{\'e}},
  {Rodr{\'\i}guez-Gil}, {Shahbaz}, {Torres}, {de Ugarte Postigo},
  {Cabrera-Lavers}, {Corradi}, {Dom{\'\i}nguez}, \&
  {Garc{\'\i}a-Alvarez}}]{dhillon2021hipercam}
{Dhillon}, V.~S., {Bezawada}, N., {Black}, M., {et~al.} 2021, \mnras, 507, 350

\bibitem[{{Freudling} {et~al.}(2013){Freudling}, {Romaniello}, {Bramich},
  {Ballester}, {Forchi}, {Garc{\'{\i}}a-Dabl{\'o}}, {Moehler}, \&
  {Neeser}}]{freudling2013esoreflex}
{Freudling}, W., {Romaniello}, M., {Bramich}, D.~M., {et~al.} 2013, \aap, 559,
  A96

\bibitem[{{Gizis} {et~al.}(1997){Gizis}, {Scholz}, {Irwin}, \&
  {Jahreiss}}]{gizis1997pm_subdwarf}
{Gizis}, J.~E., {Scholz}, R.~D., {Irwin}, M., \& {Jahreiss}, H. 1997, \mnras,
  292, L41

\bibitem[{{Goodman}(2021)}]{goodman2021W0523}
{Goodman}, S.~J. 2021, Research Notes of the American Astronomical Society, 5,
  178

\bibitem[{{Greco} {et~al.}(2019){Greco}, {Schneider}, {Cushing}, {Kirkpatrick},
  \& {Burgasser}}]{greco2019neowise}
{Greco}, J.~J., {Schneider}, A.~C., {Cushing}, M.~C., {Kirkpatrick}, J.~D., \&
  {Burgasser}, A.~J. 2019, \aj, 158, 182

\bibitem[{{Kellogg} {et~al.}(2018){Kellogg}, {Kirkpatrick}, {Metchev},
  {Gagn{\'e}}, \& {Faherty}}]{kellog2018_W0711}
{Kellogg}, K., {Kirkpatrick}, J.~D., {Metchev}, S., {Gagn{\'e}}, J., \&
  {Faherty}, J.~K. 2018, \aj, 155, 87

\bibitem[{{Kirkpatrick} {et~al.}(2012){Kirkpatrick}, {Gelino}, {Cushing},
  {Mace}, {Griffith}, {Skrutskie}, {Marsh}, {Wright}, {Eisenhardt}, {McLean},
  {Mainzer}, {Burgasser}, {Tinney}, {Parker}, \&
  {Salter}}]{kirkpatrick2012furtherY}
{Kirkpatrick}, J.~D., {Gelino}, C.~R., {Cushing}, M.~C., {et~al.} 2012, \apj,
  753, 156

\bibitem[{{Kirkpatrick} {et~al.}(2021{\natexlab{a}}){Kirkpatrick}, {Gelino},
  {Faherty}, {Meisner}, {Caselden}, {Schneider}, {Marocco}, {Cayago}, {Smart},
  {Eisenhardt}, {Kuchner}, {Wright}, {Cushing}, {Allers}, {Bardalez Gagliuffi},
  {Burgasser}, {Gagn{\'e}}, {Logsdon}, {Martin}, {Ingalls}, {Lowrance},
  {Abrahams}, {Aganze}, {Gerasimov}, {Gonzales}, {Hsu}, {Kamraj}, {Kiman},
  {Rees}, {Theissen}, {Ammar}, {Andersen}, {Beaulieu}, {Colin}, {Elachi},
  {Goodman}, {Gramaize}, {Hamlet}, {Hong}, {Jonkeren}, {Khalil}, {Martin},
  {Pendrill}, {Pumphrey}, {Rothermich}, {Sainio}, {Stenner}, {Tanner},
  {Th{\'e}venot}, {Voloshin}, {Walla}, {W{\k{e}}dracki}, \& {Backyard Worlds:
  Planet 9 Collaboration}}]{kirkpatrick2021census20pc}
{Kirkpatrick}, J.~D., {Gelino}, C.~R., {Faherty}, J.~K., {et~al.}
  2021{\natexlab{a}}, \apjs, 253, 7

\bibitem[{{Kirkpatrick} {et~al.}(2021{\natexlab{b}}){Kirkpatrick}, {Marocco},
  {Caselden}, {Meisner}, {Faherty}, {Schneider}, {Kuchner}, {Casewell},
  {Gelino}, {Cushing}, {Eisenhardt}, {Wright}, \&
  {Schurr}}]{kirkpatrick2021accident}
{Kirkpatrick}, J.~D., {Marocco}, F., {Caselden}, D., {et~al.}
  2021{\natexlab{b}}, \apjl, 915, L6

\bibitem[{{Kirkpatrick} {et~al.}(2019){Kirkpatrick}, {Martin}, {Smart},
  {Cayago}, {Beichman}, {Marocco}, {Gelino}, {Faherty}, {Cushing}, {Schneider},
  {Mace}, {Tinney}, {Wright}, {Lowrance}, {Ingalls}, {Vrba}, {Munn}, {Dahm}, \&
  {McLean}}]{kirkpatrick2019parallax184TY}
{Kirkpatrick}, J.~D., {Martin}, E.~C., {Smart}, R.~L., {et~al.} 2019, \apjs,
  240, 19

\bibitem[{{Kirkpatrick} {et~al.}(2014){Kirkpatrick}, {Schneider},
  {Fajardo-Acosta}, {Gelino}, {Mace}, {Wright}, {Logsdon}, {McLean}, {Cushing},
  {Skrutskie}, {Eisenhardt}, {Stern}, {Balokovi{\'c}}, {Burgasser}, {Faherty},
  {Lansbury}, {Rich}, {Skrzypek}, {Fowler}, {Cutri}, {Masci}, {Conrow},
  {Grillmair}, {McCallon}, {Beichman}, \& {Marsh}}]{kirkpatrick2014allwise_sd}
{Kirkpatrick}, J.~D., {Schneider}, A., {Fajardo-Acosta}, S., {et~al.} 2014,
  \apj, 783, 122

\bibitem[{{Kov{\'a}cs} {et~al.}(2004){Kov{\'a}cs}, {Mall}, {Bizenberger},
  {Baumeister}, \& {R{\"o}ser}}]{charact_omega2000}
{Kov{\'a}cs}, Z., {Mall}, U., {Bizenberger}, P., {Baumeister}, H., \&
  {R{\"o}ser}, H.-J. 2004, in Society of Photo-Optical Instrumentation
  Engineers (SPIE) Conference Series, Vol. 5499, Optical and Infrared Detectors
  for Astronomy, ed. J.~D. {Garnett} \& J.~W. {Beletic}, 432--441

\bibitem[{{Kuchner} {et~al.}(2017){Kuchner}, {Faherty}, {Schneider}, {Meisner},
  {Filippazzo}, {Gagn{\'e}}, {Trouille}, {Silverberg}, {Castro}, {Fletcher},
  {Mokaev}, \& {Stajic}}]{kuchner2017backyard}
{Kuchner}, M.~J., {Faherty}, J.~K., {Schneider}, A.~C., {et~al.} 2017, \apjl,
  841, L19

\bibitem[{{Lang} {et~al.}(2010){Lang}, {Hogg}, {Mierle}, {Blanton}, \&
  {Roweis}}]{dustin2010astrometry.net}
{Lang}, D., {Hogg}, D.~W., {Mierle}, K., {Blanton}, M., \& {Roweis}, S. 2010,
  \aj, 139, 1782

\bibitem[{{Leggett} {et~al.}(2021){Leggett}, {Tremblin}, {Phillips}, {Dupuy},
  {Marley}, {Morley}, {Schneider}, {Caselden}, {Guillaume}, \&
  {Logsdon}}]{leggett2021coldestSED}
{Leggett}, S.~K., {Tremblin}, P., {Phillips}, M.~W., {et~al.} 2021, \apj, 918,
  11

\bibitem[{{L{\'e}pine} {et~al.}(2007){L{\'e}pine}, {Rich}, \&
  {Shara}}]{lepine2007sdM}
{L{\'e}pine}, S., {Rich}, R.~M., \& {Shara}, M.~M. 2007, \apj, 669, 1235

\bibitem[{{Liu} {et~al.}(2012){Liu}, {Dupuy}, {Bowler}, {Leggett}, \&
  {Best}}]{liu2012W1217_TYbinary}
{Liu}, M.~C., {Dupuy}, T.~J., {Bowler}, B.~P., {Leggett}, S.~K., \& {Best}, W.
  M.~J. 2012, \apj, 758, 57

\bibitem[{{Lodders}(1999)}]{lodders1999alkali}
{Lodders}, K. 1999, \apj, 519, 793

\bibitem[{{Lodieu} {et~al.}(2019){Lodieu}, {Allard}, {Rodrigo}, {Pavlenko},
  {Burgasser}, {Lyubchik}, {Kaminsky}, \& {Homeier}}]{lodieu2019sdM}
{Lodieu}, N., {Allard}, F., {Rodrigo}, C., {et~al.} 2019, \aap, 628, A61

\bibitem[{{Lodieu} {et~al.}(2013){Lodieu}, {B{\'e}jar}, \&
  {Rebolo}}]{lodieu2013Yoptical}
{Lodieu}, N., {B{\'e}jar}, V.~J.~S., \& {Rebolo}, R. 2013, \aap, 550, L2

\bibitem[{{Lodieu} {et~al.}(2022){Lodieu}, {Zapatero Osorio}, {Mart{\'\i}n},
  {Rebolo L{\'o}pez}, \& {Gauza}}]{lodieu2022W1810}
{Lodieu}, N., {Zapatero Osorio}, M.~R., {Mart{\'\i}n}, E.~L., {Rebolo
  L{\'o}pez}, R., \& {Gauza}, B. 2022, \aap, 663, A84

\bibitem[{{Lodieu} {et~al.}(2010){Lodieu}, {Zapatero Osorio}, {Mart{\'\i}n},
  {Solano}, \& {Aberasturi}}]{lodieu2010GTCsdL}
{Lodieu}, N., {Zapatero Osorio}, M.~R., {Mart{\'\i}n}, E.~L., {Solano}, E., \&
  {Aberasturi}, M. 2010, \apjl, 708, L107

\bibitem[{{Luhman}(2014)}]{luhman2014W0855}
{Luhman}, K.~L. 2014, \apjl, 786, L18

\bibitem[{{Luhman} \& {Sheppard}(2014)}]{luhman2014WISEhighpm}
{Luhman}, K.~L. \& {Sheppard}, S.~S. 2014, \apj, 787, 126

\bibitem[{{Luhman} {et~al.}(2024){Luhman}, {Tremblin}, {Alves de Oliveira},
  {Birkmann}, {Baraffe}, {Chabrier}, {Manjavacas}, {Parker}, \&
  {Valenti}}]{luhman2024W0855}
{Luhman}, K.~L., {Tremblin}, P., {Alves de Oliveira}, C., {et~al.} 2024, \aj,
  167, 5

\bibitem[{{Mace} {et~al.}(2013{\natexlab{a}}){Mace}, {Kirkpatrick}, {Cushing},
  {Gelino}, {Griffith}, {Skrutskie}, {Marsh}, {Wright}, {Eisenhardt}, {McLean},
  {Thompson}, {Mix}, {Bailey}, {Beichman}, {Bloom}, {Burgasser}, {Fortney},
  {Hinz}, {Knox}, {Lowrance}, {Marley}, {Morley}, {Rodigas}, {Saumon},
  {Sheppard}, \& {Stock}}]{mace2013WISE_Tpopulation}
{Mace}, G.~N., {Kirkpatrick}, J.~D., {Cushing}, M.~C., {et~al.}
  2013{\natexlab{a}}, \apjs, 205, 6

\bibitem[{{Mace} {et~al.}(2013{\natexlab{b}}){Mace}, {Kirkpatrick}, {Cushing},
  {Gelino}, {McLean}, {Logsdon}, {Wright}, {Skrutskie}, {Beichman},
  {Eisenhardt}, \& {Kulas}}]{mace2013Wolf1130C}
{Mace}, G.~N., {Kirkpatrick}, J.~D., {Cushing}, M.~C., {et~al.}
  2013{\natexlab{b}}, \apj, 777, 36

\bibitem[{{Mace} {et~al.}(2018){Mace}, {Mann}, {Skiff}, {Sneden},
  {Kirkpatrick}, {Schneider}, {Kidder}, {Gosnell}, {Kim}, {Mulligan}, {Prato},
  \& {Jaffe}}]{mace2018Wolf1130}
{Mace}, G.~N., {Mann}, A.~W., {Skiff}, B.~A., {et~al.} 2018, \apj, 854, 145

\bibitem[{{Mainzer} {et~al.}(2014){Mainzer}, {Bauer}, {Cutri}, {Grav},
  {Masiero}, {Beck}, {Clarkson}, {Conrow}, {Dailey}, {Eisenhardt}, {Fabinsky},
  {Fajardo-Acosta}, {Fowler}, {Gelino}, {Grillmair}, {Heinrichsen}, {Kendall},
  {Kirkpatrick}, {Liu}, {Masci}, {McCallon}, {Nugent}, {Papin}, {Rice},
  {Royer}, {Ryan}, {Sevilla}, {Sonnett}, {Stevenson}, {Thompson}, {Wheelock},
  {Wiemer}, {Wittman}, {Wright}, \& {Yan}}]{mainzer2014neowise}
{Mainzer}, A., {Bauer}, J., {Cutri}, R.~M., {et~al.} 2014, \apj, 792, 30

\bibitem[{{Marley} {et~al.}(2021){Marley}, {Saumon}, {Visscher}, {Lupu},
  {Freedman}, {Morley}, {Fortney}, {Seay}, {Smith}, {Teal}, \&
  {Wang}}]{marley2021SONORA}
{Marley}, M.~S., {Saumon}, D., {Visscher}, C., {et~al.} 2021, \apj, 920, 85

\bibitem[{{Marocco} {et~al.}(2021){Marocco}, {Eisenhardt}, {Fowler},
  {Kirkpatrick}, {Meisner}, {Schlafly}, {Stanford}, {Garcia}, {Caselden},
  {Cushing}, {Cutri}, {Faherty}, {Gelino}, {Gonzalez}, {Jarrett}, {Koontz},
  {Mainzer}, {Marchese}, {Mobasher}, {Schlegel}, {Stern}, {Teplitz}, \&
  {Wright}}]{marocco2021catwise}
{Marocco}, F., {Eisenhardt}, P. R.~M., {Fowler}, J.~W., {et~al.} 2021, \apjs,
  253, 8

\bibitem[{{Mart{\'\i}n} {et~al.}(2021){Mart{\'\i}n}, {Zhang}, {Esparza},
  {Gracia}, {Rasilla}, {Masseron}, \& {Burgasser}}]{martin2021ch4nh3}
{Mart{\'\i}n}, E.~L., {Zhang}, J.~Y., {Esparza}, P., {et~al.} 2021, \aap, 655,
  L3

\bibitem[{{Mart{\'\i}n} {et~al.}(2024){Mart{\'\i}n}, {Zhang}, {Lanchas},
  {Lodieu}, {Shahbaz}, \& {Pavlenko}}]{martin2024Yoptical}
{Mart{\'\i}n}, E.~L., {Zhang}, J.~Y., {Lanchas}, H., {et~al.} 2024, \aap, 686,
  A73

\bibitem[{{Meisner} {et~al.}(2023{\natexlab{a}}){Meisner}, {Bardalez
  Gagliuffi}, {Bejar}, {Burgasser}, {Caselden}, {Casewell}, {Cushing},
  {Eisenhardt}, {Faherty}, {Gagne}, {Gauza}, {Gelino}, {Gonzales},
  {Kirkpatrick}, {Kuchner}, {Leggett}, {Line}, {Lodieu}, {Logsdon}, {Marocco},
  {Martin}, {Phillips}, {Schneider}, {Tremblin}, {Wright}, {Zapatero Osorio},
  \& {Zhang}}]{meisner2023jwst_proposal_acc}
{Meisner}, A., {Bardalez Gagliuffi}, D.~C., {Bejar}, V., {et~al.}
  2023{\natexlab{a}}, {The First Spectrum of the Coldest Halo Brown Dwarf},
  JWST Proposal. Cycle 2, ID. \#3558

\bibitem[{{Meisner} {et~al.}(2020{\natexlab{a}}){Meisner}, {Caselden},
  {Kirkpatrick}, {Marocco}, {Gelino}, {Cushing}, {Eisenhardt}, {Wright},
  {Faherty}, {Koontz}, {Marchese}, {Khalil}, {Fowler}, \&
  {Schlafly}}]{meisner2020Yspitzer}
{Meisner}, A.~M., {Caselden}, D., {Kirkpatrick}, J.~D., {et~al.}
  2020{\natexlab{a}}, \apj, 889, 74

\bibitem[{{Meisner} {et~al.}(2023{\natexlab{b}}){Meisner}, {Caselden},
  {Schlafly}, {Zelko}, {Kirkpatrick}, \& {Marocco}}]{meisner2023accident_Y}
{Meisner}, A.~M., {Caselden}, D., {Schlafly}, E.~F., {et~al.}
  2023{\natexlab{b}}, Research Notes of the American Astronomical Society, 7,
  36

\bibitem[{{Meisner} {et~al.}(2020{\natexlab{b}}){Meisner}, {Faherty},
  {Kirkpatrick}, {Schneider}, {Caselden}, {Gagn{\'e}}, {Kuchner}, {Burgasser},
  {Casewell}, {Debes}, {Artigau}, {Bardalez Gagliuffi}, {Logsdon}, {Kiman},
  {Allers}, {Hsu}, {Wisniewski}, {Allen}, {Beaulieu}, {Colin}, {Durantini
  Luca}, {Goodman}, {Gramaize}, {Hamlet}, {Hinckley}, {Kiwy}, {Martin},
  {Pendrill}, {Rothermich}, {Sainio}, {Sch{\"u}mann}, {Andersen}, {Tanner},
  {Thakur}, {Th{\'e}venot}, {Walla}, {W{\k{e}}dracki}, {Aganze}, {Gerasimov},
  {Theissen}, \& {Backyard Worlds: Planet 9
  Collaboration}}]{meisner2020extremecoldBD}
{Meisner}, A.~M., {Faherty}, J.~K., {Kirkpatrick}, J.~D., {et~al.}
  2020{\natexlab{b}}, \apj, 899, 123

\bibitem[{{Meisner} {et~al.}(2023{\natexlab{c}}){Meisner}, {Leggett},
  {Logsdon}, {Schneider}, {Tremblin}, \& {Phillips}}]{meisner2023coldoldBD}
{Meisner}, A.~M., {Leggett}, S.~K., {Logsdon}, S.~E., {et~al.}
  2023{\natexlab{c}}, \aj, 166, 57

\bibitem[{{Meisner} {et~al.}(2021){Meisner}, {Schneider}, {Burgasser},
  {Marocco}, {Line}, {Faherty}, {Kirkpatrick}, {Caselden}, {Kuchner}, {Gelino},
  {Gagn{\'e}}, {Theissen}, {Gerasimov}, {Aganze}, {Hsu}, {Wisniewski},
  {Casewell}, {Bardalez Gagliuffi}, {Logsdon}, {Eisenhardt}, {Allers}, {Debes},
  {Allen}, {Stevnbak Andersen}, {Goodman}, {Gramaize}, {Martin}, {Sainio},
  {Cushing}, \& {Backyard Worlds: Planet 9 Collaboration}}]{meisner2021esdT}
{Meisner}, A.~M., {Schneider}, A.~C., {Burgasser}, A.~J., {et~al.} 2021, \apj,
  915, 120

\bibitem[{{Morganson} {et~al.}(2018){Morganson}, {Gruendl}, {Menanteau},
  {Carrasco Kind}, {Chen}, {Daues}, {Drlica-Wagner}, {Friedel}, {Gower},
  {Johnson}, {Johnson}, {Kessler}, {Paz-Chinch{\'o}n}, {Petravick}, {Pond},
  {Yanny}, {Allam}, {Armstrong}, {Barkhouse}, {Bechtol}, {Benoit-L{\'e}vy},
  {Bernstein}, {Bertin}, {Buckley-Geer}, {Covarrubias}, {Desai}, {Diehl},
  {Goldstein}, {Gruen}, {Li}, {Lin}, {Marriner}, {Mohr}, {Neilsen}, {Ngeow},
  {Paech}, {Rykoff}, {Sako}, {Sevilla-Noarbe}, {Sheldon}, {Sobreira}, {Tucker},
  {Wester}, \& {DES Collaboration}}]{morganson2018DESpipeline}
{Morganson}, E., {Gruendl}, R.~A., {Menanteau}, F., {et~al.} 2018, \pasp, 130,
  074501

\bibitem[{Mukherjee {et~al.}(2023{\natexlab{a}})Mukherjee, Fortney, Morley,
  Batalha, Marley, Karalidi, Visscher, Lupu, Freedman, \&
  Gharib-Nezhad}]{mukherjee2023sonora_T}
Mukherjee, S., Fortney, J., Morley, C., {et~al.} 2023{\natexlab{a}}, {The
  Sonora Substellar Atmosphere Models. IV. Elf Owl: Atmospheric Mixing and
  Chemical Disequilibrium with Varying Metallicity and C/O Ratios (T- type
  Models)}

\bibitem[{Mukherjee {et~al.}(2023{\natexlab{b}})Mukherjee, Fortney, Morley,
  Batalha, Marley, Karalidi, Visscher, Lupu, Freedman, \&
  Gharib-Nezhad}]{mukherjee2023sonora_Y}
Mukherjee, S., Fortney, J., Morley, C., {et~al.} 2023{\natexlab{b}}, {The
  Sonora Substellar Atmosphere Models. IV. Elf Owl: Atmospheric Mixing and
  Chemical Disequilibrium with Varying Metallicity and C/O Ratios (Y- type
  Models)}

\bibitem[{{Mukherjee} {et~al.}(2024){Mukherjee}, {Fortney}, {Morley},
  {Batalha}, {Marley}, {Karalidi}, {Visscher}, {Lupu}, {Freedman}, \&
  {Gharib-Nezhad}}]{mukherjee2024sonora}
{Mukherjee}, S., {Fortney}, J.~J., {Morley}, C.~V., {et~al.} 2024, \apj, 963,
  73

\bibitem[{{Murray} {et~al.}(2011){Murray}, {Burningham}, {Jones}, {Pinfield},
  {Lucas}, {Leggett}, {Tinney}, {Day-Jones}, {Weights}, {Lodieu}, {P{\'e}rez
  Prieto}, {Nickson}, {Zhang}, {Clarke}, {Jenkins}, \&
  {Tamura}}]{murray_burningham2011blueT}
{Murray}, D.~N., {Burningham}, B., {Jones}, H.~R.~A., {et~al.} 2011, \mnras,
  414, 575

\bibitem[{Nakajima {et~al.}(1995)Nakajima, Oppenheimer, Kulkarni, Golimowski,
  Matthews, \& Durrance}]{nakajima1995discovery}
Nakajima, T., Oppenheimer, B., Kulkarni, S., {et~al.} 1995, nature, 378, 463

\bibitem[{{Newton} {et~al.}(2014){Newton}, {Charbonneau}, {Irwin},
  {Berta-Thompson}, {Rojas-Ayala}, {Covey}, \& {Lloyd}}]{newton2014feh447dM}
{Newton}, E.~R., {Charbonneau}, D., {Irwin}, J., {et~al.} 2014, \aj, 147, 20

\bibitem[{{Oke}(1974)}]{oke1974ABsystem}
{Oke}, J.~B. 1974, \apjs, 27, 21

\bibitem[{{Onken} {et~al.}(2024){Onken}, {Wolf}, {Bessell}, {Chang}, {Luvaul},
  {Tonry}, {White}, \& {Da Costa}}]{onken2024skymapper_arxiv}
{Onken}, C.~A., {Wolf}, C., {Bessell}, M.~S., {et~al.} 2024, arXiv e-prints,
  arXiv:2402.02015

\bibitem[{{Park} {et~al.}(2021){Park}, {Folkner}, {Williams}, \&
  {Boggs}}]{park2021JPL_DE441}
{Park}, R.~S., {Folkner}, W.~M., {Williams}, J.~G., \& {Boggs}, D.~H. 2021,
  \aj, 161, 105

\bibitem[{{Pavlenko} {et~al.}(2007){Pavlenko}, {Zhukovska}, \&
  {Volobuev}}]{pavlenko2007alkali}
{Pavlenko}, Y.~V., {Zhukovska}, S.~V., \& {Volobuev}, M. 2007, Astronomy
  Reports, 51, 282

\bibitem[{{Phillips} {et~al.}(2020){Phillips}, {Tremblin}, {Baraffe},
  {Chabrier}, {Allard}, {Spiegelman}, {Goyal}, {Drummond}, \&
  {H{\'e}brard}}]{phillips2020TYatm}
{Phillips}, M.~W., {Tremblin}, P., {Baraffe}, I., {et~al.} 2020, \aap, 637, A38

\bibitem[{{Pinfield} {et~al.}(2014){Pinfield}, {Gomes}, {Day-Jones}, {Leggett},
  {Gromadzki}, {Burningham}, {Ruiz}, {Kurtev}, {Cattermole}, {Cardoso},
  {Lodieu}, {Faherty}, {Littlefair}, {Smart}, {Irwin}, {Clarke}, {Smith},
  {Lucas}, {G{\'a}lvez-Ortiz}, {Jenkins}, {Jones}, {Rebolo}, {B{\'e}jar}, \&
  {Gauza}}]{pinfield2014subdwarf}
{Pinfield}, D.~J., {Gomes}, J., {Day-Jones}, A.~C., {et~al.} 2014, \mnras, 437,
  1009

\bibitem[{Rebolo {et~al.}(1995)Rebolo, Osorio, \&
  Mart{\'\i}n}]{rebolo1995discovery}
Rebolo, R., Osorio, M., \& Mart{\'\i}n, E. 1995, Nature, 377, 129

\bibitem[{{Rodrigo} \& {Solano}(2020)}]{rodrigo2020SVOfilter}
{Rodrigo}, C. \& {Solano}, E. 2020, in XIV.0 Scientific Meeting (virtual) of
  the Spanish Astronomical Society, 182

\bibitem[{{Rodrigo} {et~al.}(2012){Rodrigo}, {Solano}, \&
  {Bayo}}]{rodrigo2012SVOfilter}
{Rodrigo}, C., {Solano}, E., \& {Bayo}, A. 2012, {SVO Filter Profile Service
  Version 1.0}, IVOA Working Draft 15 October 2012

\bibitem[{{Rojas-Ayala} {et~al.}(2012){Rojas-Ayala}, {Covey}, {Muirhead}, \&
  {Lloyd}}]{rojas-ayala2012metallicity_Kband_M}
{Rojas-Ayala}, B., {Covey}, K.~R., {Muirhead}, P.~S., \& {Lloyd}, J.~P. 2012,
  \apj, 748, 93

\bibitem[{{Sanghi} {et~al.}(2023){Sanghi}, {Liu}, {Best}, {Dupuy}, {Siverd},
  {Zhang}, {Hurt}, {Magnier}, {Aller}, \&
  {Deacon}}]{sanghi2023hawaii_parallax_ucd}
{Sanghi}, A., {Liu}, M.~C., {Best}, W. M.~J., {et~al.} 2023, \apj, 959, 63

\bibitem[{{Saumon} {et~al.}(1994){Saumon}, {Bergeron}, {Lunine}, {Hubbard}, \&
  {Burrows}}]{saumon1994cia}
{Saumon}, D., {Bergeron}, P., {Lunine}, J.~I., {Hubbard}, W.~B., \& {Burrows},
  A. 1994, \apj, 424, 333

\bibitem[{{Schneider} {et~al.}(2020){Schneider}, {Burgasser}, {Gerasimov},
  {Marocco}, {Gagn{\'e}}, {Goodman}, {Beaulieu}, {Pendrill}, {Rothermich},
  {Sainio}, {Kuchner}, {Caselden}, {Meisner}, {Faherty}, {Mamajek}, {Hsu},
  {Greco}, {Cushing}, {Kirkpatrick}, {Bardalez-Gagliuffi}, {Logsdon}, {Allers},
  {Debes}, \& {Backyard Worlds: Planet 9
  Collaboration}}]{Schneider2020W0414_W1810}
{Schneider}, A.~C., {Burgasser}, A.~J., {Gerasimov}, R., {et~al.} 2020, \apj,
  898, 77

\bibitem[{{Schneider} {et~al.}(2015){Schneider}, {Cushing}, {Kirkpatrick},
  {Gelino}, {Mace}, {Wright}, {Eisenhardt}, {Skrutskie}, {Griffith}, \&
  {Marsh}}]{Schneider2015HubbleWISE_BD}
{Schneider}, A.~C., {Cushing}, M.~C., {Kirkpatrick}, J.~D., {et~al.} 2015,
  \apj, 804, 92

\bibitem[{{Schneider} {et~al.}(2021){Schneider}, {Meisner}, {Gagn{\'e}},
  {Faherty}, {Marocco}, {Burgasser}, {Kirkpatrick}, {Kuchner}, {Gramaize},
  {Rothermich}, {Brooks}, {Vrba}, {Bardalez Gagliuffi}, {Caselden}, {Cushing},
  {Gelino}, {Line}, {Casewell}, {Debes}, {Aganze}, {Ayala}, {Gerasimov},
  {Gonzales}, {Hsu}, {Kiman}, {Popinchalk}, {Theissen}, \& {Backyard Worlds:
  Planet 9 Collaboration}}]{schneider2021Ross19B}
{Schneider}, A.~C., {Meisner}, A.~M., {Gagn{\'e}}, J., {et~al.} 2021, \apj,
  921, 140

\bibitem[{{Scholz}(2010)}]{scholz2010ULAS1416}
{Scholz}, R.~D. 2010, \aap, 510, L8

\bibitem[{{Sivarani} {et~al.}(2009){Sivarani}, {L{\'e}pine}, {Kembhavi}, \&
  {Gupchup}}]{sivarani2009sdL}
{Sivarani}, T., {L{\'e}pine}, S., {Kembhavi}, A.~K., \& {Gupchup}, J. 2009,
  \apjl, 694, L140

\bibitem[{{Solano} {et~al.}(2021){Solano}, {G{\'a}lvez-Ortiz}, {Mart{\'\i}n},
  {G{\'o}mez Mu{\~n}oz}, {Rodrigo}, {Burgasser}, {Lodieu}, {B{\'e}jar},
  {Hu{\'e}lamo}, {Morales-Calder{\'o}n}, \& {Bouy}}]{solano2021virtual}
{Solano}, E., {G{\'a}lvez-Ortiz}, M.~C., {Mart{\'\i}n}, E.~L., {et~al.} 2021,
  \mnras, 501, 281

\bibitem[{{Stetson}(1987)}]{stetson1987daophot}
{Stetson}, P.~B. 1987, \pasp, 99, 191

\bibitem[{{Tody}(1986)}]{tody1986iraf}
{Tody}, D. 1986, in Society of Photo-Optical Instrumentation Engineers (SPIE)
  Conference Series, Vol. 627, Instrumentation in astronomy VI, ed. D.~L.
  {Crawford}, 733

\bibitem[{{Tody}(1993)}]{tody1993iraf}
{Tody}, D. 1993, in Astronomical Society of the Pacific Conference Series,
  Vol.~52, Astronomical Data Analysis Software and Systems II, ed. R.~J.
  {Hanisch}, R.~J.~V. {Brissenden}, \& J.~{Barnes}, 173

\bibitem[{{Weinberg} {et~al.}(1987){Weinberg}, {Shapiro}, \&
  {Wasserman}}]{weinberg1987binary_fate}
{Weinberg}, M.~D., {Shapiro}, S.~L., \& {Wasserman}, I. 1987, \apj, 312, 367

\bibitem[{{Woolf} \& {Wallerstein}(2006)}]{woolf_wallerstein2006feh_M_molecule}
{Woolf}, V.~M. \& {Wallerstein}, G. 2006, \pasp, 118, 218

\bibitem[{{Zhang} {et~al.}(2023){Zhang}, {Lodieu}, \&
  {Mart{\'\i}n}}]{zhangjerry2023optical_sdT}
{Zhang}, J.~Y., {Lodieu}, N., \& {Mart{\'\i}n}, E.~L. 2023, \aap, 678, A105

\bibitem[{{Zhang} {et~al.}(2019){Zhang}, {Burgasser}, {G{\'a}lvez-Ortiz},
  {Lodieu}, {Zapatero Osorio}, {Pinfield}, \&
  {Allard}}]{zhangzenghua2019metal-poor_sdT}
{Zhang}, Z.~H., {Burgasser}, A.~J., {G{\'a}lvez-Ortiz}, M.~C., {et~al.} 2019,
  \mnras, 486, 1260

\bibitem[{{Zhang} {et~al.}(2018){Zhang}, {Galvez-Ortiz}, {Pinfield},
  {Burgasser}, {Lodieu}, {Jones}, {Mart{\'\i}n}, {Burningham}, {Homeier},
  {Allard}, {Zapatero Osorio}, {Smith}, {Smart}, {L{\'o}pez Mart{\'\i}},
  {Marocco}, \& {Rebolo}}]{zhangzenghua2018subL}
{Zhang}, Z.~H., {Galvez-Ortiz}, M.~C., {Pinfield}, D.~J., {et~al.} 2018,
  \mnras, 480, 5447

\bibitem[{{Zhang} {et~al.}(2017){Zhang}, {Pinfield}, {G{\'a}lvez-Ortiz},
  {Burningham}, {Lodieu}, {Marocco}, {Burgasser}, {Day-Jones}, {Allard},
  {Jones}, {Homeier}, {Gomes}, \& {Smart}}]{zhang2017six_sdL_classification}
{Zhang}, Z.~H., {Pinfield}, D.~J., {G{\'a}lvez-Ortiz}, M.~C., {et~al.} 2017,
  \mnras, 464, 3040

\end{thebibliography}
% your references Yourfile.bib

\onecolumn

\begin{appendix}
\section{Astrometry for five metal-poor T dwarf candidates.}
\begin{figure}[htbp]
     \centering
     \begin{subfigure}[b]{0.497\textwidth}
         \centering
         \includegraphics[width=\textwidth]{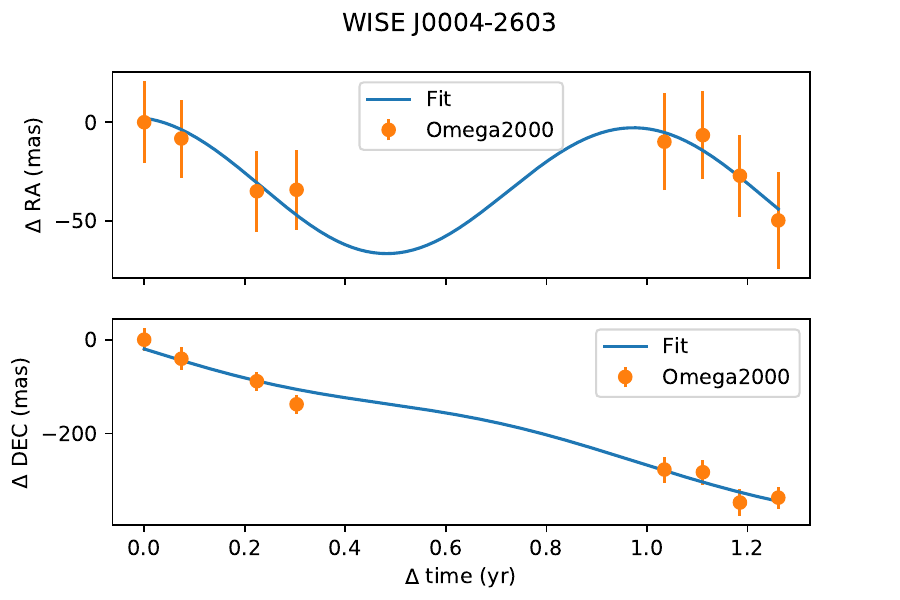}
         \label{W0004_ast1}
     \end{subfigure}
     \begin{subfigure}[b]{0.497\textwidth}
         \centering
         \includegraphics[width=\textwidth]{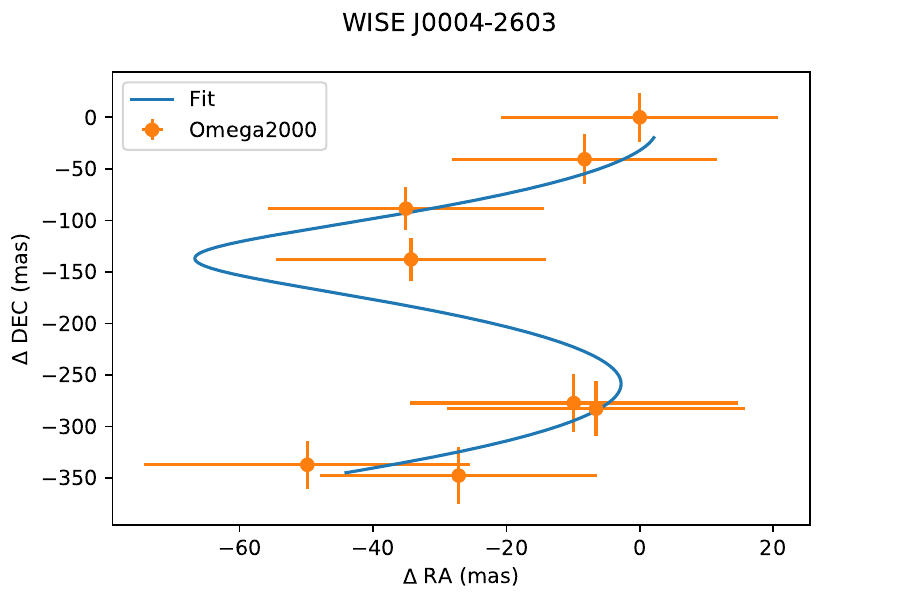}
         \label{W0004_ast2}
     \end{subfigure}
     \\
          \begin{subfigure}[b]{0.497\textwidth}
         \centering
         \includegraphics[width=\textwidth]{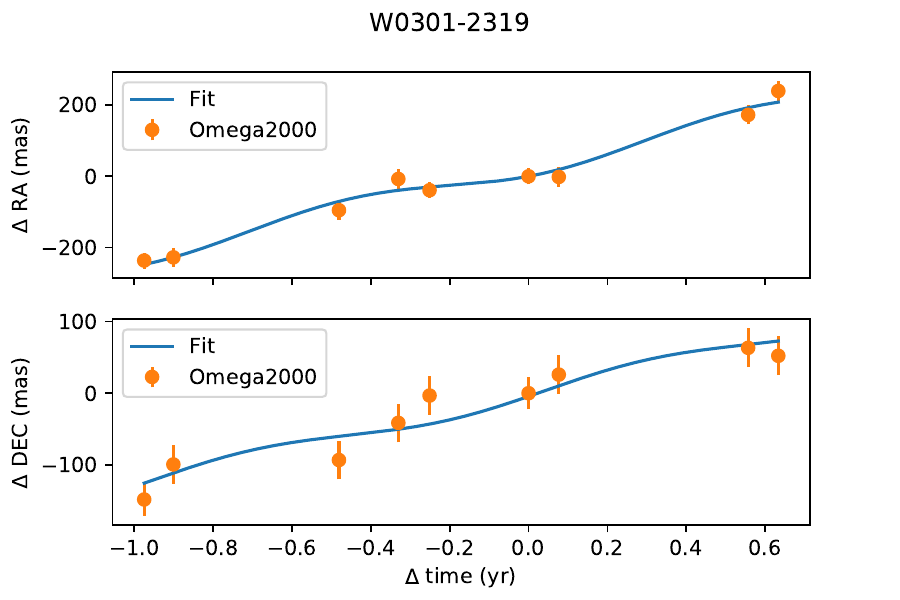}
         \label{W0301_ast1}
     \end{subfigure}
     \begin{subfigure}[b]{0.497\textwidth}
         \centering
         \includegraphics[width=\textwidth]{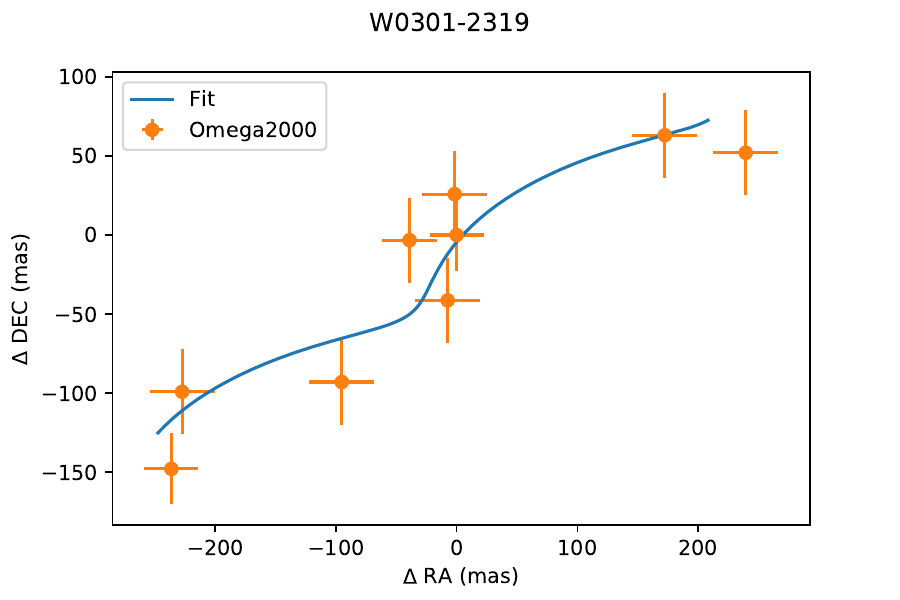}
         \label{W0301_ast2}
     \end{subfigure}
     \\
     \begin{subfigure}[b]{0.497\textwidth}
         \centering
         \includegraphics[width=\textwidth]{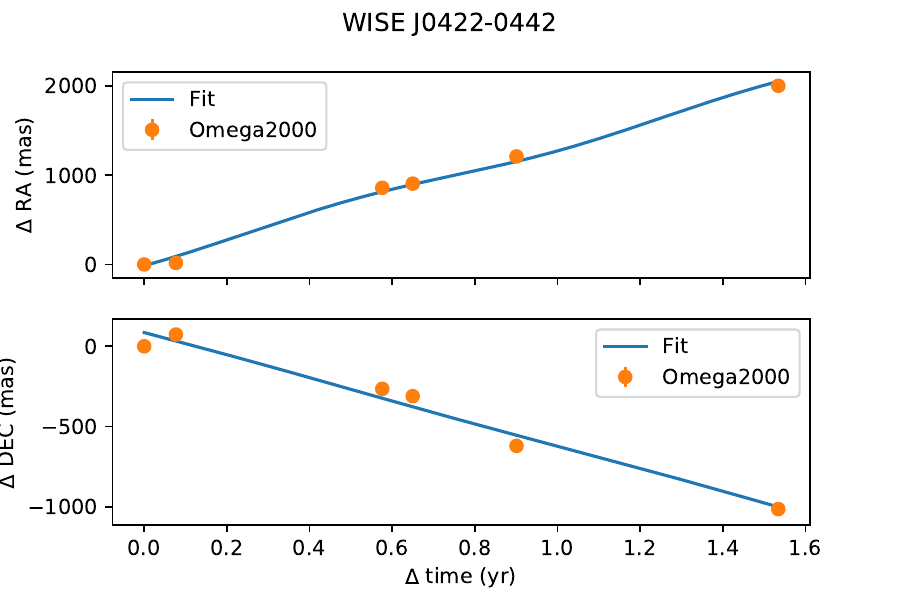}
         \label{W0422_ast1}
     \end{subfigure}
     \begin{subfigure}[b]{0.497\textwidth}
         \centering
         \includegraphics[width=\textwidth]{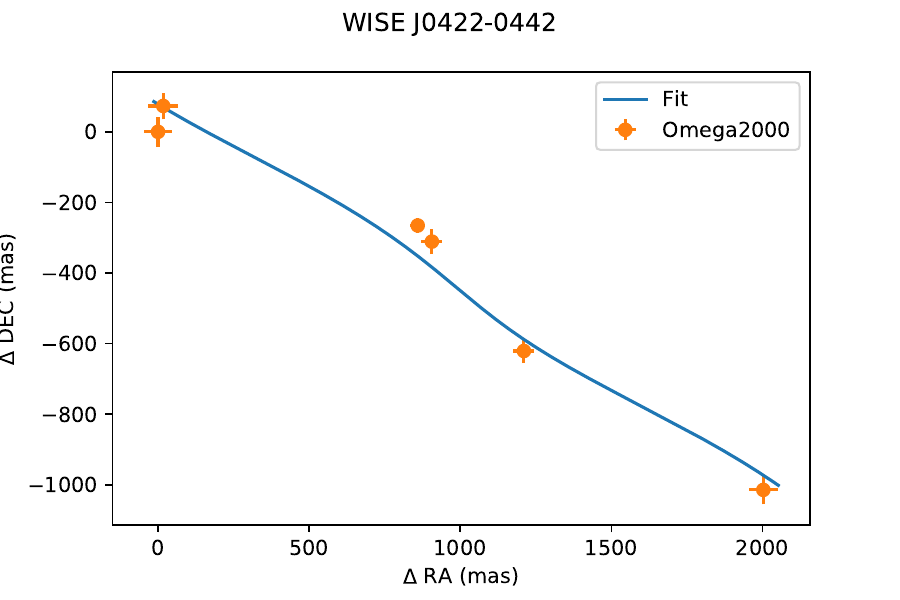}
         \label{W0422_ast2}
     \end{subfigure}

        \caption{%Observations obtained from different instrument configurations are denoted in different colours.   
        \textit{Left:} observed relative astrometry separated in RA and Dec against the relative time of the reference epoch (0.0 year). \textit{Right:} observed relative astrometry together in RA and Dec. }
        \label{fig:parallax}
\end{figure}

\begin{figure}[htbp]
\ContinuedFloat
     \centering
     \begin{subfigure}[b]{0.497\linewidth}
         \centering
         \includegraphics[width=\textwidth]{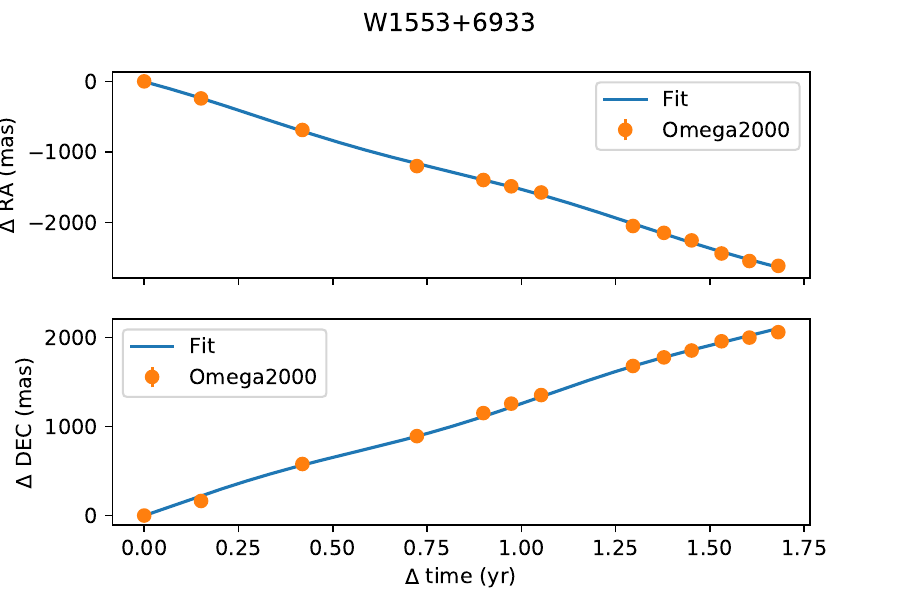}
         \label{W1553_ast1}
     \end{subfigure}
     \begin{subfigure}[b]{0.497\textwidth}
         \centering
         \includegraphics[width=\textwidth]{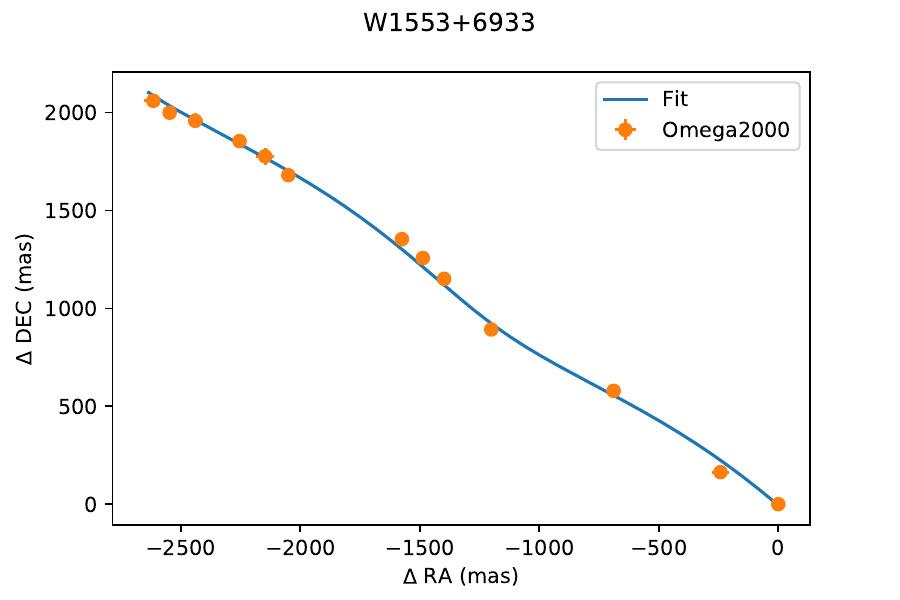}
         \label{W1553_ast2}
     \end{subfigure}
     \\
     \begin{subfigure}[b]{0.497\textwidth}
         \centering
         \includegraphics[width=\textwidth]{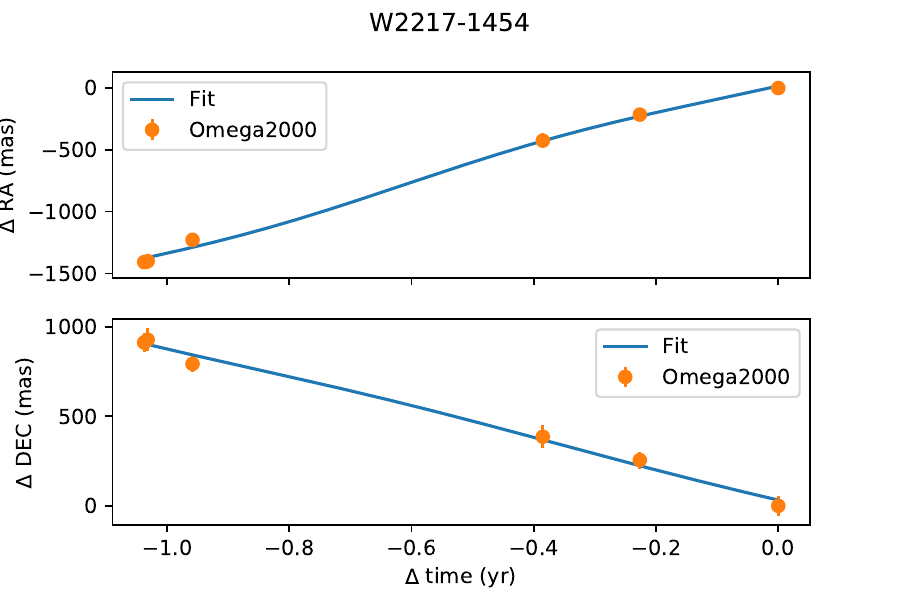}
         \label{W2217_ast1}
     \end{subfigure}
     \begin{subfigure}[b]{0.497\textwidth}
         \centering
         \includegraphics[width=\textwidth]{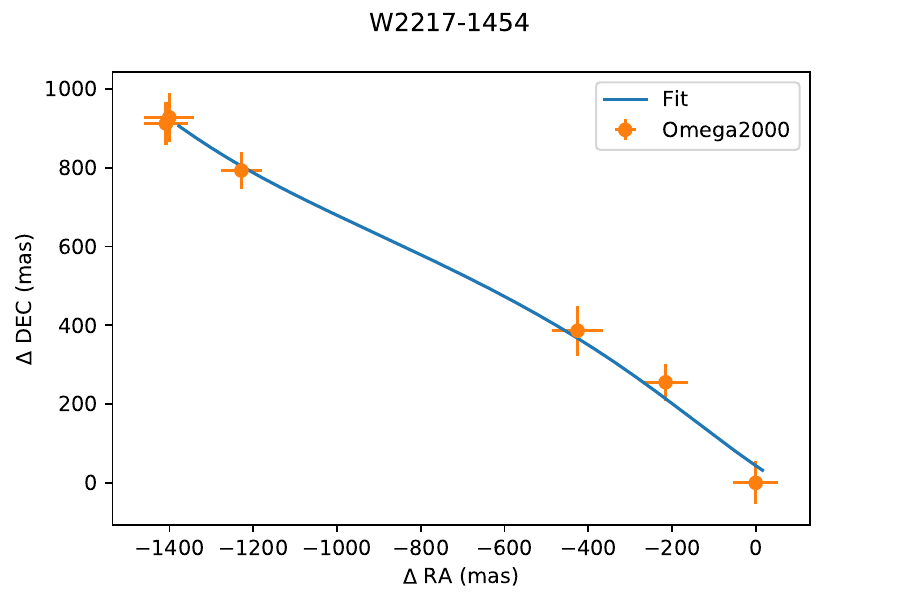}
         \label{W2217_ast2}
     \end{subfigure}
        \caption{Continued.}
        \label{fig:parallax2}
\end{figure}

\clearpage 
\section{GTC \& VLT optical images of metal-poor T and Y dwarf candidates.}
\begin{figure}[htbp]
     \centering
     \begin{subfigure}[b]{0.326\textwidth}
         \centering
         \includegraphics[width=\textwidth]{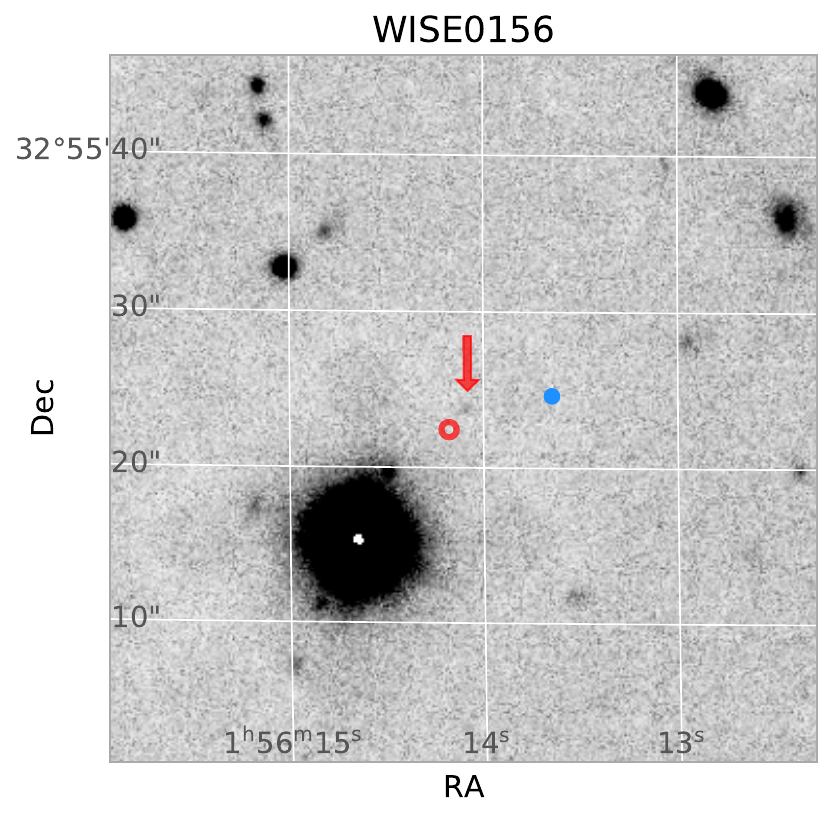}
         \label{W0156_illu}
        \end{subfigure}
    \begin{subfigure}[b]{0.326\textwidth}
         \centering
         \includegraphics[width=\textwidth]{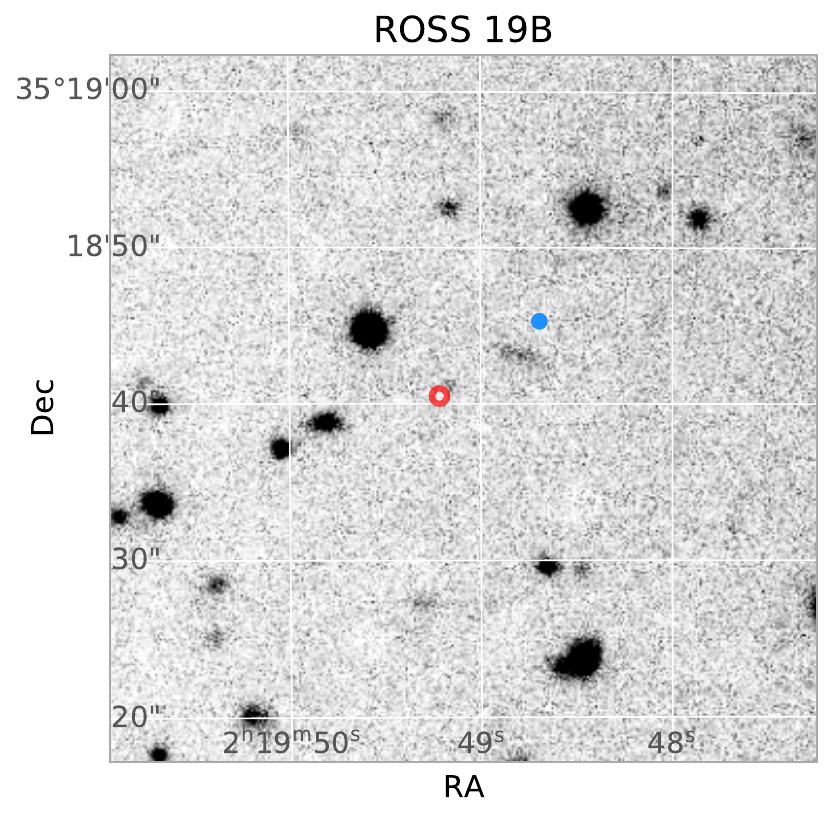}
         \label{ROSS19B_illu}
        \end{subfigure}
     \begin{subfigure}[b]{0.33\textwidth}
         \centering
         \includegraphics[width=\textwidth]{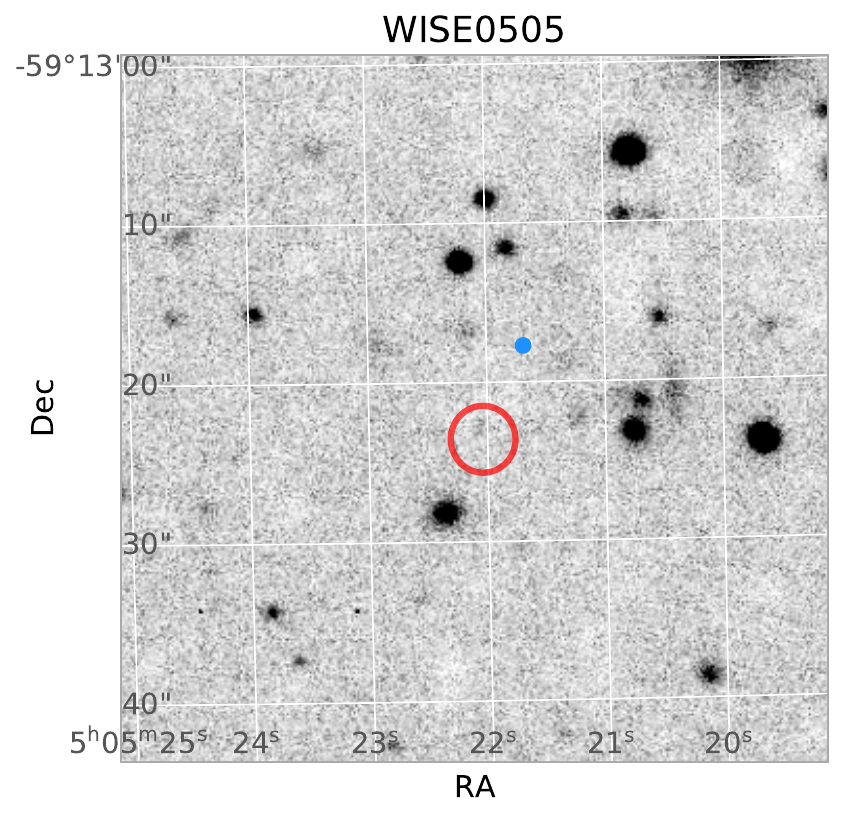}
         \label{W0505_illu}
        \end{subfigure}
        \\
     \begin{subfigure}[b]{0.326\textwidth}
         \centering
         \includegraphics[width=\textwidth]{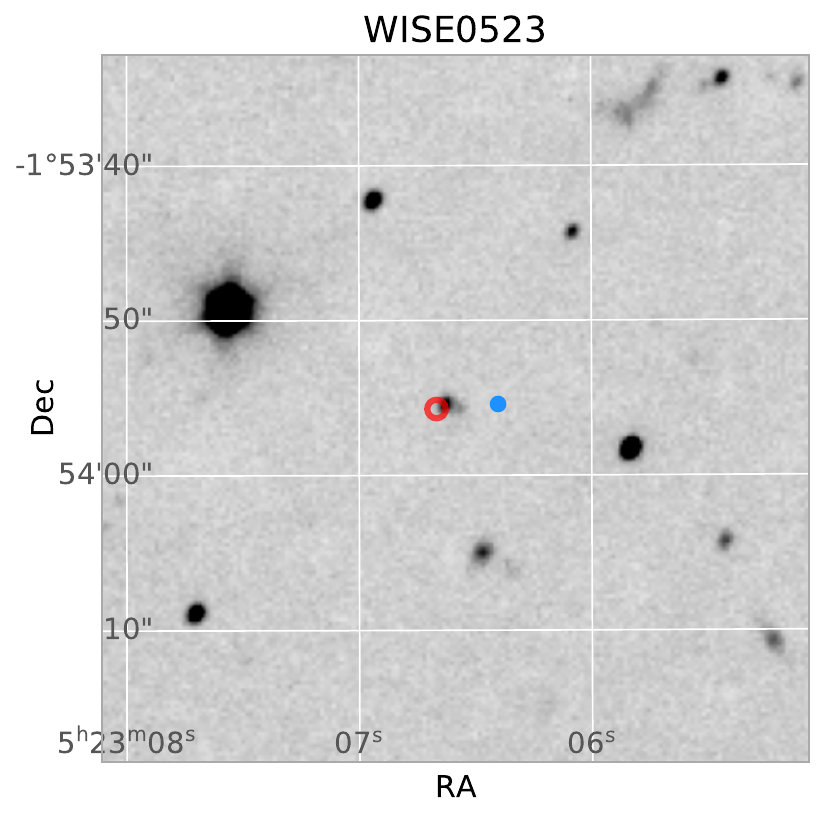}
         \label{W0523_illu}
     \end{subfigure}
    \begin{subfigure}[b]{0.33\textwidth}
         \centering
         \includegraphics[width=\textwidth]{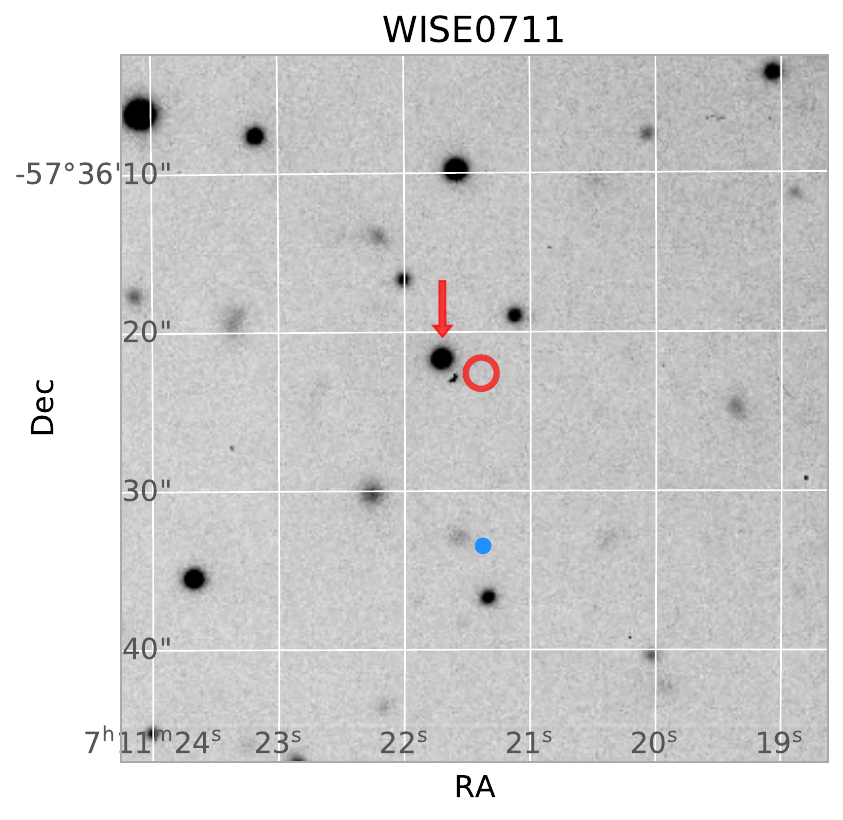}
         \label{W0711_illu}
     \end{subfigure}
    \begin{subfigure}[b]{0.33\textwidth}
         \centering
         \includegraphics[width=\textwidth]{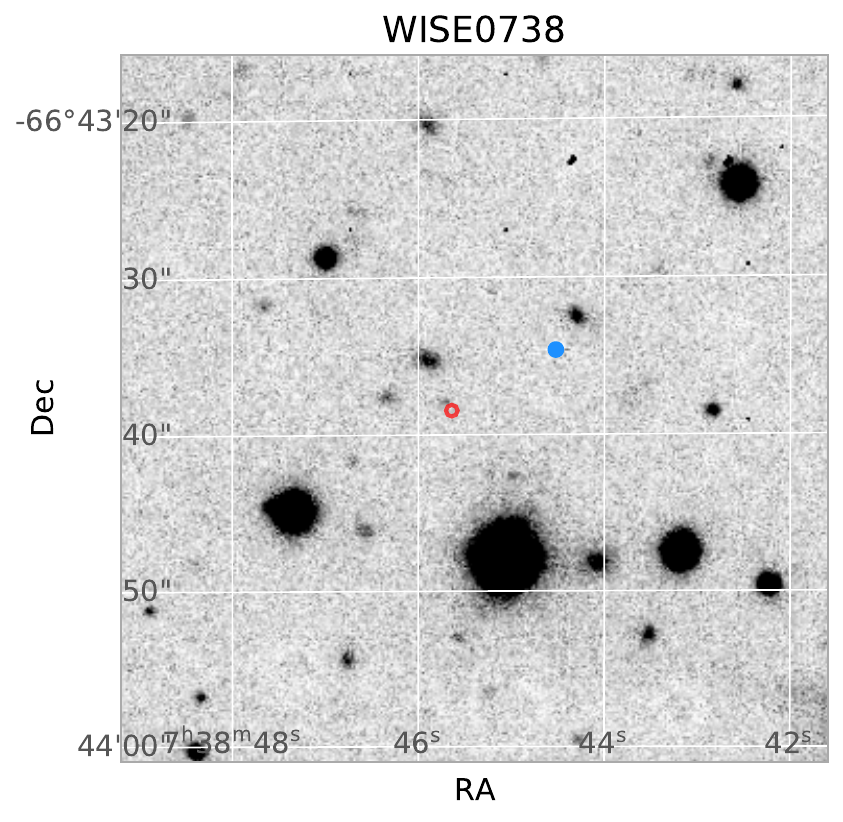}
         \label{W0738_illu}
     \end{subfigure}
      \\
     \begin{subfigure}[b]{0.33\textwidth}
         \centering
         \includegraphics[width=\textwidth]{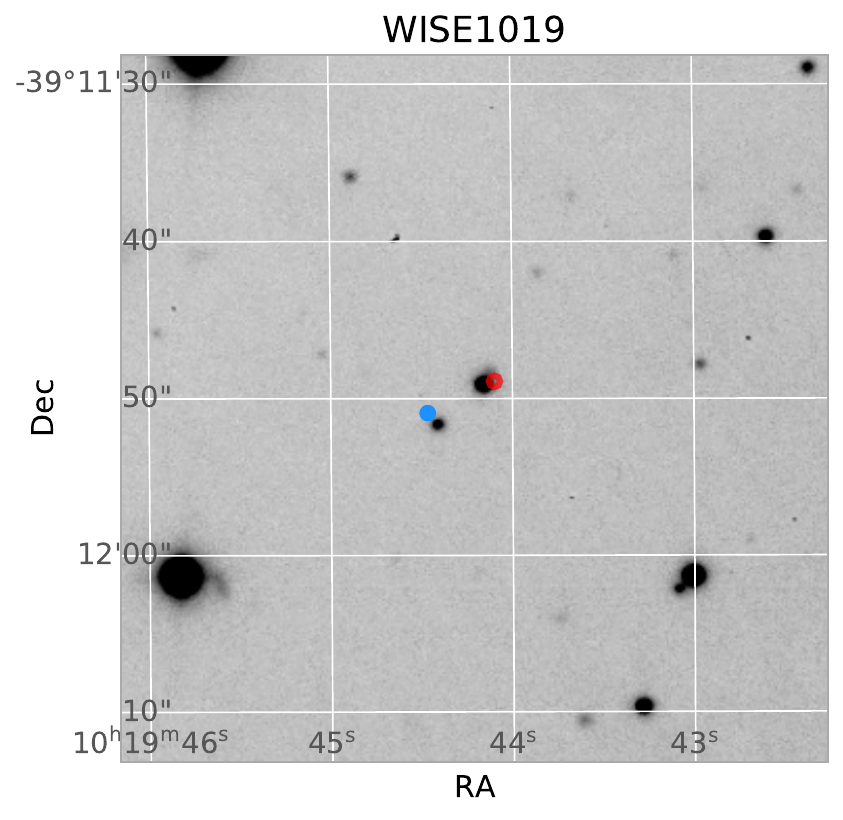}
         \label{W1019_illu}
     \end{subfigure}
     \begin{subfigure}[b]{0.33\textwidth}
         \centering
         \includegraphics[width=\textwidth]{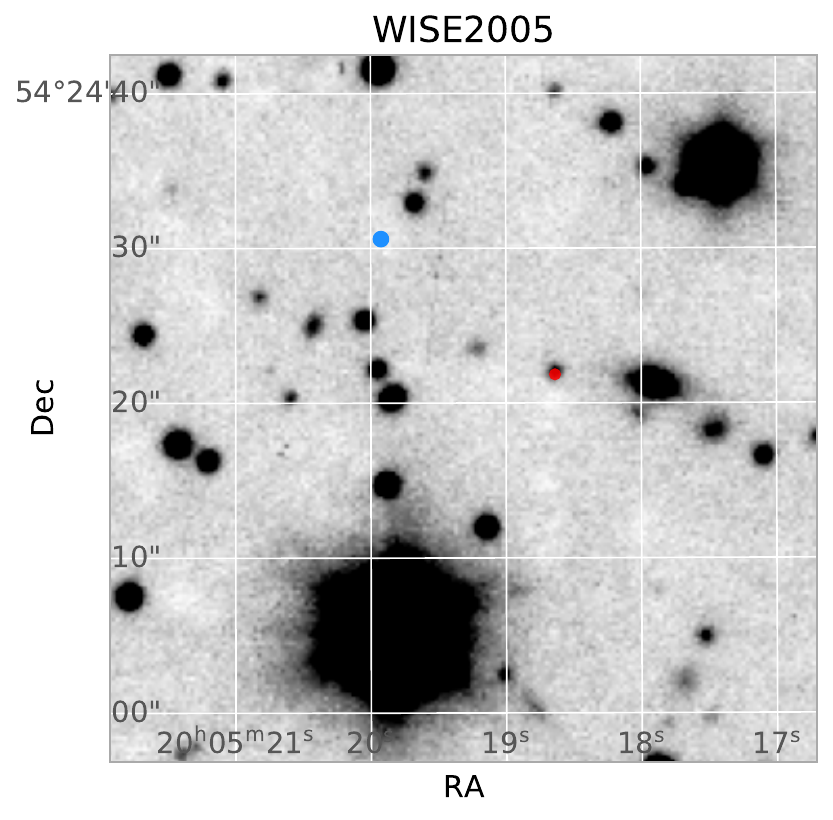}
         \label{W2005_illu}
     \end{subfigure}
     \begin{subfigure}[b]{0.33\textwidth}
         \centering
         \includegraphics[width=\textwidth]{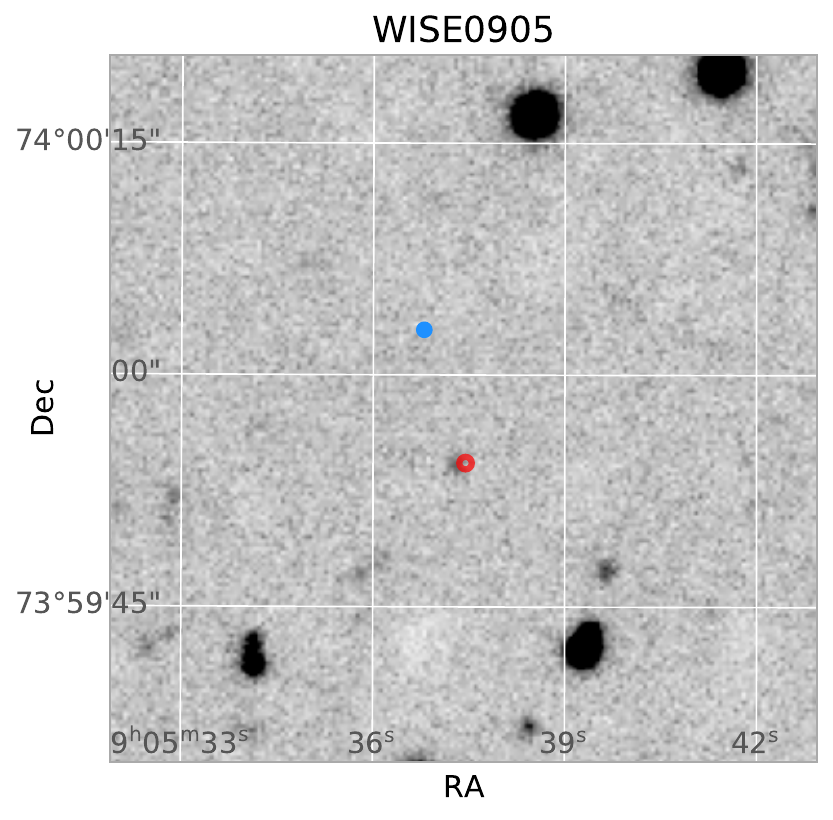}
         \label{W0905_illu}
     \end{subfigure}
        \caption{45\arcsec$\times$45\arcsec DES, GTC/OSIRIS, and VLT/FORS2 $z$-band images of the fields of all the targets with conventional orientation (up is North and left is East). The blue dots are the object positions at the previous epochs published by other authors (Table~\ref{tg}), and the red ellipses are the projected positions where those targets are supposed to be on the dates of the observations according to their proper motions from the literature in Table~\ref{tg}. Parallax effects are negligible. The two semi-axes of the ellipses show the errors of projected position in RA and Dec, based on proper motion uncertainties and the time difference between the two epochs. All the objects have been detected. The correct positions of W0156 and W0711 are pointed out by red arrows.}
        \label{fig:projection}
\end{figure}

\begin{figure}[htbp]
\ContinuedFloat
     \centering
     \begin{subfigure}[b]{0.323\textwidth}
         \centering
         \includegraphics[width=\textwidth]{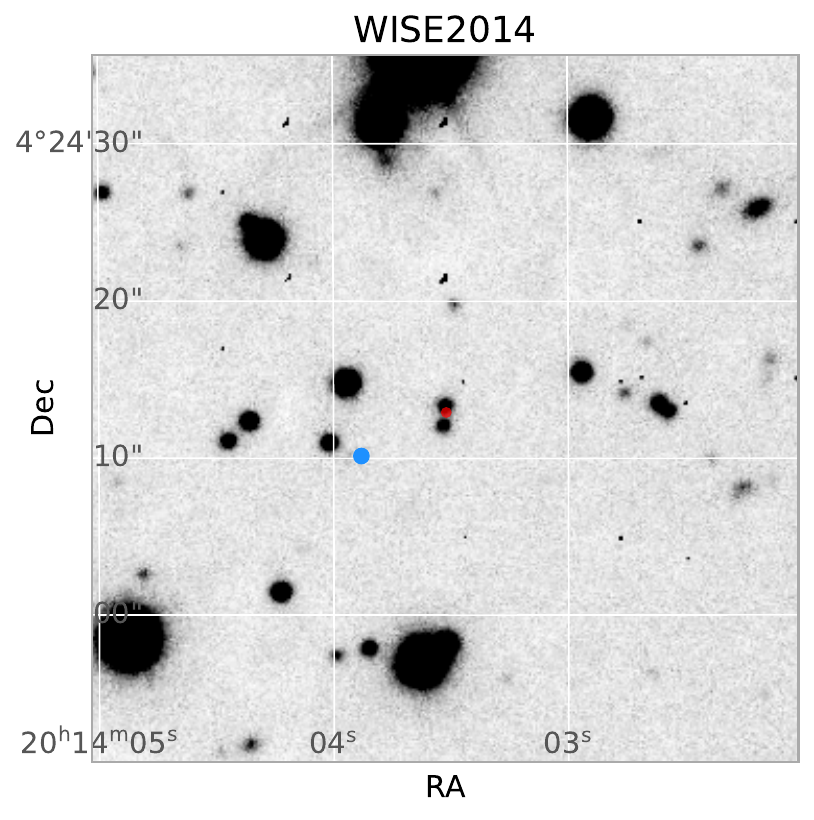}
         \label{W2014_illu}
     \end{subfigure}
    \begin{subfigure}[b]{0.33\textwidth}
         \centering
         \includegraphics[width=\textwidth]{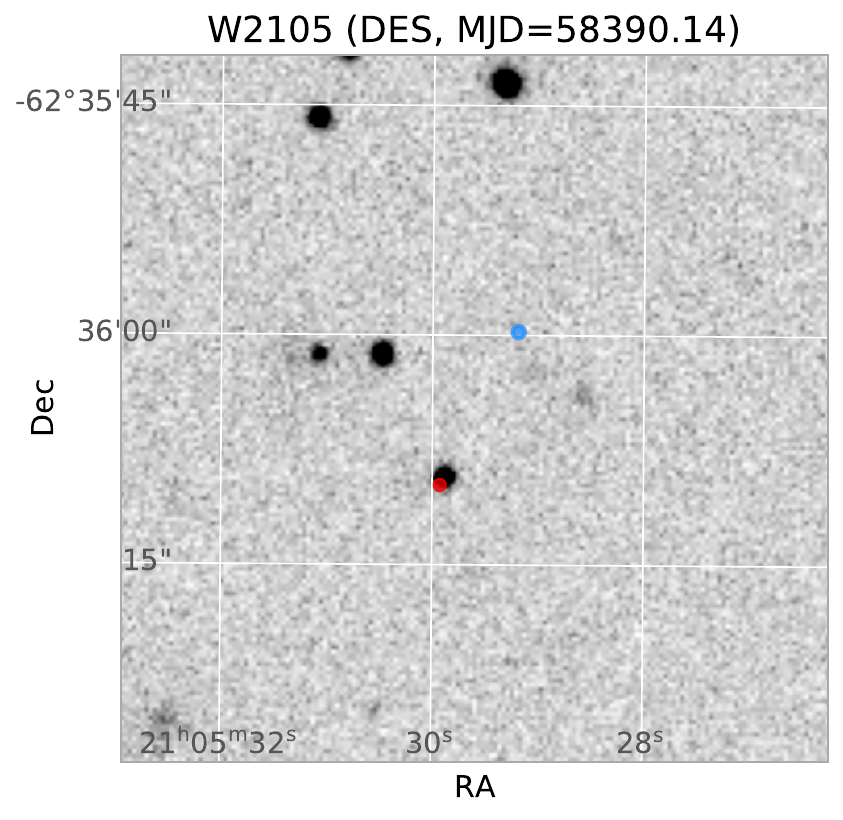}
         \label{W2105_illu}
     \end{subfigure}
     \begin{subfigure}[b]{0.33\textwidth}
         \centering
         \includegraphics[width=\textwidth]{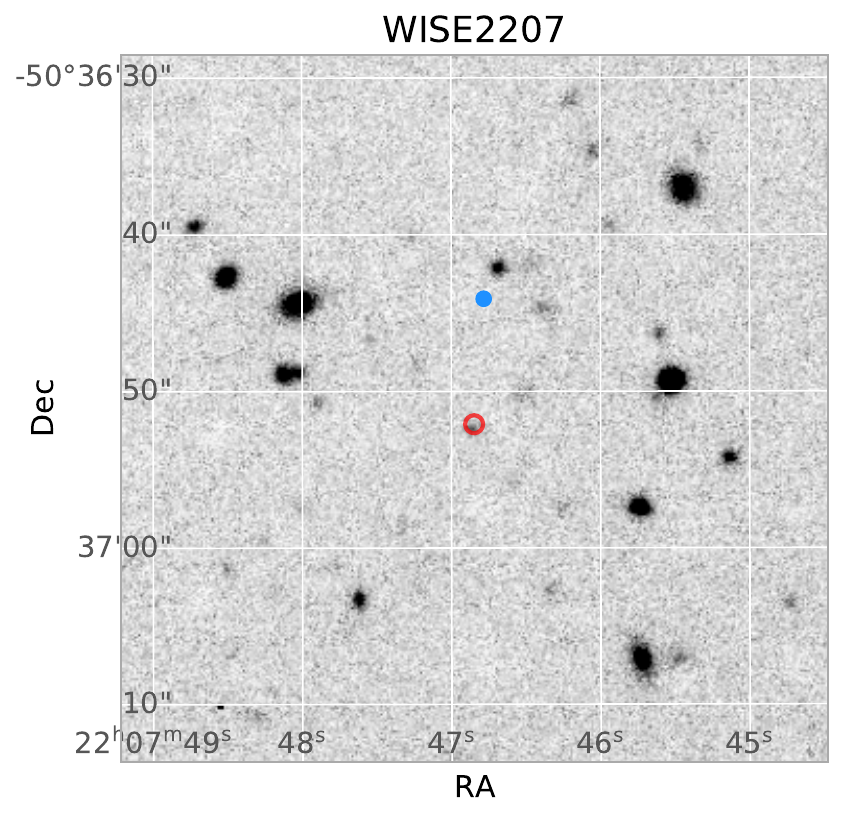}
         \label{W2217_illu}
     \end{subfigure}
        \caption{Continued.}
        \label{fig:projection2}
\end{figure}

\begin{figure}[htbp]
\centering
\includegraphics[width=1\textwidth]{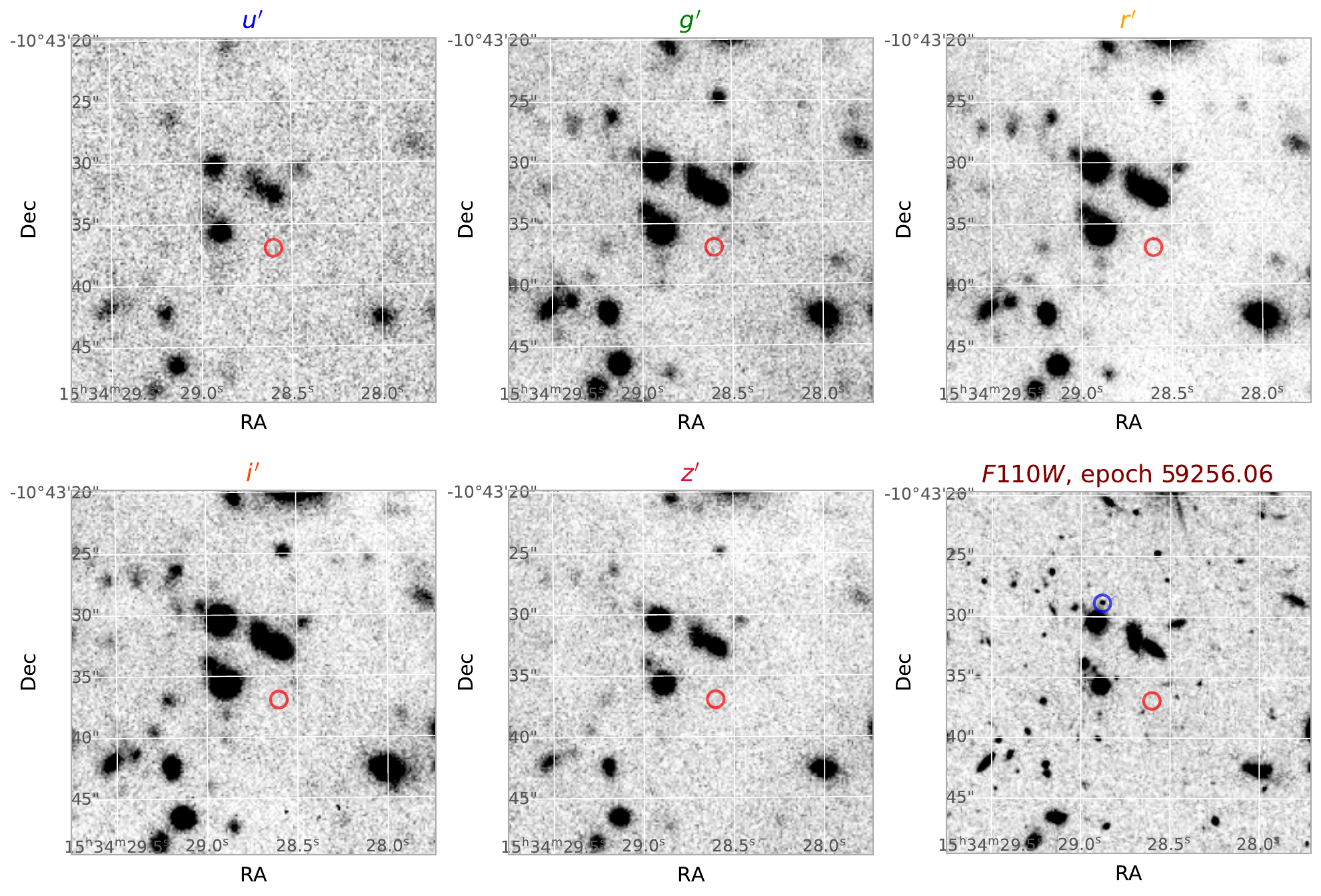}
\caption{30\arcsec$\times$30\arcsec GTC/HiPERCAM quintuple-band optical images of the Accident field with red circles indicating the expected position at the HiPERCAM epoch, compared with HST F110W image (right bottom) with blue circle indicating the Accident at the HST observation epoch. The Accident was not detected in any of the bands.}
\label{fig:acc_projection}
\end{figure}

\end{appendix}

\end{document}